\documentclass{aa}

\usepackage{natbib}
\bibpunct{(}{)}{;}{a}{}{,} 
\usepackage{graphics}   
\usepackage{graphicx}   
\usepackage{epstopdf}
\usepackage{amsmath}
\usepackage{longtable}   
\usepackage{url}      
\usepackage{bm}        
\usepackage[varg]{txfonts}
\usepackage{pdflscape}
\usepackage{caption}
\usepackage{supertabular}
\usepackage{hyperref}   
\usepackage{siunitx} 
\usepackage{textgreek} 
\usepackage{upgreek}
\usepackage{subcaption}

\usepackage{tikz}

\usetikzlibrary{arrows, decorations.markings}
      \tikzstyle{vecArrow} = [thick, decoration={markings,mark=at position
      1 with {\arrow[semithick]{open triangle 60}}},
      double distance=1.4pt, shorten >= 5.5pt,
      preaction = {decorate},
      postaction = {draw,line width=1.4pt, white,shorten >= 4.5pt}]
     \tikzstyle{innerWhite} = [semithick, white,line width=1.4pt, shorten >= 4.5pt]

\usetikzlibrary{positioning}

\newcommand{\commentaire}[1]{}

\hypersetup{
    bookmarks=true,         
    unicode=false,          
    colorlinks=true,       
    linkcolor=blue,          
    citecolor=blue,        
    filecolor=blue,      
    urlcolor=blue           
}
\usepackage{lipsum,graphicx,subcaption}
\makeatletter
\newcommand\appendix@section[1]{%
\refstepcounter{section}%
\orig@section*{Appendix \@Alph\c@section: #1}%
\addcontentsline{toc}{section}{Appendix \@Alph\c@section: #1}%
}
\let\orig@section\section
\g@addto@macro\appendix{\let\section\appendix@section}
\makeatother

\begin{document}

\title{A new wavelength calibration for echelle spectrographs using Fabry-P\'erot etalons\\}
\subtitle{}

\author{F. Cersullo\inst{1,*} \& A. Coffinet\inst{1,**}, B. Chazelas\inst{1}, C. Lovis\inst{1}, and F. Pepe \inst{1}}

 \titlerunning{A new wavelength calibration for echelle spectrograph using Fabry-P\'erot etalons}

 \authorrunning{F. Cersullo \& A. Coffinet et al.}

 \institute{\inst{1} Observatoire Astronomique de l'Universit\'e de Gen\`eve, 51 Ch. des Maillettes, 1290 Versoix, Switzerland\\
 * \email{maria.cersullo@unige.ch}, 
 ** \email{adrien.coffinet@unige.ch}
}

\date{Received /
Accepted}

\abstract  {The study of Earth-mass extrasolar planets via the radial-velocity technique and the measurement of the potential cosmological variability of fundamental constants call for very-high-precision spectroscopy at the level of $\updelta\lambda/\lambda<10^{-9}$. Only an accurate wavelength calibration of the spectrograph can guarantee that the aimed precision is achieved over a multi-exposure and multi-epoch data set. Wavelength accuracy is obtained by providing two fundamental ingredients: 1) an absolute and information-rich wavelength source and 2) the ability of the spectrograph and its data reduction of transferring the reference scale (wavelengths) to a measurement scale (detector pixels) in a repeatable manner.}
{The goal of this work is to improve the wavelength calibration accuracy of the HARPS spectrograph by combining the \emph{absolute} spectral reference provided by the emission lines of a thorium-argon hollow-cathode lamp (HCL) with the spectrally rich and \emph{precise} spectral information of a Fabry-P\'erot-based calibration source.}
{On the basis of calibration frames acquired each night since the Fabry-P\'erot etalon was installed on HARPS in 2011, we construct a combined wavelength solution which fits simultaneously the thorium emission lines and the Fabry-P\'erot lines. The combined fit is anchored to the absolute thorium wavelengths, which provide the ``zero-point'' of the spectrograph, while the Fabry-P\'erot lines are used to improve the (spectrally) local precision. The obtained wavelength solution is verified for auto-consistency and tested against a solution obtained using the HARPS Laser-Frequency Comb (LFC).}
{The combined thorium+Fabry-P\'erot wavelength solution shows significantly better performances compared to the thorium-only calibration. In both cases, the residuals of the LFC line positions to the fitted wavelength solution follow a Gaussian distribution with an \emph{rms} value of about \SI{14}{\m\per\s} for the combined solution, and twice as large for the thorium-only solution (\SI{29}{\m\per\s}). Given these positive results, we have applied the new calibrations to scientific frames and tested the radial-velocity residual on three well-known stars: HD~10700, HD~20794 and HD~69830. In all three cases the RV scatter could be reduced compared to the measurements using the previous calibration.} 
{The richness of the Fabry-P\'erot spectrum helps improving the wavelength calibration using thorium-argon lamps or extending the wavelength domain of LFCs with limited operational range. The presented techniques will therefore be used in the new HARPS and HARPS-N pipeline, and will be exported to the ESPRESSO spectrograph.}

\maketitle

\section{Introduction}
\indent  
The \textit{High-Accuracy Radial-velocity Planet Searcher} (HARPS) \citep{Pepe+2000,Mayor2003} is a visible-wavelength high-resolution fiber-fed echelle spectrograph located at ESO's 3.6-m telescope and is one of the most precise instruments for the measure of radial velocities in the context of exoplanet research, reaching a precision better than \SI{1}{\m\per\s}. New instruments aim at even better precision. The challenge for the next generation of spectrographs is to go towards the \si{\cm\per\s} level of precision, e. g. \SI{10}{\cm\per\s}  for ESPRESSO \citep{Pepe2014}, GHOST \citep{ghost}, EXPRES \citep{EXPRES}, G-CLEF \citep{gmt};  \SI{2}{\cm\per\s}  for HIRES, \citep{Marconi2016}.\\

In order to achieve this extreme precision, these spectrographs will have to rely on accurate wavelength calibration, i. e. to assign to each detector pixel uniquely a wavelength that allows us to determine precisely the photocenter of a spectral line. For this purpose, one has to employ spectral sources that provide a rich spectrum of lines that can be used as absolute spectral reference. \\

Today, hollow-cathode thorium lamps are employed for wavelength calibration and drift measurements in most echelle spectrographs for precise radial velocities (PRV). Indeed, they offer numerous spectral features over the whole visible and near-infrared range, narrow highly symmetric line profiles and a very stable spectrum. A large electric potential difference, applied between anode and cathode (thorium), ionizes the gas (argon) and transforms it into plasma. Under the effect of the electric field, the positive ions of the gas are accelerated towards the cathode and hit it. Following these collisions, some atoms of the cathode are sputtered and mixed with the inert gas. These atoms pass to an excited state colliding with the other particles produced during ionization and return  into the ground state, then emit photons of characteristic wavelengths that are independent from the environment. By the same process also the atoms of the inert gas can emit photons at different characteristic wavelengths. The result is a rich spectrum of thorium and argon emission lines \citep{kerber2007}.\\

The \textit{Atlas of the Thorium Spectrum} obtained by \citet{PE83} and more recently the thorium atlas made by \cite{RNS14} provide a complete list of accurate thorium and argon wavelengths that are used for wavelength calibration.
The accurate Th wavelengths of the \textit{Atlas of the Thorium Spectrum} were obtained with the McMath-Pierce \SI{1}{\m} Fourier Transform Spectrometer (FTS) of the National Solar Observatory at Kitt Peak. The more recent \cite{RNS14} combines the lines measured with the 2 m FTS at the National Institute of Standards and Technology (NIST) with several other studies. \citet{RNS14}'s list contains two sets of wavelengths: the Ritz wavelengths computed from globally optimized energy levels, and the measured wavelengths. For the same reasons as explained in \cite{Coffinet+2018}, we used these Ritz wavelengths as our reference.\\

However, several limitations hinder us to reach  precision of \si{\cm\per\s} measurements such as:
\begin{itemize}
\item{high dynamical range of line intensities}
\item{line blending in finite resolution spectrographs}
\item{aging} i.e. line shifts due to gas-pressure changes inside the HCL over its life cycle \citep{kerber2007, Seemann2014}
\item{flux variability and pollution by ``strong'' argon lines}
\item{sparse coverage of the spectral range}
\end{itemize}

Because of the aforementioned limiting factors, other calibration sources are currently under development.\\
Laser frequency combs (LFC) are undeniably the most accurate calibration source  \citep{wilken2010, Locurto2012, ravi2017}. However, the current generation of LFC has still shortcomings in term of reliability, wavelength coverage and usability in an observatory environment. For this reasons, we had proposed in the past a solution based on a Fabry-P\'erot etalon (FP) \citep{wildi2010, Chazelas2012, cersullo2017}. \\

These Fabry-P\'erot etalons have been employed mainly for drift measurements providing nevertheless internal precision of \si{\cm\per\s} and repeatability of about $\SI{10}{\cm\per\s}$ over an observing night \citep{wildi2010}. We cannot consider a passive Fabry-P\'erot as an absolute calibrator, however. Indeed, the wavelength of each FP line is not known a priori, because it is related to the etalon gap (spacing of the mirrors) being a mechanical characteristics rather than anchored to fundamental physics. Therefore, an exterior absolute reference is necessary to anchor the spectral lines. \cite{wildi2010} proposed to link the FP spectrum to an external absolute spectral reference such as an LFC or HCL. \cite{Bauer} adopted such an approach and presented algorithms that combine the HCL and the Fabry P\'erot spectra within a single wavelength calibration of echelle spectrographs. This approach has been tested on HARPS calibration spectra and eventually implemented on the Carmenes spectrograph \citep{Caballero}.
\\

In the case of HARPS, the FP module guarantees nearly equally spaced spectral lines of uniform intensity over an extremely wide spectral range. The mirror coatings introduce slow variations of the phase (causing group-delay dispersion, GDD, \citealt{wildi2010}). Nevertheless, we have shown in \citet{cersullo2017} that the GDD remains stable with time and that only one global parameter $D_0$, the average etalon gap, must be determined by using an external absolute reference. The line separation and position being perfectly smooth and continuous in wavelength, the FP lines can be used to improve the wavelength solution, especially to determine the higher-order terms of the polynomials used to describe the pixel-to-wavelength relationship (wavelength solution), which is otherwise poorly constrained by the sparse lines of hollow-cathode lamps only.

Inspired by the aforementioned works \citep{wildi2010, wildi2011, Bauer}, we present in this paper the development of the technique to obtain a locally more accurate wavelength solution by combining the (accurate) information provided by a hollow-cathode lamp with the (precise) information given by a Fabry-P\'erot etalon. Section~2 describes the characterization of the Fabry-P\'erot and its main operating principles in the view of using it for wavelength calibration. In Section~3, we describe our concept and algorithms of wavelength calibration combining thorium and Fabry-P\'erot lines. Section~4 finally presents results obtained by applying the new calibration to a few standard stars of the HARPS radial-velocity program.

\section{Characterizing the Fabry-P\'erot}
The HARPS Fabry-P\'erot (FP) used for drift measurements consists of two flat and partially reflective parallel mirrors separated by an optical gap (cavity spacing in vacuum). HARPS's Fabry-P\'erot has a cavity of $D=\SI{7.3}{\mm}$ and a measured effective finesse of 4.3. Temperature and pressure are stabilized inside a vacuum tank. Details on the employed Fabry-P\'erot, its design and its performances can be found in \citet{wildi2010}.\\
Multiple interferences occur and create a transmission spectrum that consists of a series of
discrete peaks. At normal incidence and in vacuum, a transmission maximum is reached every time the phase difference between the interfering waves is an integral multiple of 2\textpi, which is equivalent to: 
\begin{equation}
2 D = m \lambda
\label{equ:main}
\end{equation}
where $m$ is an integer number defined as the interference order and $\lambda$ is the wavelength of the transmission peak. The theory and the characterization of such a kind of Fabry P\'erot is described in details in \cite{cersullo2017}.\\

The Fabry-P\'erot equation (Eq. \ref{equ:main}) allows us to associate a wavelength to every transmission peak of order $m$. If we wanted to use the Fabry-P\'erot as an absolute calibrator, these wavelengths $\lambda_m$ would have to be known with an accuracy of better than \SI{1}{\m\per\s}. By consequence, the cavity spacing $D$ would have to be known to better than $\updelta\lambda/\lambda=\num{3e-9}$. It means that, for the FP system of HARPS, we would need to know, at each moment in time, $D$ with an accuracy of better than \SI{0.2}{\angstrom}. In practice, the cavity spacing is known to about \SI{1}{\micro\meter}, i.e. at a level \SI{10000}{times} worse than needed for absolute-calibration purposes. Furthermore, the used Fabry-P\'erot system is fully passive, although stabilized in temperature and pressure, and may show drifts of the order of \SI{0.2}{\m\per\s} over 24\,hours.

Finally, and probably most importantly, the cavity spacing is not constant across the wavelength range of the spectrograph. This is due to the Fabry-P\'erot mirrors that are coated with multi-layer dielectric films. The optical depth of the coating and thus the phase shift introduced by its reflection will result in a variable cavity spacing $D(\lambda)$, which however we expect to be a smooth function $\lambda$ and can therefore be modelled with a low-order polynomial. If this model function was known, it could be used to assign to every transmission peak an accurate wavelength and subsequently use the Fabry-P\'erot spectrum to perform the wavelength calibration of the spectrograph and obtain the corresponding wavelength solution. For this reason, the first step was to characterize the Fabry-P\'erot and determine $D(\lambda)$.\\

\begin{figure}[htbp]
 \centering
\includegraphics[width=9.3cm,height=7.3cm]{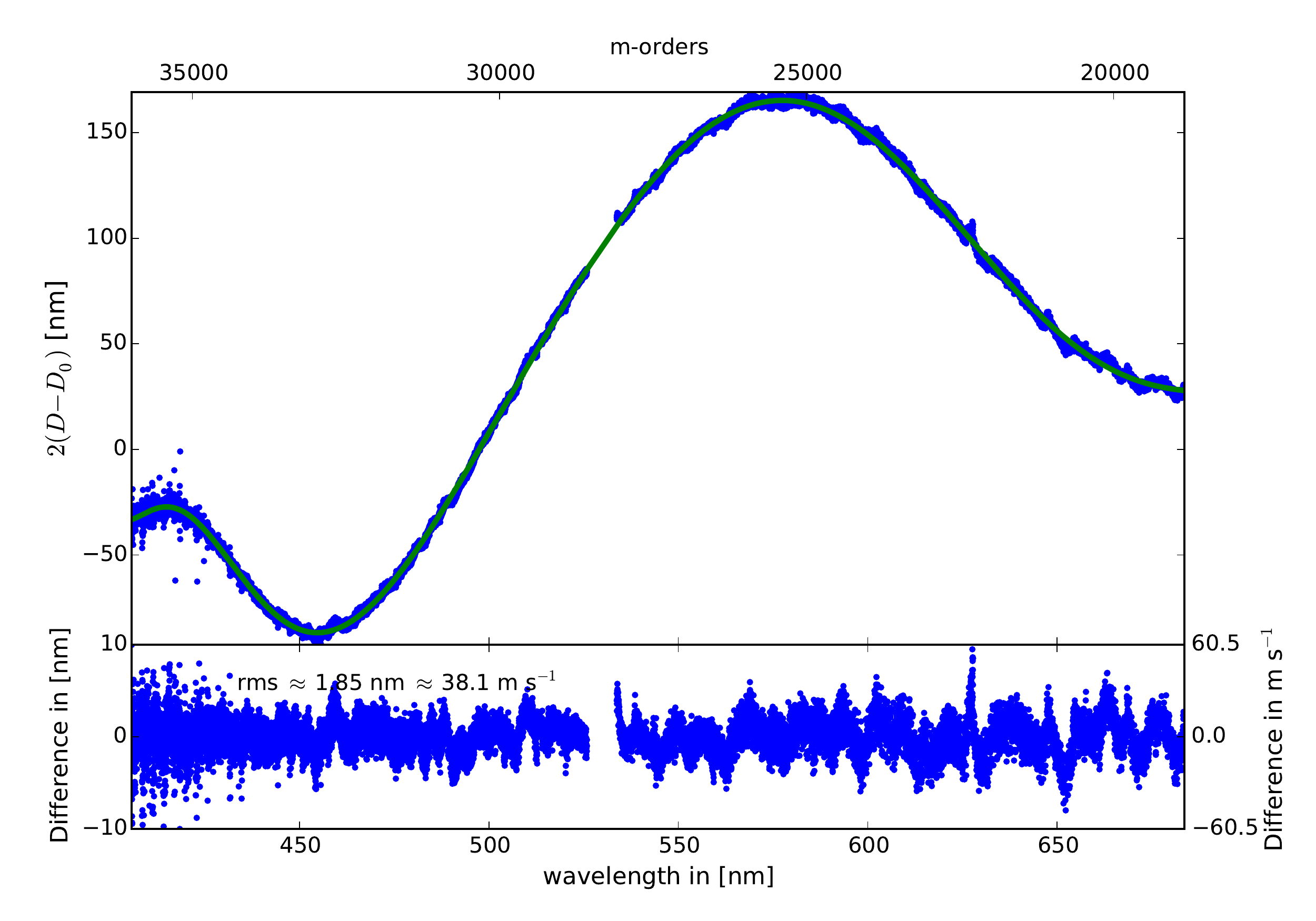}
\caption{Effective etalon gap variation $2D(\lambda)-2D_0$ as a function of wavelength as computed for each Fabry-P\'erot Peak using the initial thorium wavelength solution (blue line, upper panel). The green line shows the best-fit polynomial. The residuals to the fit are shown in the lower panel.}
\label{fig:1}
\end{figure}

The initial wavelength $\lambda_i$ of every transmission peak $i$ of order $m_i$ of the Fabry-P\'erot are determined from the calibration using the thorium spectrum. This means in practice that the spectrograph is calibrated by injecting the thorium spectrum and computing, for each spectral line and from the association of its wavelength $\lambda_i$ with the pixel coordinate $x$ of its photocenter and a wavelength solution $\lambda(x)$, where $x$ is the pixel coordinate of the detector. Then, the spectrograph is fed with the Fabry-P\'erot spectrum. Using the previously determined thorium wavelength solution, an initial wavelength is assigned to the pixel coordinate $x_i$ of each Fabry-P\'erot peak determined by a Gaussian fit on the extracted spectrum. We determine the order number $m_i$ of the peak with the formula $m_i=2D_0/\lambda_i$, assuming a gap $2D_0\,=\,\SI{14.6}{\mm}$ and rounding to the closest integer value. Inverting this same formula and adopting the integer value for $m$, we determine the effective gap $D_i(m_i)=m_i\lambda_i/2$ of the etalon. Due to errors on the wavelength solution and the uncertainty on $D_0$, the computed $D_i(m_i)$ will be affected by errors and even discontinuities induced by misnumbering. Such a misnumbering is unlikely to happen within an echelle order of the spectrograph, since the subsequent peaks \emph{must} correspond to adjacent Fabry-P\'erot orders (continuous numbering). However, a discontinuity may occur at the boundary between two echelle orders, in particular if the initial $D_0$, assumed constant, is sufficiently different from the effective $D_{\lambda}$ at this very wavelength. On the other hand we can assume that $D(m)$ must be continuous and smooth. The order numbering must be corrected, if necessary, to achieve a continuous function and minimize its slope; any discontinuity or residual average slope would be an indication for a wrong order numbering. Fig.\ref{fig:1} shows the gap variation $2D_i(m_i)-2D_0$ for each peak $m_i$ as a function of wavelength (or $m$). Opposite to \citet{Bauer} we chose to fit a polynomial function $D(m)$ to the data pairs [$D_i,m_i$]. By comparing the Bayesian information criterion (BIC) for different degrees of the function we concluded that the polynomial of degree~9 best fits the data. The model parameters of $D(m)$ were determined through a least-square fit. The mean effective cavity spacing that minimizes the slope and discontinuities is $2D_0 = 14.6008851$\,mm. The rms scatter to the model is 1.85 nm equivalent to about 38 m s$^{-1}$.\\

The model function $D(m)$ provides us a very precise prediction of the cavity gap as a function of Fabry-P\'erot order. From this function, we can determine for each Fabry-P\'erot peak the precise wavelength, i.e. a wavelengths which are not affected by local errors of the initial wavelength solution determined using thorium lines or other local instrumental effects (e.g. pixel errors). The so-obtained wavelengths table for the Fabry-P\'erot peaks can be used to construct a new wavelength solution.\\

\section{Constructing new wavelength solutions for  HARPS}
\subsection{Data set}
HARPS has a spectral resolution of \num{115000} and it covers a wavelength range from 380 to \SI{690}{\nm} for which we have ThAr, Fabry-P\'erot and Laser Frequency Comb spectra. The spectral format is split between the lower, blue CCD (orders $0-43$, from 378 to \SI{530}{\nm}) and the upper, red CCD (orders $44-71$, from 533 to \SI{690}{\nm}). HARPS's CCDs are made of ``blocks'' of $512\times1024$ pixels stitched \citep{wilken2010} together. The junction between these blocks is not perfect, such that the pixels at the junction may have sizes slightly different from the nominal size, producing ``gaps'' in the pixel coordinates and/or wavelength solution whenever the pixel number is used as the coordinate. A pixel size map is used to correct for the mismatch between pixel number and ``physical'' coordinate. This pixel size map has been determined on the basis of flat-field exposures \citep{Coffinet+2018}. Exactly in the same way, the flat-field map (pixel sensitivity) is corrected for the effective pixel size. Hereafter we will follow the recipe described in \citet{Coffinet+2018} to extract and wavelength calibrate the spectra, and to correct them for the effect of block-stitching errors. The weighting of the equations is also the same as in \citet{Coffinet+2018}, with a line-dependent weight for the wavelength-solution equations one for the Littrow equations. The only difference will consist in using also FP-lines to better constrain the wavelength solution. The resulting spectra consist of gap-corrected, flat-fielded 2-D spectra of 71 lines (orders of the echelle spectrum) and 4096~columns (CCD pixels along the dispersion direction).\\

In this section, we describe how to determine the wavelength solution of the HARPS high-resolution echelle spectrograph. We will show its architecture and configuration. The strategy starts from the wavelength frames produced by the thorium and Fabry-P\'erot calibration exposures.

\subsection{Wavelength-calibration plan}
$D(\lambda)$ is fully characterized every time we do a FP exposure. Whenever a FP exposure on the fiber to be calibrated is not available (e.g. in the early times of the instrument), we use the FP-line Table issued from the FP exposure the closest in time and by allowing an offset for $D_0$. In fact, we can assume that $D(\lambda)$ does not change in time apart from the global $D_0$. Therefore at each calibration only $D_0$ is refitted in this case.\
The ideal (normal) calibration sequence is shown in Tab.\ref{tab:fibre}:\\

\begin{table}[h]
\centering 
\begin{tabular}{l l l}
\hline\hline
&\textbf{Fiber A} & \textbf{Fiber B} \\ 
\hline
1. & TH & TH\\
2. & FP & FP\\
3. & TH & FP\\
4. & OB & FP \\
\hline\hline
\end{tabular}
\caption{HARPS calibration sequence.}
\label{tab:fibre}
\end{table}

The TH-TH + FP-FP exposures, serve to define the FP and TH line tables for which we compute the (joint) wavelength solution. However, since the spectrograph may drift, we make an TH-FP exposure by which we define the nightly zero-point, i.e. on fiber A we redetermine the wavelength solution for exposure 3), this time fitting the lines TH and FP from 2) and 3), while with fiber B we set the "zero-drift" point for exposure 3), that will serve as reference for the scientific observation 4) of the night. On this latter, possible instrumental drifts are measured by comparing the FP lines position on fiber B following the so-called "simultaneous-reference" concept \citep{baranne96, Queloz2001}. This latter concept assumes that the FP remains stable to the required instrumental precision between to the two  calibration sequences, i.e. on time scales of typical 24\,hours.\\
In addition, this new method requires FP-FP exposures. The whole set of wavelength calibration is TH-TH + FP-FP + TH-FP. It is not possible to obtain a good calibration on fiber A if it has not been exposed with FP. Fiber B is much less critical, since it is used only for simultaneous reference.

\subsection{Wavelength-calibration process}

\begin{figure*}
\centering
\begin{tikzpicture}

\definecolor{backcolor}{HTML}{EEEEEE}  
\definecolor{boxcolor}{HTML}{000000}   
\definecolor{arrowcolor}{HTML}{000000} 
\definecolor{textcolor}{HTML}{000000}  
\definecolor{outercolor}{HTML}{F7F7F7}  

\node[fill=outercolor,draw=black,text=blue,text width=5.20cm,minimum height=15.30cm] at (-5.45,0.0) {\commentaire{du texte}};
\node[fill=outercolor,draw=black,text=blue,text width=5.20cm,minimum height=5.70cm] at (5.555,4.85) {\commentaire{du texte}};

\matrix [column sep=13mm, row sep=9mm, every node/.style={
    shape=rectangle,
    text width=3.90cm,
    minimum height=1.00cm,
    text centered,
    font=\sffamily\small,
    very thick,
    color=textcolor,
    draw=boxcolor,
    fill=backcolor,
}] {
  \node (a1) {FP spectrum}; &
  &
  \node (a3) {ThAr spectrum}; \\
  \node (b1) {$x_{i,\mathrm{FP}}$}; &
  &
  \node (b3) {$x_{i,\mathrm{Th}}$}; \\
  \node (c1) {$\lambda_{i,\mathrm{FP}}$};&
  &
  \node (c3) {$\lambda_{i,\mathrm{Th}}$}; \\
  \node (d1) {$m_{i,\mathrm{FP}}$};&
  \node (d2) {Table Th}; \\
  \node (e1) {$D_{i,\mathrm{FP}}$};&
  &
  \node (e3) {Table ThFP}; \\
  \node (f1) {$D_{\mathrm{fit}}$};&
  \node (f2) {Table FP}; \\
  \node (g1) {$D_{f,i}$}; \\
  \node (h1) {$\lambda_{f,i}$};&
  &
  \node (h3) {$\lambda(x)$}; \\
};


\begin{scope}[->, very thick, arrowcolor]
  \draw (a1) -- node [left, midway] {Fit} node [right, midway] {(4)} (b1);
  \draw (b1) -- node [left, midway] {$\lambda_i=\lambda(x_i)$} node [right, midway] {(5)} (c1);
  \draw (c1) -- node [left, midway] {$m_i=\mathrm{nint}\left(\frac{2D_0}{\lambda_i}\right)$} node [right, midway] {(6)} (d1);
  \draw (d1) -- node [left, midway] {$D_i=\frac{\lambda_i m_i}{2}$} node [right, midway] {(7)} (e1);
  \draw (e1) -- node [left, midway] {Fit $D_i(m_i)$} node [right, midway] {(8)} (f1);
  \draw (f1) -- node [left, midway] {$D_{f,i}=D_\mathrm{fit}(m_i)$} node [right, midway] {(9)} (g1);
  \draw (g1) -- node [left, midway] {$\lambda_{f,i}=\frac{D_{f,i}}{m_i}$} node [right, midway] {(10)} (h1);
  \draw (b1) -- (f2);
  \draw (d1) -- (f2);
  \draw (g1) -- (f2);
  \draw (h1) -- (f2);
  \draw (a3) -- node [right, midway] {Fit} node [left, midway] {(1)} (b3);
  \draw (b3) -- node [right, midway] {Atlas} node [left, midway] {(2)} (c3);
  \draw (b3) -- (d2);
  \draw (c3) -- node [left, yshift=2mm] {(3)} (d2);
  \draw (d2) -- (e3);
  \draw (f2) node [xshift=25mm, yshift=12mm] {(11)} -- (e3);
  \draw (e3) node [xshift=-30mm, yshift=7mm] {(12)} -- node [right, midway] {Fit} node [left, midway] {(13)} (h3);

\end{scope}
\end{tikzpicture}
\caption{Structure of the thorium+FP combined calibration process.}
\label{fig:3}
\end{figure*}
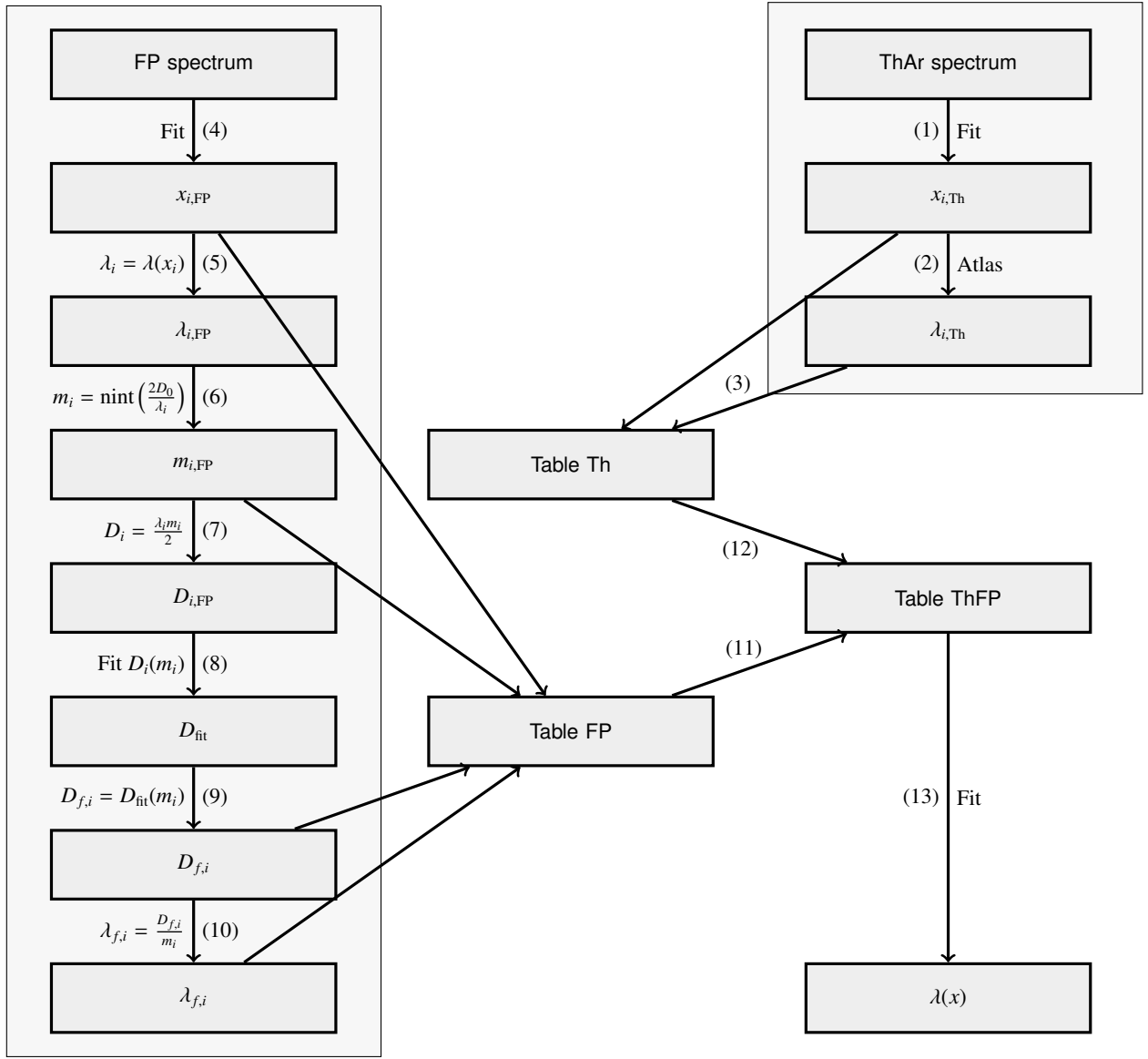

The diagram shown in Fig. \ref{fig:3} represents the structure of the new reduction process. The left part of the diagram (steps~4 to 10) corresponds to the treatment of the FP files. As described in the previous section, the goal is to obtain a model $D(m)$ from which, by knowing the order $m_i$ of each FP line, we can compute the corresponding peak wavelength $\lambda_i$ of the same line. The upper right part (steps~1 and 2) corresponds to the first part of the standard thorium calibration. From these two parts, we get independently (steps~11 and 3, respectively) two input tables $[x_i,\lambda_i,m_i,D_i,o_i,s_i]$ where $x_i$ is the position of the line, $o_i$ the echelle order this line belongs to and $s_i$ its type (thorium or FP line). These two tables are merged in step~12 (zeros are added in the table for the $m_i$ and $D_i$ of the thorium lines simply to keep the correct number of columns, but these values are not used later) and serve as input to compute the final wavelength solution, which is obtained by a global solution (in the sense of least squares) to the ThAr, the FP and the Littrow equations (step~13) given hereafter:

\begin{align}
\sum_{k=0}^K a_{k,o} x_{i,TH}^k & &= \lambda_{i,TH} && \text{ThAr $\lambda(x)$ equations} \\
\left( \sum_{k=0}^K a_{k,o} x_{j,FP}^k \right) &+ \frac{2\Delta D}{m_{j,FP}} &= \lambda_{j,FP} && \text{FP $\lambda(x)$ equations} \\
\left( \sum_{k=0}^K a_{k,o} x_{n,LIT}^k \right) &- \left( \sum_{l=0}^L b_{l,x}\left(\frac{1}{o_{n,LIT}}\right)^l \right) &= 0 && \text{Littrow equations}
\end{align}

where $x$ is the physical (stitching-corrected) pixel coordinate along the diffraction order $o$ of the echelle spectrograph. The weighting of the equations is the same as in \cite{Coffinet+2018}.

The echelle order number ranges in HARPS from 161 (blue end) to 89 (red end), but for practical reasons they are numbered from 0 to 71 in the data-reduction software and described in this way hereafter. The parameter $\Delta D$ in the second equation is used to allow for a possible temporal change in the etalon spacing, since we do not have any guarantee (and actually no requirement) that it remains stable over timescale much longer than 24\,hours. It is important to note that the zero-point of the calibration is locked by the thorium lines, while from the FP lines we shall use only the (spectrally) local information. Since $D(m)$ of a given FP exposure is based on any (previously reduced) thorium exposure, the $\Delta D$ not only accounts for the actual cavity spacing but also for possible drifts of the spectrograph in the time between the previous thorium exposure and the TH-FP exposure used as the zero-point reference for the night. With this latter exposure the previously determined FP wavelengths stored in the line list are ``realigned'' on the absolute wavelength scale given by the thorium lines and the so-determined wavelength solution will be used for the following scientific exposures. For this reason $\Delta D$ can be redetermined independently for every wavelength solution.\\


The echelle spectrum consists of a number of spectral chunks (spectral orders of the echelle grating), each covering at least one free spectral range (FSR). For this reason, many of the spectral lines are present on two consecutive orders. Thanks to cross-dispersions that acts perpendicularly to the main dispersion, the spectral orders are ‘sorted’ by wavelength and distributed  side-by-side across the detector.
Furthermore, the Littrow equations can be used to link the wavelength solutions of the echelle orders together and give more ``rigidity'' to the 2-D wavelength map. In fact, if we ``cut'' the CCD frame along the cross-dispersion direction at any arbitrary pixel coordinate we know from the optical setup of the echelle grating (Littrow mounting) that the wavelength at that position must be inversely proportional to the 
 echelle order number. Therefore we compute the wavelength in every echelle order for a fix set of pixel coordinates ($x$ = 400, 1200, 2000, 2800 and 3600). Then, we use a third-degree polynomial to describe the Littrow relation across the orders. The higher degree of the polynomial is chosen in order to include also optical aberrations and other distortions of the wavelength map. The relative weight of these equations with respect to the ThAr- and FP-equations is given by the number of Littrow equations, which is turn is determined by the number of ``cuts'' in cross-dispersion direction. In order not to over- or under-constrain the wavelength solution we chose to adopt a number of cuts similar to the degree of the polynomial describing the wavelength scale in main-dispersion direction.






The output of the solution to the equation set given above consists of the coefficients of the wavelength solutions $a_{k,o}$ (4~coefficients per order), the FP-gap correction $\Delta D$ (2~coefficients, one for each CCD), and the coefficients of the Littrow solutions $b_{l,x}$ (4~coefficients for each position $x$ where these Littrow equations are computed). One has to note that the $a_{k,o}$ and $b_{l,x}$ are the same for both the thorium and Fabry-P\'erot equations, while the the FP-gap correction $\Delta D$ is defined only for the FP equations. The total number of free parameters is 330 and should be compared with the more than 22'000 spectral lines (about 15'000 FP lines and more than 7'000 thorium lines) used to constrain the model. The total number of parameter takes into account that the HARPS 4k4 detector is a mosaic of two mechanically independent 2k4 chips, i.e two Littrow equations.\\

The very first thorium calibration exposure reduced by our pipeline, a first wavelength solution based \emph{only} on the thorium file is computed. This first step is identical to the wavelength calibration process presented in \citet{Coffinet+2018}. The preliminary solution is then used for the Fabry-P\'erot part at step~5 (see the diagram), after what we compute the combined TH+FP wavelength solution. For the thorium calibrations reduced after the first one, the preliminary wavelength solution used at step~5 is the combined TH+FP wavelength solution of the previous thorium calibration that was reduced. This additional iteration produces a more accurate wavelength solution compared to the preliminary thorium-only based solution. This new wavelength solution is used to redetermine a new model $D(m)$, which, given the iteration, is also more accurate than the very first model based only on the thorium solution.

For practical reasons, we decided to reduce the Fabry-P\'erot calibration frames inside the reduction recipe for the TH-calibration frame. The idea is to be able to use optimally, but freely, whatever FP calibration exposure in the computation of the combined TH+FP wavelength solution. The FP frame is nonetheless not chosen randomly. The combined solution shall use TH and FP frames taken as close in time as possible, ideally one after the other, such that possible drifts of the spectrograph are close to zero.

\begin{figure*}
  \centering
 \subcaptionbox*{25th order index}[.3\linewidth][c]{%
    \includegraphics[width=.32\linewidth]{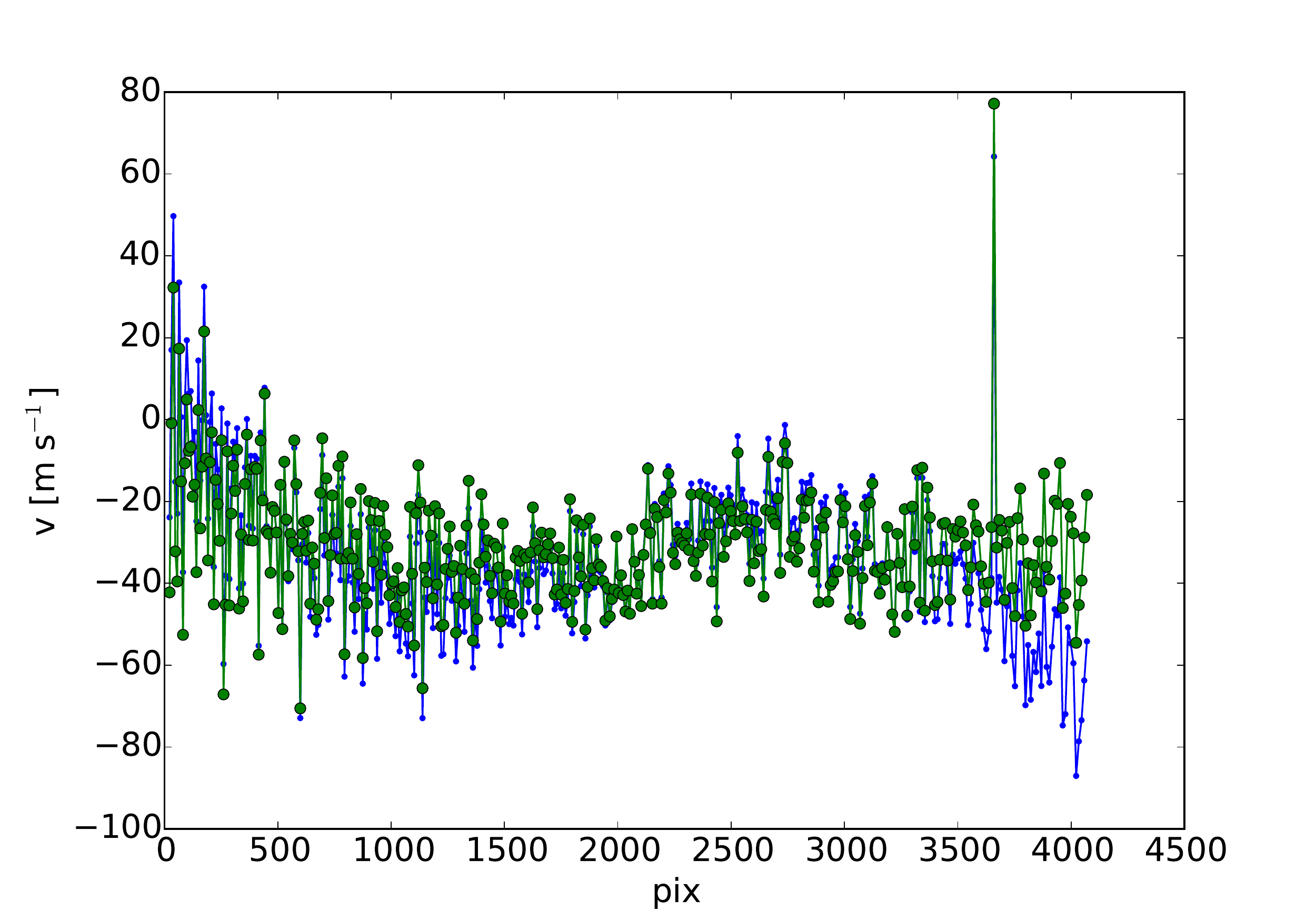}}\quad
  \subcaptionbox*{35th order index}[.3\linewidth][c]{%
    \includegraphics[width=.32\linewidth]{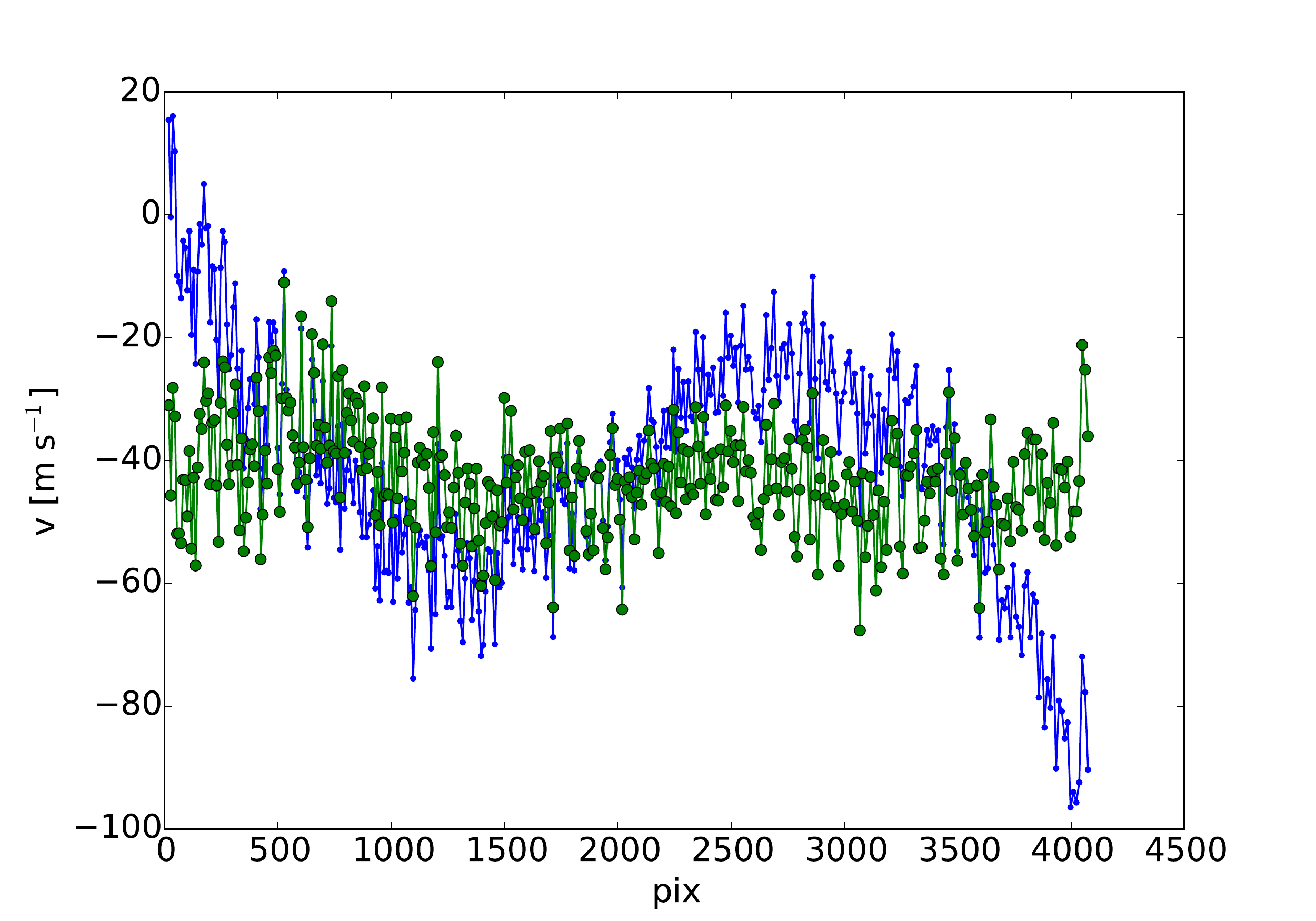}}\quad
  \subcaptionbox*{45th order index}[.3\linewidth][c]{%
    \includegraphics[width=.32\linewidth]{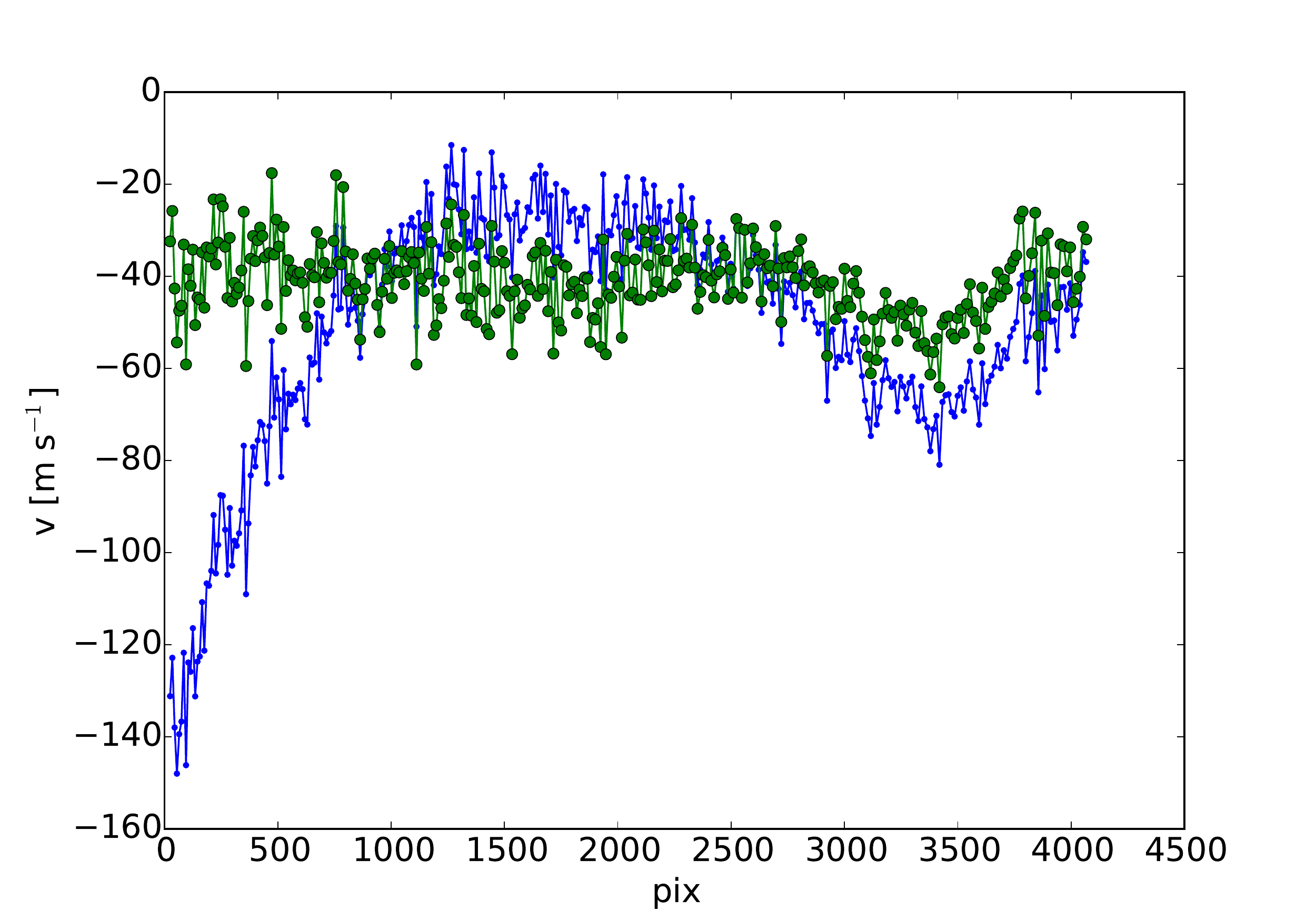}}

  \bigskip

   \subcaptionbox*{55th order index}[.3\linewidth][c]{%
    \includegraphics[width=.32\linewidth]{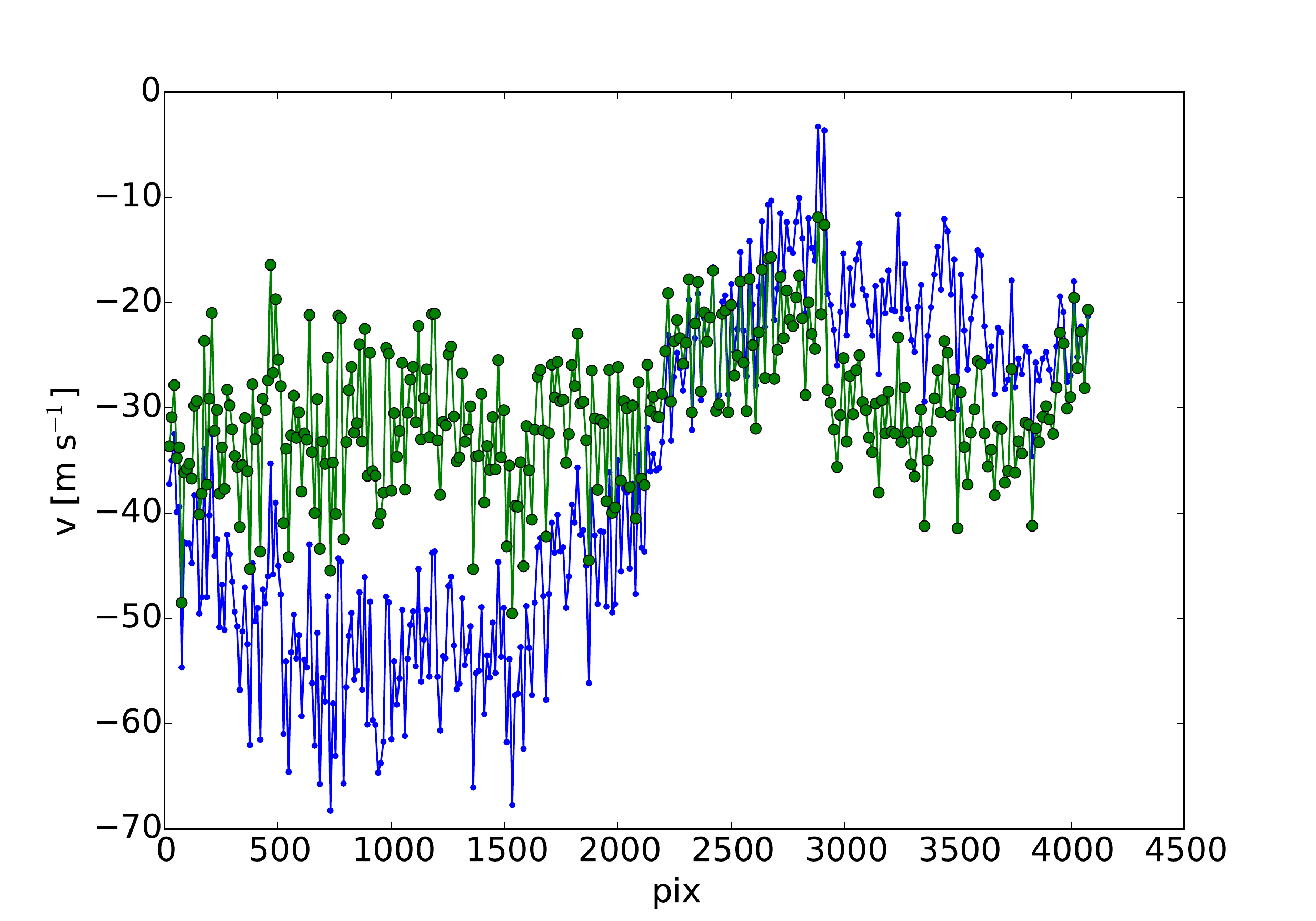}}\quad
  \subcaptionbox*{65th order index}[.3\linewidth][c]{%
    \includegraphics[width=.32\linewidth]{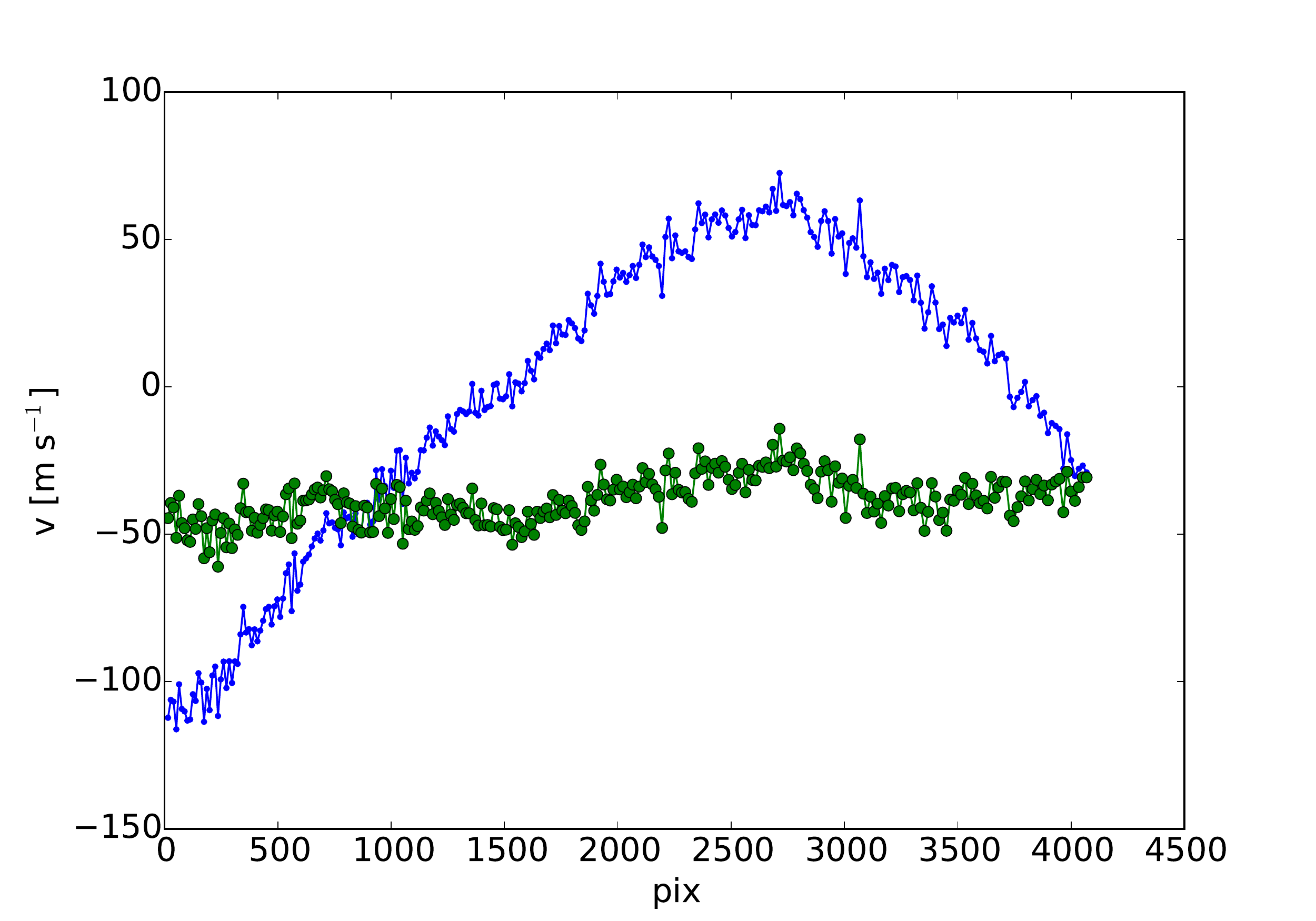}}\quad
  \subcaptionbox*{69th order index}[.3\linewidth][c]{%
    \includegraphics[width=.32\linewidth]{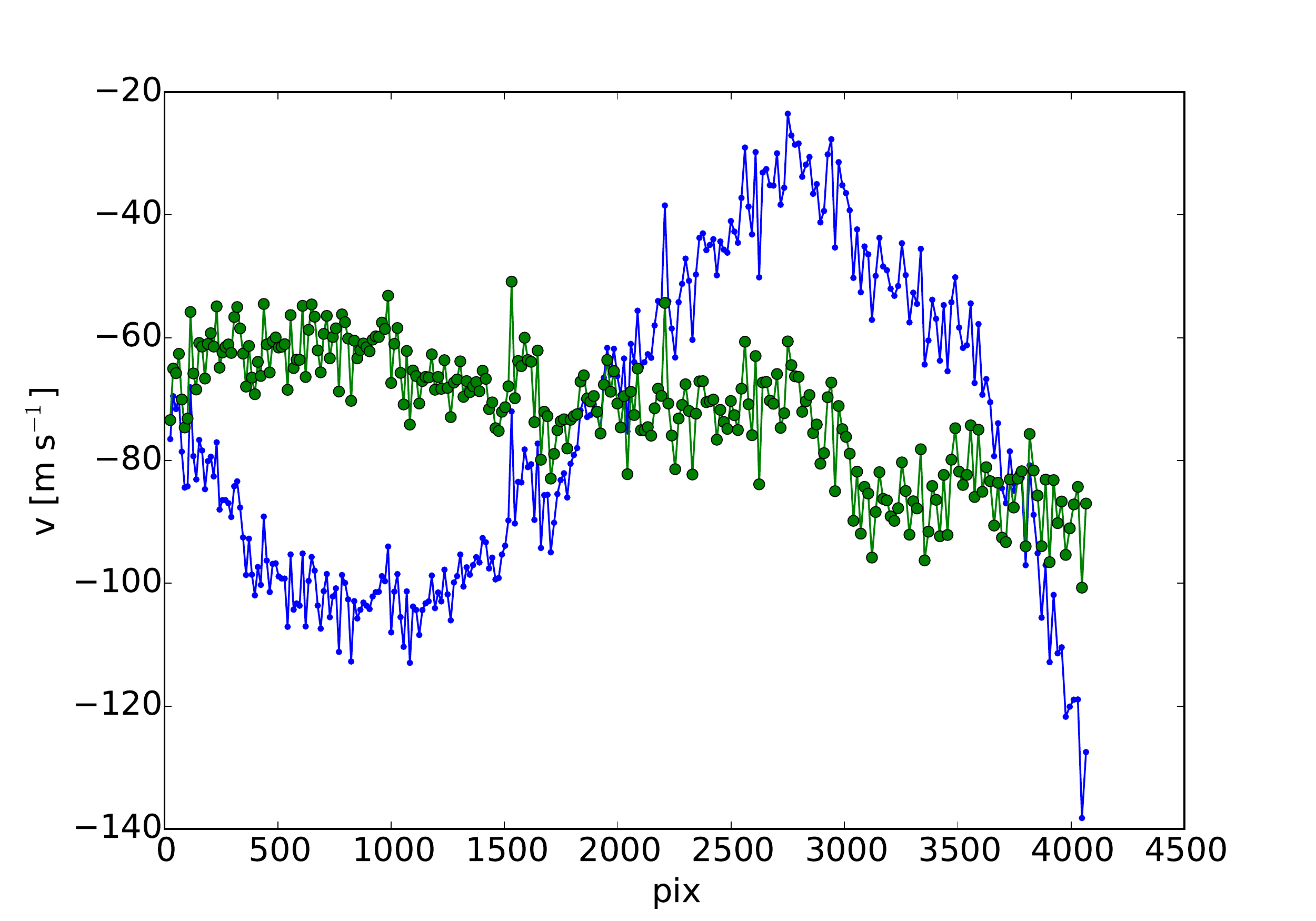}}
 
    \bigskip
      \caption{Difference between the theoretical wavelengths of the LFC lines and the wavelengths along a subset of spectral orders for HARPS obtained after calibration with jumps gaps corrected \citep{Coffinet+2018} using the "pure" thorium-wavelength solution (blue curve) and the TH+FP wavelength solution of the new DRS (green curve). The remaining orders are shown in the appendix \ref{app:c}}
      \label{fig:3}
\end{figure*}

It must be recalled here that $\Delta D$ was not introduced to account for a possible drift of the spectrograph, but as the variable (fitted) part of the ‘unknown’ etalon cavity spacing $D_0$. In the frame of the proposed wavelength-calibration concept, the absolute value $D$ of the etalon cavity spacing is not relevant, since we simply fit for it by adjusting $\Delta D$ that will take into account possible variations of $D$ with time. Nevertheless, we cannot exclude that small drifts of the spectrograph or the CCD may occur between the time we recorded the TH exposure and the time of the FP exposure. In the proposed method we would then interprete the spectrograph drift as a change of the FP cavity, and spectrograph drifts would simply be absorbed by the free parameter $\Delta D$ that was foreseen to fit the actual FP etalon cavity at the time of the FP calibration. In fact, a change in $\Delta D$ is strictly proportional to $\Delta \rm{v}$, which in turn is proportional to a motion $\Delta x$ (pixel position) of a spectral line located in the central part (at blaze) of an echelle order. On the other hand, spectrograph drifts are supposed to be approximately constant in $\Delta x$ across the chip but not in $\Delta \rm{v}$, mainly because along the echelle orders the linear dispersion $dx/d\lambda$ varies by almost a factor of 2. Therefore, a spectrograph drift is not fully modelled by a single velocity change $\Delta \rm{v}$ and thus by simply adjusting $\Delta D$. For this reasons, we have finally limited the application of the presented wavelength-calibration concept to the period after 2011, i.e. from the date starting from which FP-calibration frames have started to be taken systematically on HARPS, and for which we can assume that the spectrograph drifts between the TH and the FP expsoures are small, i.e. much smaller than the pixel size or typically m\, s$^{-1}$.\\

Nevertheless, we would like to briefly provide an idea of how to extend this concept retro-actively to a time when FP frames were not yet available, and thus even in presence of significant instrumental drifts. Let's assume that we have recorded at any moment of the lifetime $T_1$ of HARPS during the same night TH-calibration spectra and stellar observations. Let's also suppose that we have taken both TH and FP calibrations at a much later time $T_2 >> T_1$, and we want to use these later FP-spectra to refine the wavelength solution at $T_1$. Then, we can use the TH spectra to determine a motion $\Delta x(x,o)$ of the spectra on the CCD between $T_1$ and $T_2$ as a function of order $o$ and pixel $x$ along each order. These shifts can directly be applied to the FP line-list obtained from the FP-calibration at time $T_2$ and transposed to the time $T_1$, such that it can be combined with the TH-calibration of time $T_1$ and so obtain a combined wavelength solution. The only assumption we made is that the 'local' geometry of the detector and the spectral format did not change with time (apart from very low-order effects) and that the information of the FP frames taken later on is applicable. For clarity, we shall however again recall that this modified concept has not been applied to the the tests and results presented hereafter.

\subsection{Comparison with LFC data}
Laser frequency combs (LFC) are the best sources for absolute wavelength calibration of astronomical spectrographs \citep{Locurto2012}. The spectrum of an optical LFC consists of an optical spectrum of equidistant spectral lines. Since, on the one hand, this spectrum is related to a regular train of ultra-short pulses that are synchronized with an absolute radio frequency reference e.g. a global positioning system (GPS) signal or atomic clock, and, on the other hand, the offset frequency is controlled by octave-spanning self-referencing, the frequencies (or wavelengths) of all spectral lines are known in absolute terms. Therefore, such a kind of absolute source, with dense and perfectly regular spectral lines is most suitable for the accurate calibration of astronomical spectrographs.\\
HARPS is using an LFC since the first commissioning run held on 2015-04-22. It provides spectral emission lines with a mode spacing (i.e. repetition rate) that can be resolved by the spectrograph but dense enough to maximize the information. Typical values are of $\sim 20$GHz.\\
Optical frequencies of frequency comb lines are uniquely described by this equation: $f_n = n \cdot f_\textrm{rep}+f_0$.\\
We define the $f_\textrm{rep}$ repetition rate as the frequency difference between two neighboring lines and $f_0$ as the carrier-envelope offset frequency. The frequency $f_n$ of each line is characterized by a unique integer $n$ . 
Then, the nominal wavelength is derived in a standard way as $\lambda_n = c/f_n$.\\
To determine these $\lambda_n$, a first, approximate wavelength is determined based on the previous thorium-based wavelength solution. This is necessary to get the absolute wavelengths. From these approximate wavelengths, one can compute an approximate $n$ for each LFC line based on the known $f_\textrm{rep}$ and $f_0$ of the laser frequency comb, $f_0 = 288.0598$ THz et $f_{rep} = 18.0$ GHz (private communication with Gaspare Lo Curto). These $n$, that should be very close to an integer value, are then rounded to the closest integer. The theoretical LFC wavelength is then computed from this rounded $n$.\\
Given the nature of the LFC and the well-known wavelength of its lines, we decided to test our wavelength calibration method on LFC lines. The method is as follows: we calibrate the spectrograph using either the standard DRS (only thorium lines) or the new DRS (TH+FP). Then we measure the position of the LFC lines in pixels and apply the two wavelength solutions to transform, for each line, the pixel position in wavelength. Finally, we compare the so-obtained (measured) wavelength with the nominal wavelength of the LFC line.




Fig.~\ref{fig:3} shows the difference between the theoretical wavelengths and the ones of the laser's lines along a subset of spectral order of HARPS covered by the LFC. In green we show the solution obtained after calibration of the spectra using the new DRS (combination of both thorium and Fabry-P\'erot) and in blue the standard DRS (thorium only). One can remark that in all cases the residuals obtained calibration with the new DRS are much smoother and flatter than with the standard DRS. In particular it should be remarked that the underlying wavelength solution rarely ``diverges'' towards the edges of the orders, which is certainly due to the much higher density of information provided by the FP lines compared to the sparse thorium lines.\\

\begin{figure}[htbp]
 \centering
\includegraphics[width=9.35cm,height=7.6cm]{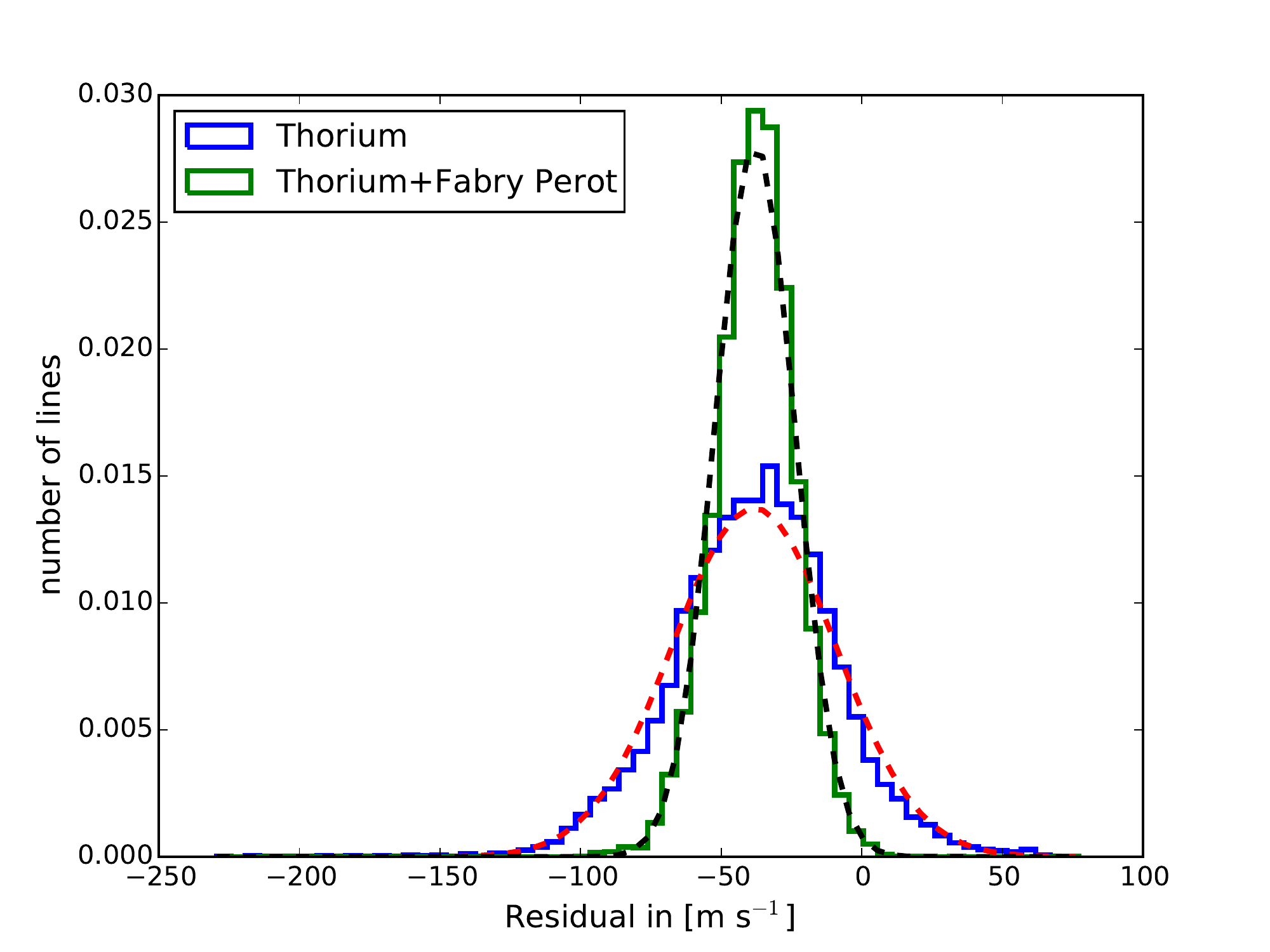}
\caption{Histogram of the differences in wavelengths between the theoretical and the calibrated (measured) wavelengths of the LFC spectral lines. The results obtained with the standard (blue) and the new (green) DRS are compared. The dashed plots (red and black) are the corresponding Gaussian fits.}
\label{fig:4}
\end{figure}

In order to further illustrate that the new method is superior, we have computed the  normalized distribution of the residuals on LFC line positions for the two wavelength solutions with respect to the nominal LFC wavelengths. (Fig.~\ref{fig:4}). The results for the standard DRS using thorium-only (blue histogram) and the new DRS using TH+FP (green histogram) are shown and compared. As expected a simple Gaussian fit to both distributions give nearly the same mean offset of \SI{-38.3}{\m\per\s} and \SI{-38.0}{\m\per\s}, respectively, while the standard deviation is significantly reduced: \SI{29.0}{\m\per\s} for the standard DRS and \SI{14.2}{\m\per\s} for the new DRS. These values can be compared to the result from the LFC wavelength solution, for which a standard deviation of about \SI{7}{\m\per\s} was obtained (see \citealt{Coffinet+2018}, Section~5.1 and Figure~5). Both curves show the same mean offset which is significantly different from zero. Given the fact that the offset is identical for the two wavelength solutions (using quite different algorithms) we assume that it is possibly due to a systematic offset of the wavelengths of either the thorium or the LFC-lines' reference wavelengths. As mentioned in \cite{Coffinet+2018}, we suspect that the anchor frequency of the laser may be wrong by $\approx\SI{100}{MHz}$. Although and admittedly not fully understood, we have to remark that, in terms of radial-velocities, this offset has no consequence.

\begin{table*}[h]
\centering
\begin{tabular}{|c|c|c|c|r|r|r|r|r|r|l|l|l|l|}
  \hline
   System & Points & $\Delta t$ & Model & \multicolumn{3}{c|}{Global rms} & \multicolumn{3}{c|}{rms before} & \multicolumn{3}{c|}{rms after}\\ 
  HD \# & & & & std & C19 & new & std & C19 & new & std & C19 & new\\ 
  \hline
  10700 & 338 & 2200 & 1 pl. & 1.22 & 1.09 &1.02 & 1.13 & 1.04 &1.03 & 1.56 &1.21  &0.98 \\
   \hline
20794 & 409 & 1860 & 3 pl. & 1.09 &1.04  &1.02 &1.05 & 1.05  &1.04 &1.36 &0.99  &0.96 \\
 
   \hline                
69830 & 59 & 2177 & 2 pl. & 1.54 & 1.57 &1.43 & 1.56 & 1.55 &1.44 &1.32 & 1.69  &1.35\\ 
\hline
  \end{tabular}
\caption{Comparison of the standard (std), the intermediate (C19) \citep{Coffinet+2018} and the new (new) versions of the DRS. The second column indicated the number of data points (nightly averages) per target, the third column the time span $\Delta t$ in days and the fourth column the number of planets assumed in the model. The dispersion (\emph{rms} in \si{\m\per\s}) of the residuals to the fit of the radial-velocity data is given for all three versions of the DRS, for the full data set (Global), and the data sets before and after the fiber change separately. }
     \label{tab:table-rms}
\end{table*}

\section{Results on verification of new calibration applied on three standard stars.}
Given these promising results, we decided to test the new wavelength calibration on a few ``stable'' stars of the HARPS long-term radial-velocity programme for which we have acquired continuously observations since the start of HARPS in October 2003. The considered objects are \object{HD~10700} (\object{tau Cet}), probably the most observed and stable star of the HARPS programme, and \object{HD~20794} (\object{e Eri}) and \object{HD~69830}, two quiet planet-hosting stars. Our goal was to confirm that the new implementation of the wavelength solution improves the radial velocities, i.e. the rms of these stable stars is reduced in the new DRS compared to the standard and the \citealt{Coffinet+2018} (C19) versions of the HARPS Pipeline.\\
The version 3.5 of the HARPS Pipeline or Data Reduction Software (DRS) was installed at the telescope on October 20th, 2010. All the data obtained with HARPS, including pre-2010 data, were (re-)reduced using this same version of the DRS. We shall call here-after the version 3.5 of the DRS the ``standard'' DRS. The C19  version of DRS is based on DRS version 3.5 but includes block-stitching correction map of the CCD. This change required quite a fundamental intervention in the data structure, although the algorithms of most of the data-reduction steps hadn't been changed. For a detailed description of the C19 DRS version, we refer to \citet{Coffinet+2018}. The latest version of the DRS, called ``New DRS'' hereafter is similar to the C19 version, with the only exception of the wavelength calibration. Given the fact that the Fabry-P\'erot was installed end of 2010 on HARPS, we applied the ``New DRS'' only on the 2011-2017 data set. \\
In May 2015, a new set of octagonal fibers were installed on HARPS \citep{Locurto2015}, which required an adaptation of the DRS since both the spectral format and the Instrumental Profile (IP) had slightly changed. All the DRS versions have been applied separately to the pre-6/2015 data set (before the fiber change) and on the post-6/2015 data set (after the fiber change), since they had to be treated as if they had been obtained from two different instruments (different configuration files in the DRS). As a consequence, all three versions of the DRS foresee an additional fit parameter, i.e. one free offset value between pre- and post-6/2015 data.\\
We compare in the following the three versions of the DRS in terms of radial velocities. To do so we look at the radial-velocity dispersion after subtraction of known or well characterized signals. Also, for all targets, the Lomb-Scargle Periodograms (LSP) of the residuals and the false alarm probability (FAP) of residual signals was computed. We define that a signal is statistically significant when its peak exceed a FAP level of 10$^{-2}$. A summary of the results for the three analyzed stars is given in Table \ref{tab:table-rms}.

\begin{figure*}[ht] 
   \begin{minipage}[b]{0.5\linewidth}
    \centering
    \includegraphics[width=6.8cm]{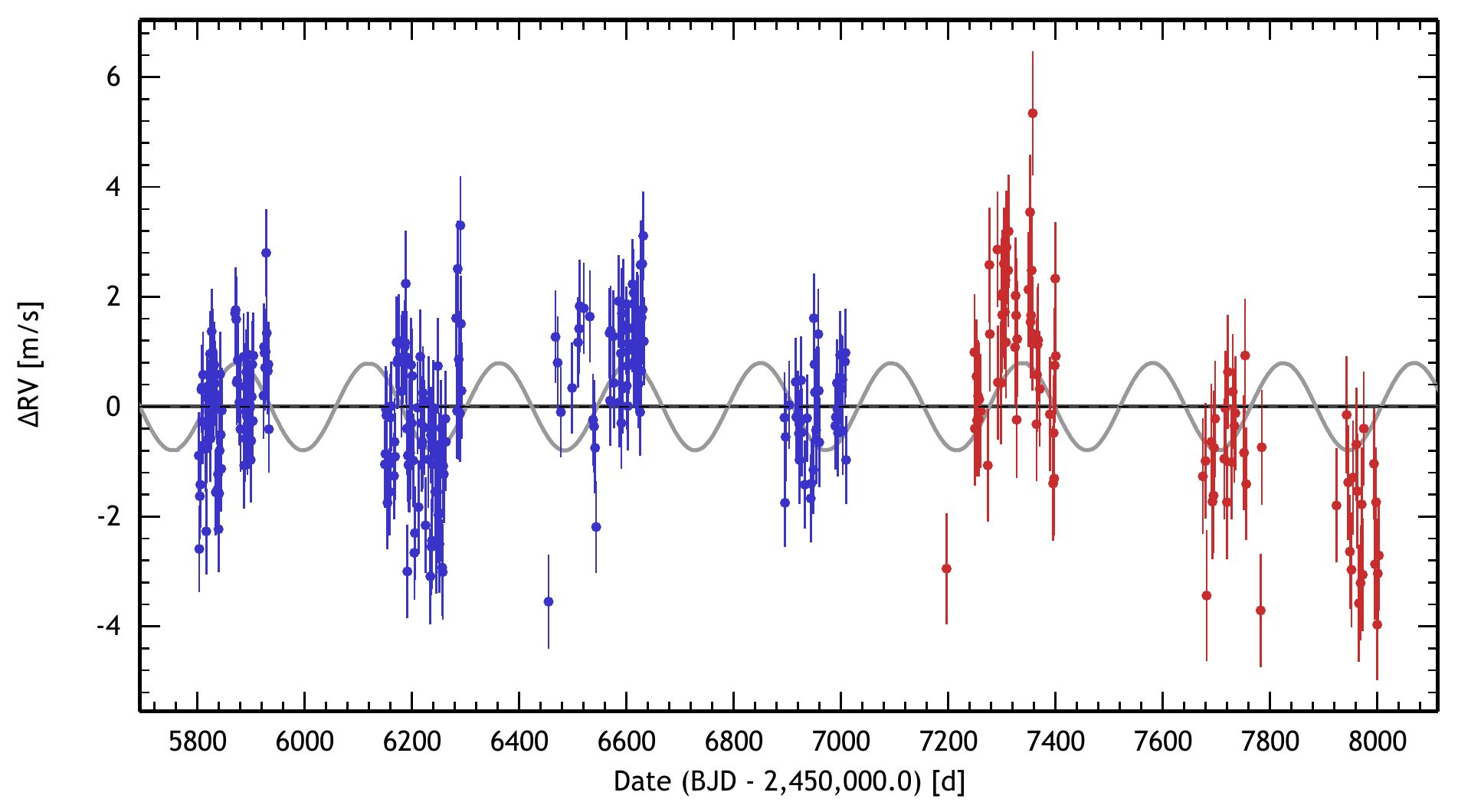} 
    \vspace{4ex}
  \end{minipage}
  \begin{minipage}[b]{0.5\linewidth}
    \centering
    \includegraphics[width=9cm]{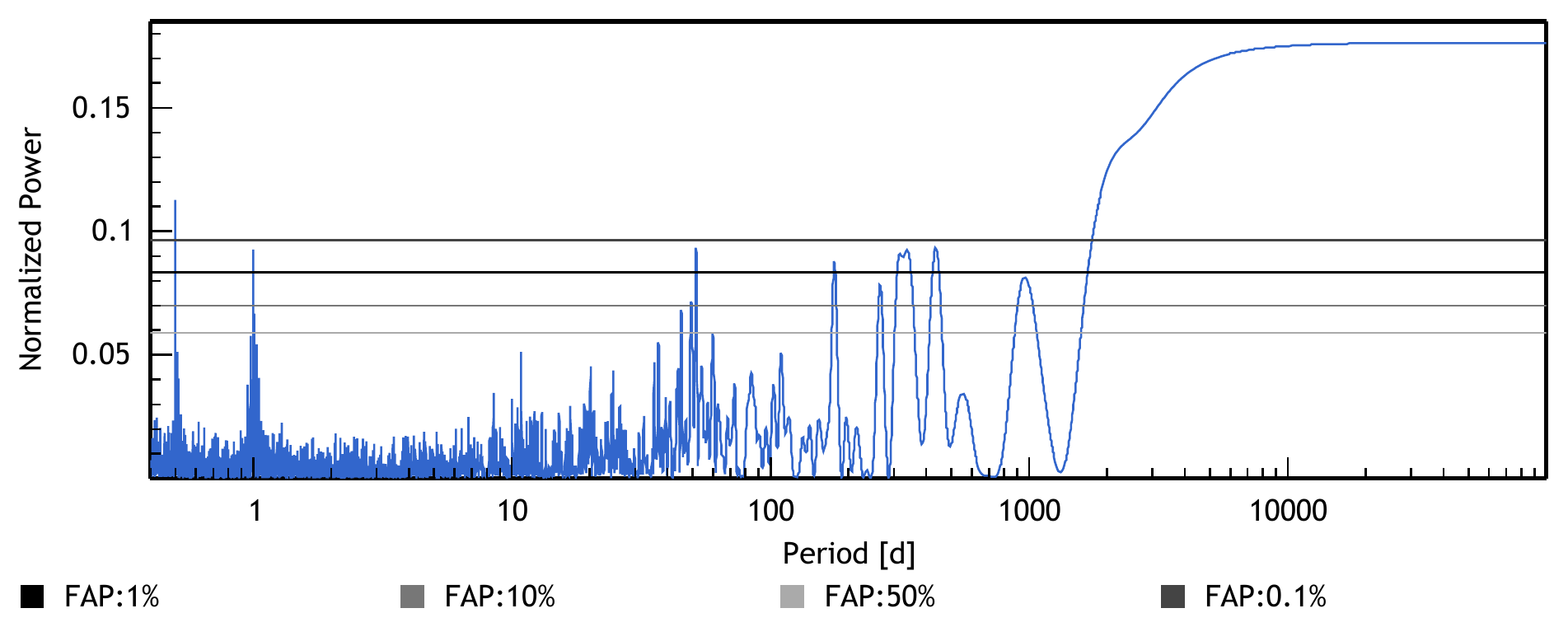} 
    \vspace{4ex}
  \end{minipage} 
  \begin{minipage}[b]{0.5\linewidth}
    \centering
    \includegraphics[width=6.8cm]{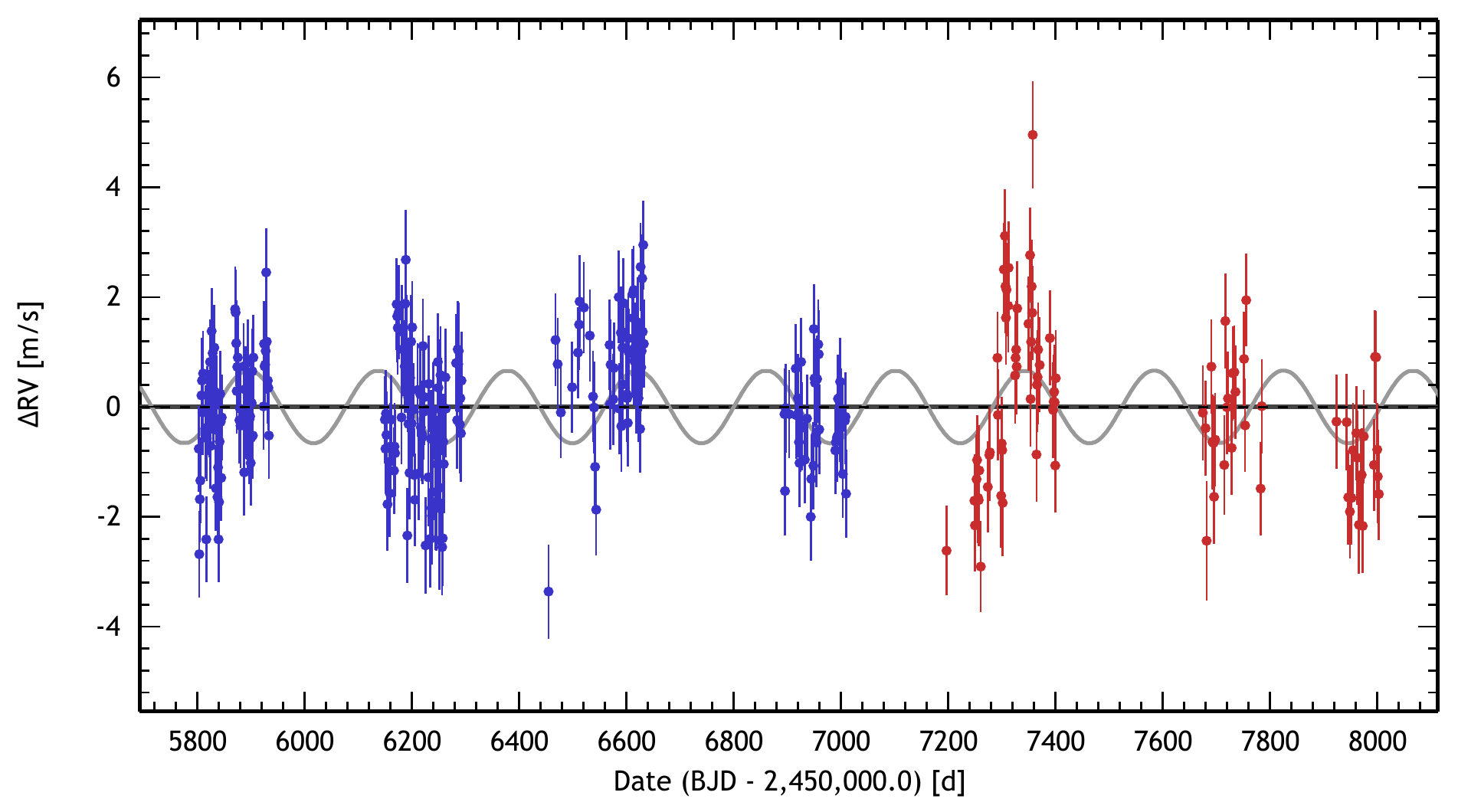} 
    \vspace{4ex}
  \end{minipage}
  \begin{minipage}[b]{0.5\linewidth}
    \centering
    \includegraphics[width=9cm]{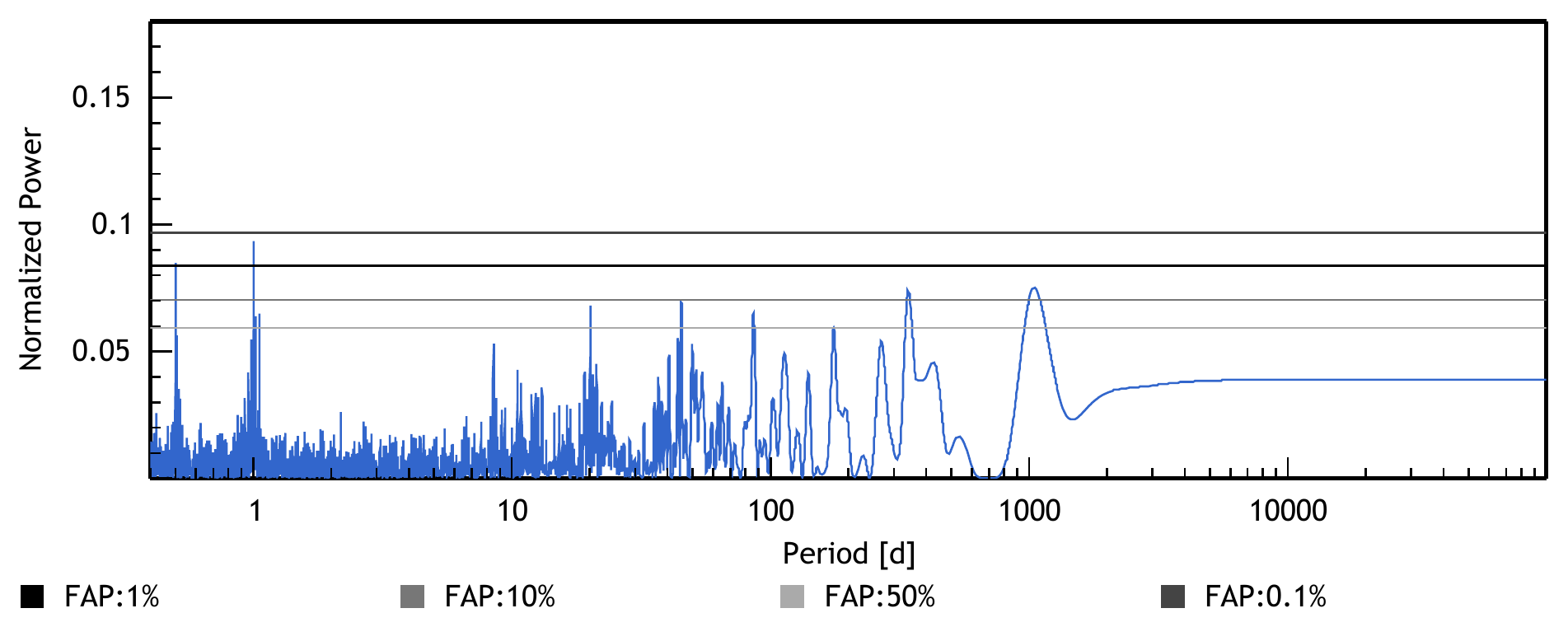} 
    \vspace{4ex}
  \end{minipage}
\begin{minipage}[b]{0.5\linewidth}
    \centering
    \includegraphics[width=6.8cm]{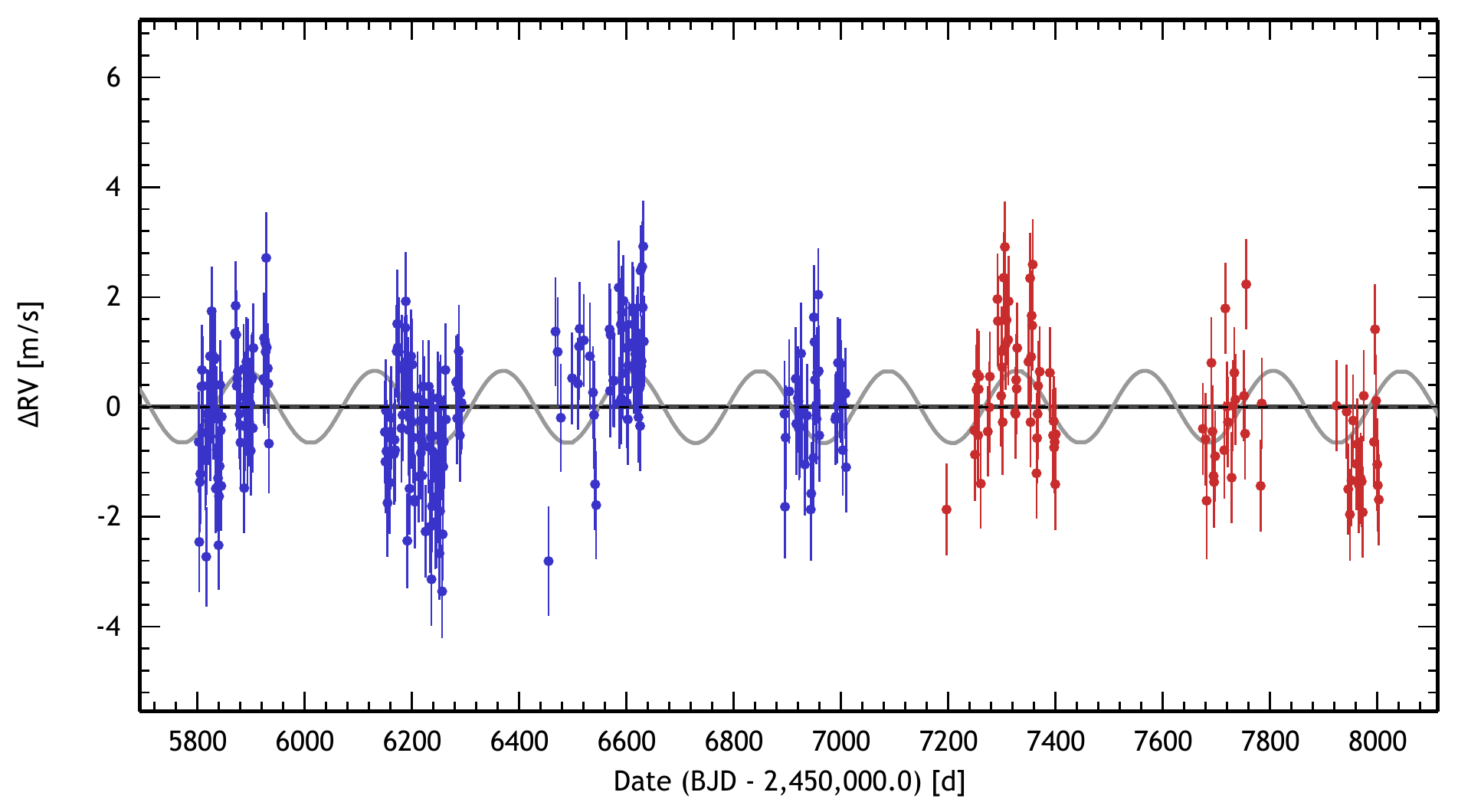} 
    \vspace{4ex}
  \end{minipage}
  \begin{minipage}[b]{0.5\linewidth}
    \centering
    \includegraphics[width=9cm]{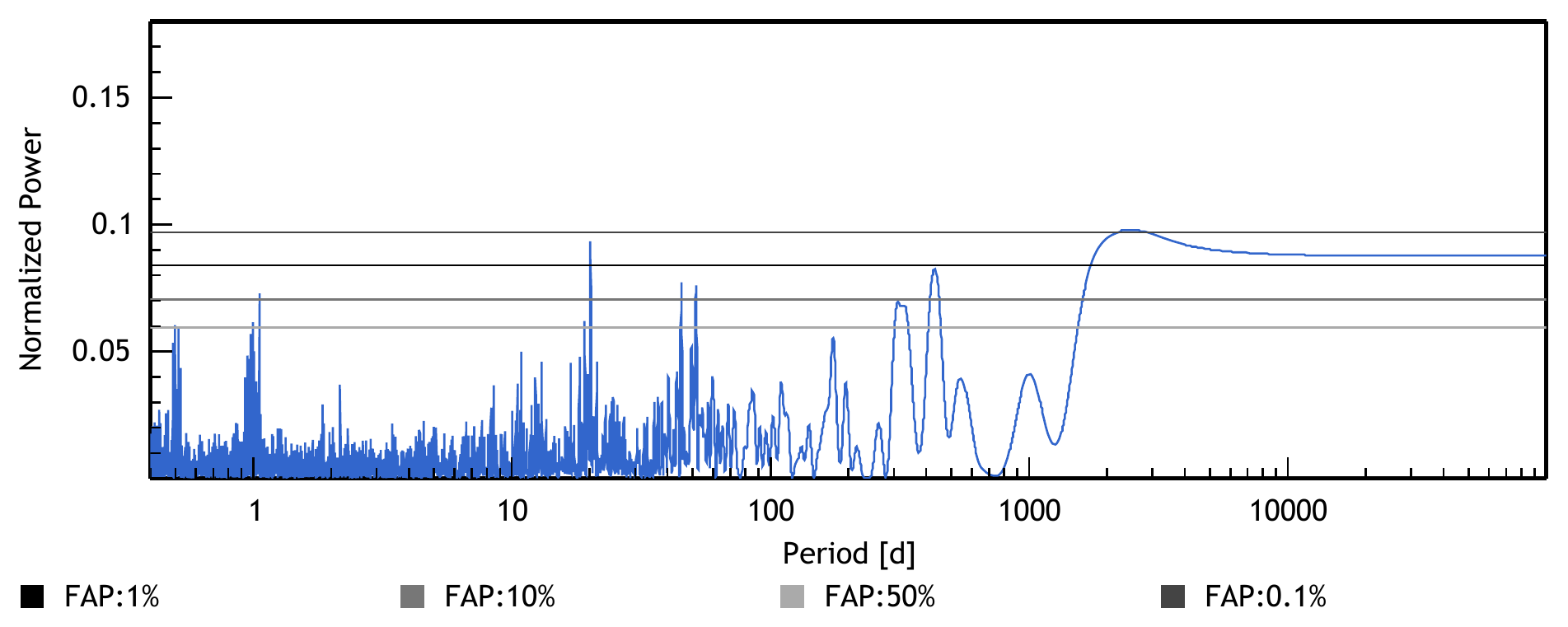} 
    \vspace{4ex}
  \end{minipage}
  \caption{On the left side, the HARPS RV data of HD~10700 obtained between 2011 to 2017 are shown being reduced with the standard (top), C19 (center) and new DRS (bottom). The data set before the change of the fibres is plotted in blue while the data set after the change of the fibres is plotted in red. The right-hand plots show the respective LSP of the residuals to the fitted Keplerian model.}
    \label{fig:7}
\end{figure*}

\subsection{HD~10700}
HD~10700, or $\uptau$~Ceti, is a G8{V} star located within 12~light-years (3.7~parsecs) from the Solar System. It has a very low ``raw'' RV dispersion just above \SI{1}{\m\per\s} \citep{Pepe2011, Tuomi2013}.\\
Comparing the data issued from the standard and new versions of the DRS, we found a few peaks in the LSP with a FAP below $1\%$. For the sake of simplicity and transparency, we fitted only the most significant peak at a period of $\sim\,240$\,days. This signal does not correspond to any of the planets claimed by \citet{Tuomi2013} and \cite{feng2017}, although it could be a one-year alias of the claimed planet f at 636.13 days (or vice-versa). Since, for the C19 DRS, we find an unlikely value for the eccentricity $e = \sim 0.96$ for this signal, we decided to fix $e = 0$ for all three data sets (standard, C19 and new DRS) in order not to penalize any of the data sets and make the most direct comparison. See Table \ref{tab:2} for the model parameters.\\
In terms of radial-velocity dispersion, after subtracting the 1-planet signal, we obtain 1.13/1.56  m\,s$^{-1}$ (before/after change of fibres) for the standard DRS and 1.03/0.98  m\,s$^{-1}$ for the new DRS, respectively. The two datasets, before/after the change of the fibres, are fitted jointly. The values obtained with the C19 DRS, which already includes the corrections of the CCD block-stitching error (see \citealt{Coffinet+2018}) but not the new combined TH+FP wavelength calibration, lie in between the results of the two other DRS.\\
The left side of Fig.\ref{fig:7} shows the radial velocities of HD~10700 and the fitted 1-planet model. The right side of the figure shows instead the LSP of the residuals to the fit for every version of the DRS, respectively, and FAPs of 50\%, 10\%, 1\% and 0.1\%. In the case of the new DRS, we have a significant signal at $\sim 20$ days, which may correspond to a signal reported by \citet{Tuomi2013, feng2017} and assigned to a potential planet HD\,10700\,g of. Similarly, the weaker signal at 51.41 days may correspond to the signal assigned to HD\,10700\,h. Based on the data obtained with the new DRS we could detect other signals reported by \citet{Tuomi2013, feng2017}. The long-period signal around $2000-2700$ days is significant but may be assigned either to stellar activity or instrumental effects. Only continued observations and additional observable (e.g. activity indicators) may discriminate between the various possibility. This study was not pursued, however, not being the scope of this paper.

\begin{table}[h]
\begin{center}
\begin{tabular}{ccccc}
\hline\hline
\multicolumn{3}{c}{HD10700 }\\
\hline
\hline
P&[day]&239.30$\pm$1.67\\
T$_{\rm p}$&[day]&55414.70\\
e&&0.00\\
$\omega$&[deg]&0.00\\
K&[m/s]&0.66$\pm$0.07\\
\hline
m\,$\sin{(i)}$&[M$_{\rm Earth}]$&6.40\\
a&[AU]&0.75\\
\hline
\end{tabular}
\end{center}
\caption{Model parameters of the Keplerian fit to the HD10700 radial-velocity data obtained with the new DRS.}
\label{tab:2}
\end{table}

\begin{table*}[h]
\begin{center}
\begin{tabular}{ccccc}
\hline\hline
\multicolumn{5}{c}{HD20794}\\
\hline
\hline
Keplerian&&HD20794\,b&HD20794\,d&HD20794\,e\\
\hline
P&[day]&18.34$\pm$0.01&90.20$\pm$0.18 &236.31$\pm$1.58\\
T$_{\rm p}$&[day]&55489.45&55439.53 &55433.82\\
e&&0.29$\pm$0.12&0.36$\pm$0.07 &0.46$\pm$0.09\\
$\omega$&[deg]&38.66$\pm$26.08&164.44$\pm$13.66 &301.89$\pm$13.87\\
K&[m/s]&0.60$\pm$0.08&0.98$\pm$0.10 &0.78$\pm$0.09\\
\hline
m\,$\sin{(i)}$&[M$_{\rm Earth}]$&2.36&6.41 &6.72\\
a&[AU]&0.14 &0.39 &0.75\\
\hline
\end{tabular}
\end{center}
\caption{Model parameters of the Keplerian fit to the HD20794 radial-velocity data obtained with the new DRS.}
\label{tab:3}
\end{table*}

\begin{figure*}[ht] 
   \begin{minipage}[b]{0.5\linewidth}
    \centering
    \includegraphics[width=6.8cm]{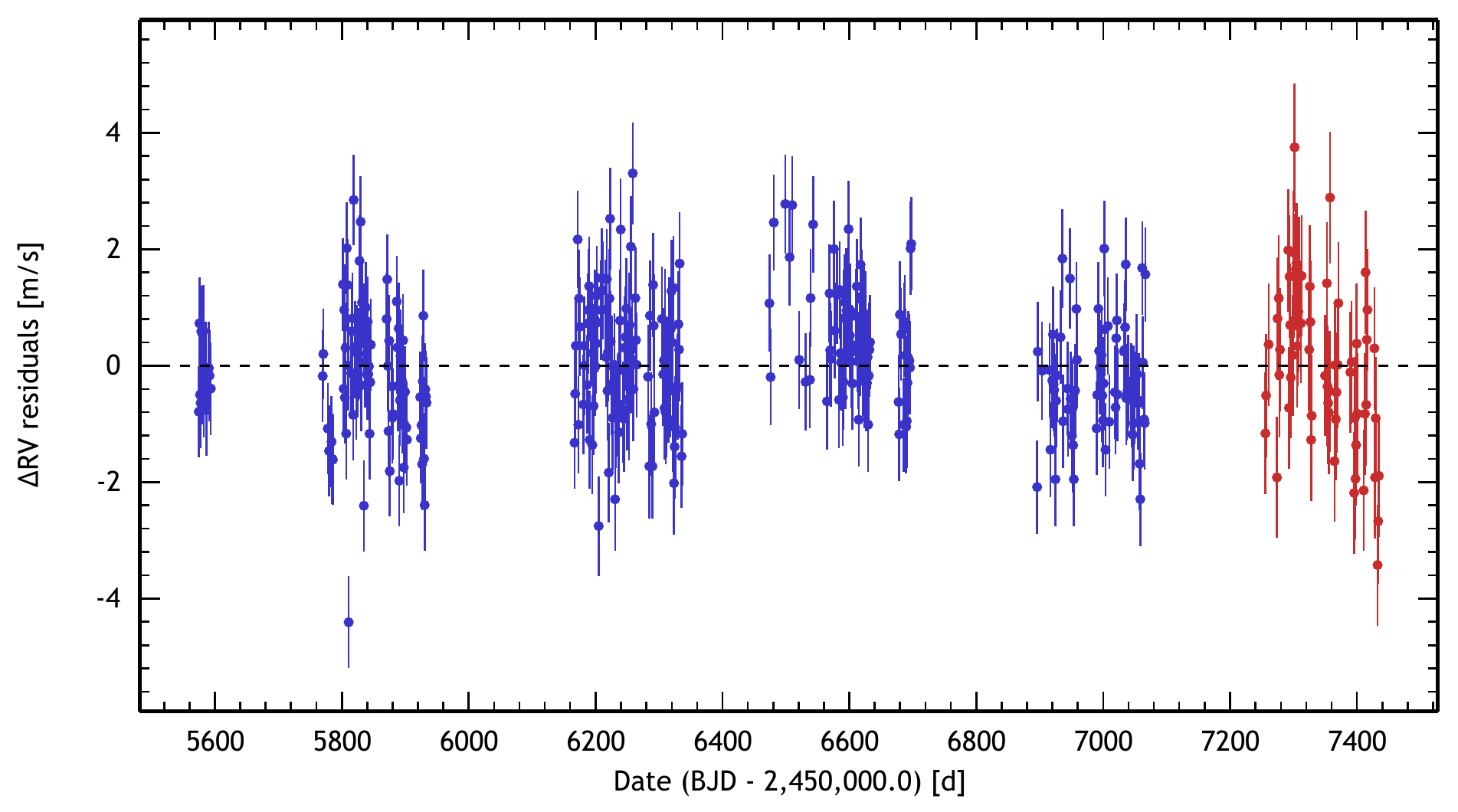} 
    \vspace{4ex}
  \end{minipage}
  \begin{minipage}[b]{0.5\linewidth}
    \centering
    \includegraphics[width=9cm]{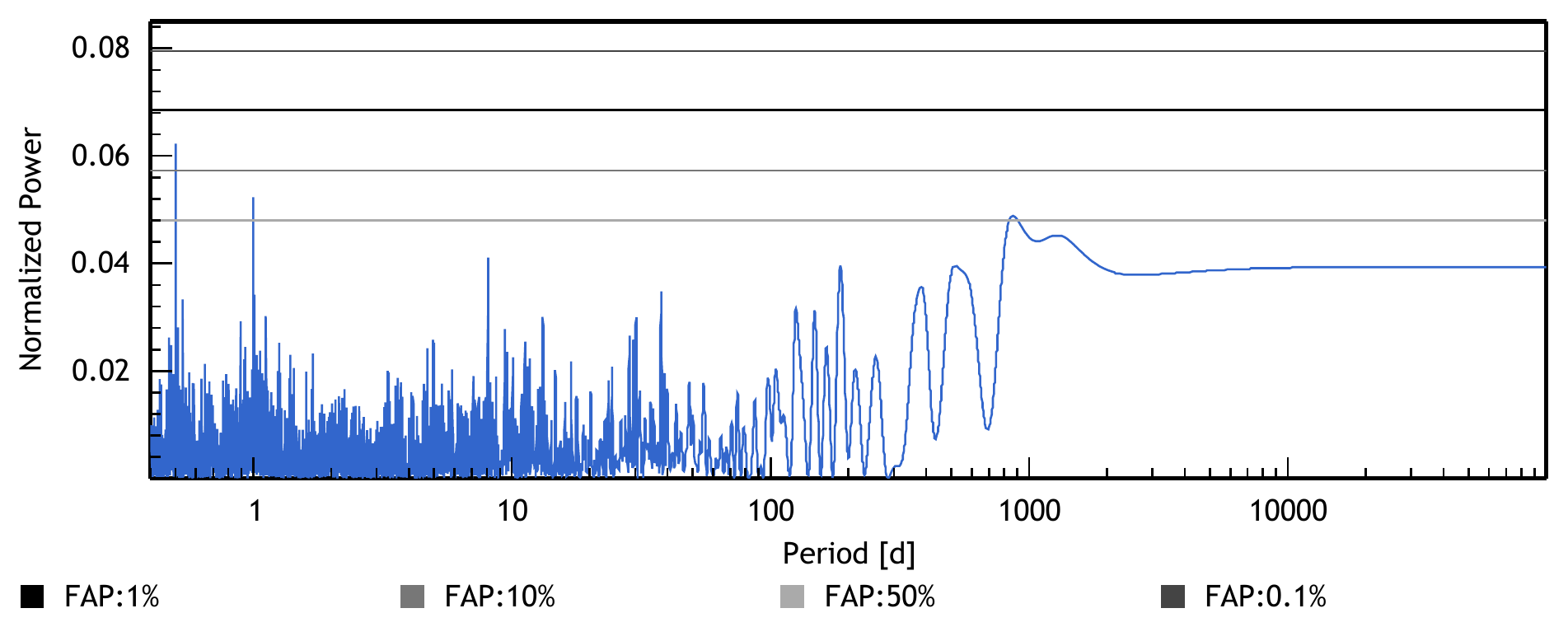} 
    \vspace{4ex}
  \end{minipage} 
  \begin{minipage}[b]{0.5\linewidth}
    \centering
    \includegraphics[width=6.8cm]{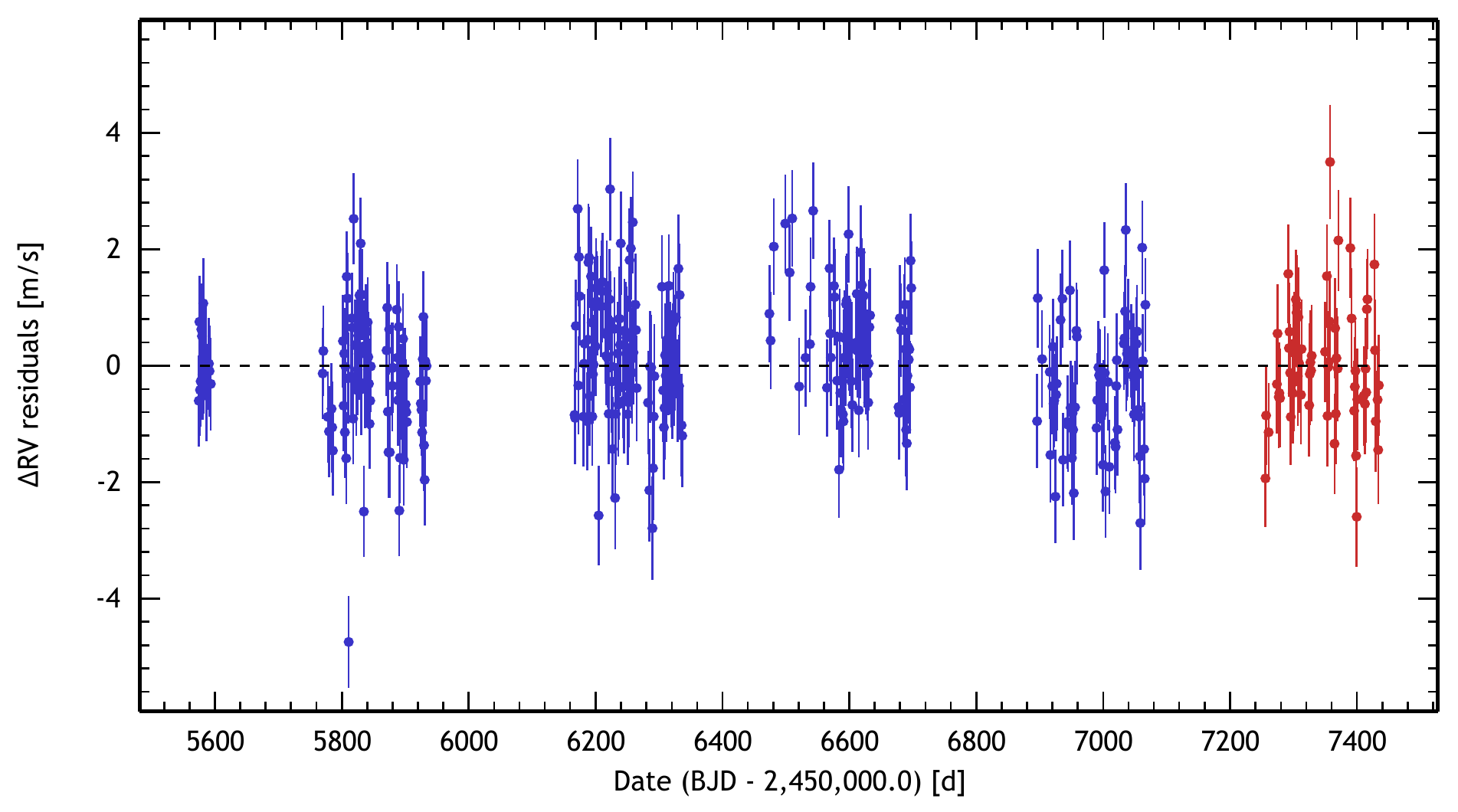} 
    \vspace{4ex}
  \end{minipage}
  \begin{minipage}[b]{0.5\linewidth}
    \centering
    \includegraphics[width=9cm]{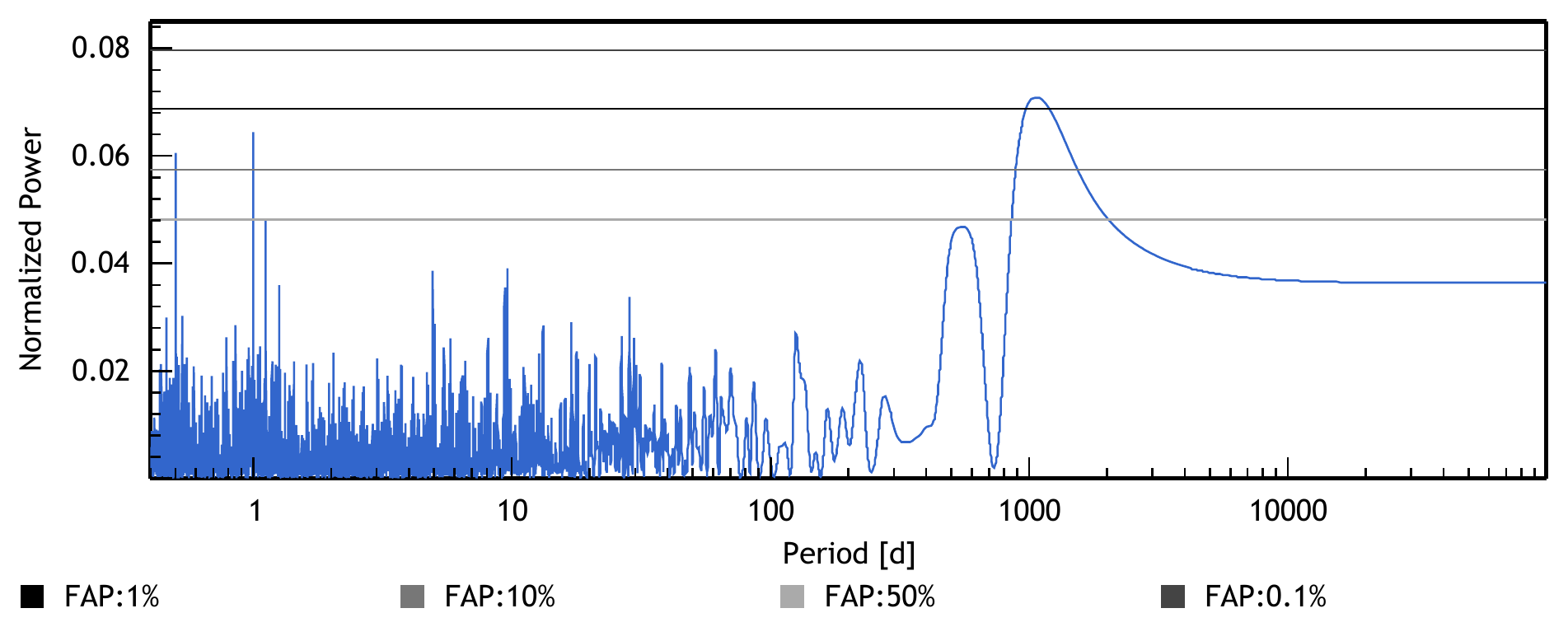} 
    \vspace{4ex}
  \end{minipage}
\begin{minipage}[b]{0.5\linewidth}
    \centering
    \includegraphics[width=6.8cm]{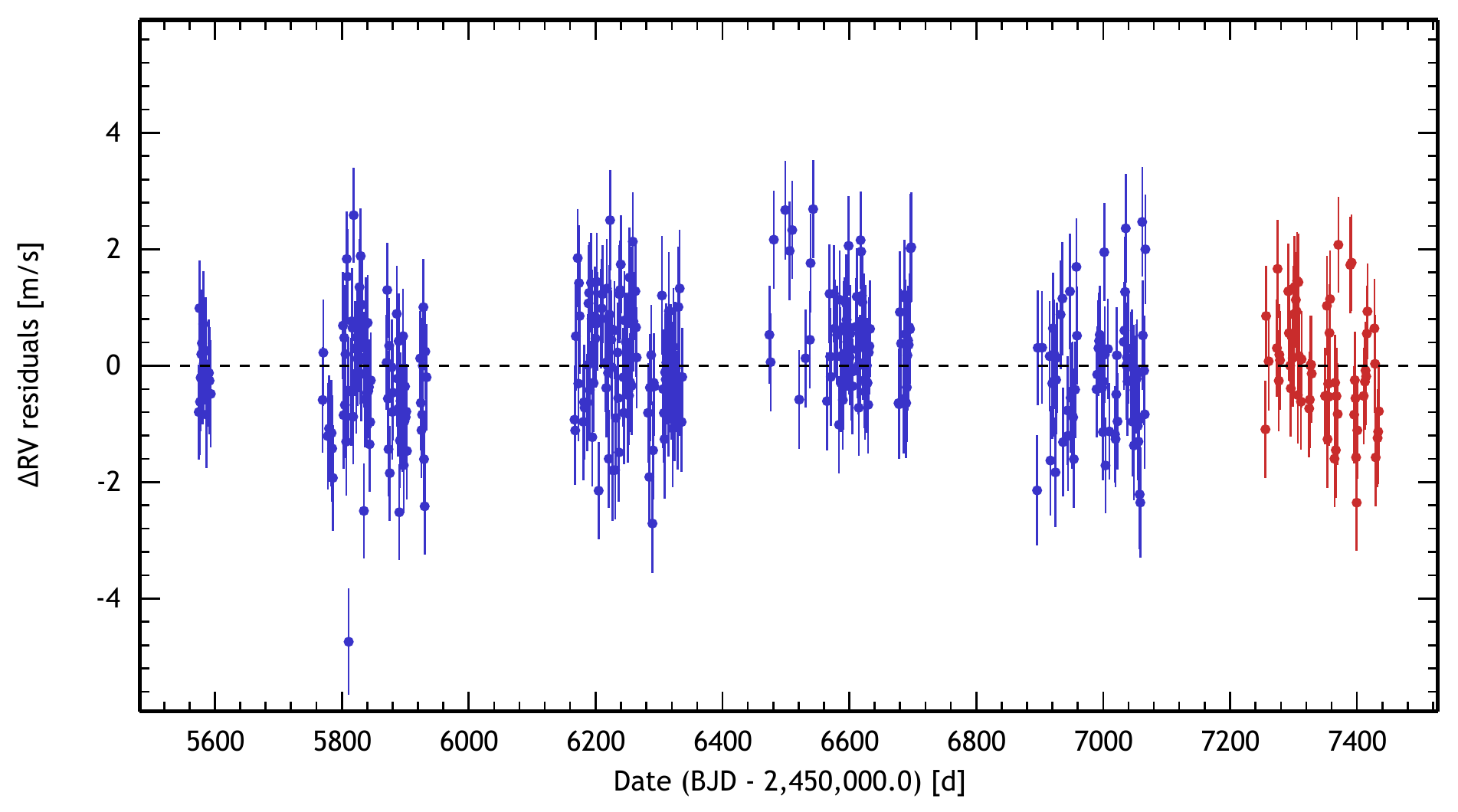} 
    \vspace{4ex}
  \end{minipage}
  \begin{minipage}[b]{0.5\linewidth}
    \centering
    \includegraphics[width=9cm]{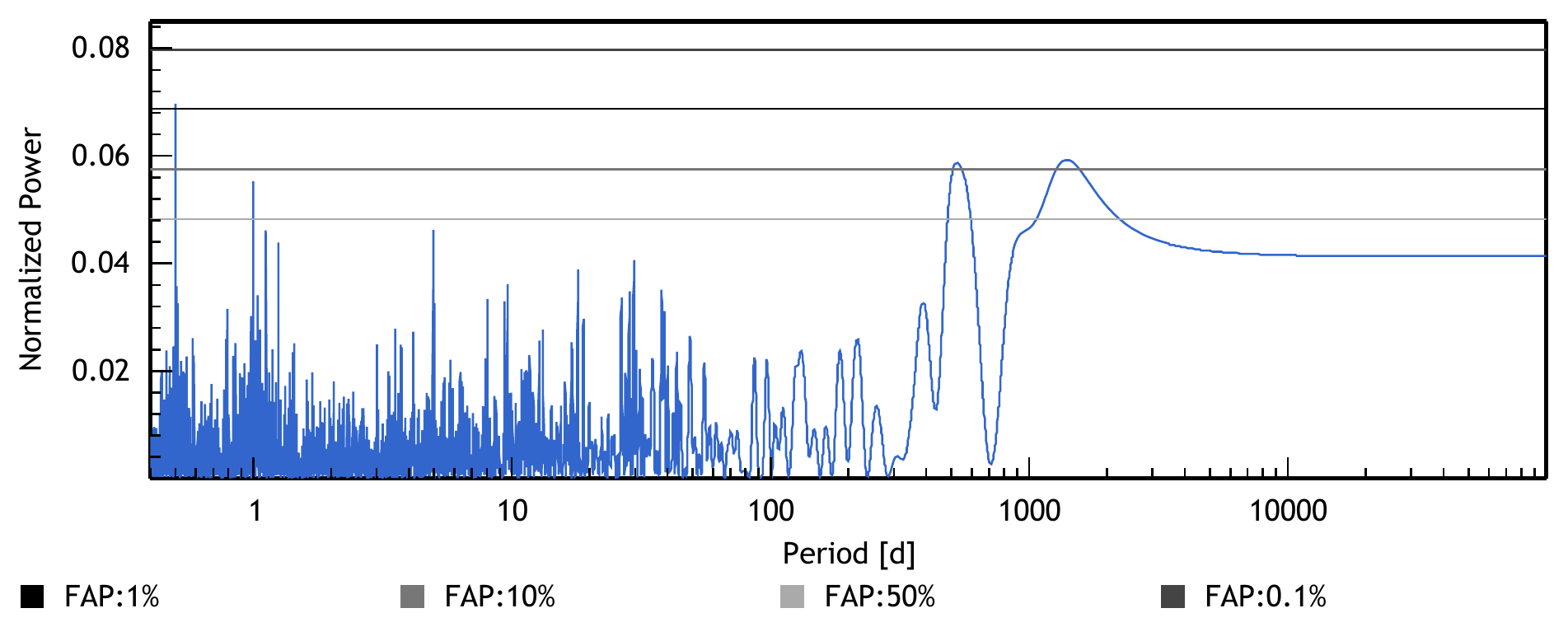} 
    \vspace{4ex}
  \end{minipage}
  \caption{On the left, the HARPS RV data of HD~20794. The plotted period, the order of the different DRS and the colour of the two different datasets (before and after the change of the fibres) is the same as in Fig.\ref{fig:7}.
  The right-hand plots show the respective LSP of the residuals to the fitted Keplerian model with the two known planets at periods of 18.32 and 89.7 days and another significant signal at 230.58\,days.}
    \label{fig:8}
\end{figure*}

\subsection{HD~20794}
HD~20794 is a bright G8 dwarf $\sim \,20$ light years away from Earth. In 2011, the discovery of two, possibly three planets orbiting this star was announced \citep{Pepe2011}. As part of our test, we re-reduced  the data set taken after 2011 with the three DRS versions and compared the result.\\
The two signals at $\sim \, 18$ and $\sim \, 89$ days period, corresponding to planets b and d, respectively, could easily be confirmed, while the signal at $\sim\,40$\,days period, which had been reported by \cite{Pepe2011} to be at the limit of significance, is not detected. Instead, another signal at $\sim\,231$ days appeared clearly in all the data sets. This signal may correspond to that around 250\,days found by \citet{Feng2017pepe} and proposed to be the yearly alias of another signal at 147\,days. We found however that fitting the 231\,days signal perfectly removed residual power at other periods, while fitting the 147\,days produced worse results. We adopted therefore a 3-planets model with the parameters presented in Table \ref{tab:3}.
The radial-velocity dispersion, after subtracting the 3-planet signal, results to 1.05/1.36 \,m\,s$^{-1}$ (before/after change of fibres) for the standard DRS and 1.04/0.96  m\,s$^{-1}$ for the new DRS, respectively. The values obtained with the C19 DRS are slightly worse but similar to the new DRS.\\
The left side of Fig.\ref{fig:8} shows the radial velocities of HD~20794 and the fitted 3-planet model. The right side of the figure shows again the LSP of the residuals to the fit for every version of the DRS, respectively, with the usual FAP levels. We would like to note that no significant signal is left in the data produced with the new DRS, and that the power distribution is amazingly uniform.
 

\subsection{HD~69830}
HD~69830 is a yellow dwarf star located $\sim \,41$ light-years away \citep{van2007}. It is a G8-K0V star with magnitude $V=5.95$) harbouring 3 super-Earth/Neptune-mass planets \citep{Lovis2006}.\\
For all the versions of the DRS we fitted the planets at 8.7 and 31\,days period (planets b and c) reported by \citet{Lovis2006}, while the signal at 200 days (planet d) did not appear in a significant way, most probably due to the lower number of observations in the period from 2011 to 2017. In addition, the star has become more active in recent years showing higher stellar jitter. Therefore we remained on the two-planets model for all versions of the DRS. Again we refer to Table \ref{tab:4} for the model parameters.

\begin{figure*}[ht] 
   \begin{minipage}[b]{0.5\linewidth}
    \centering
    \includegraphics[width=6.8cm]{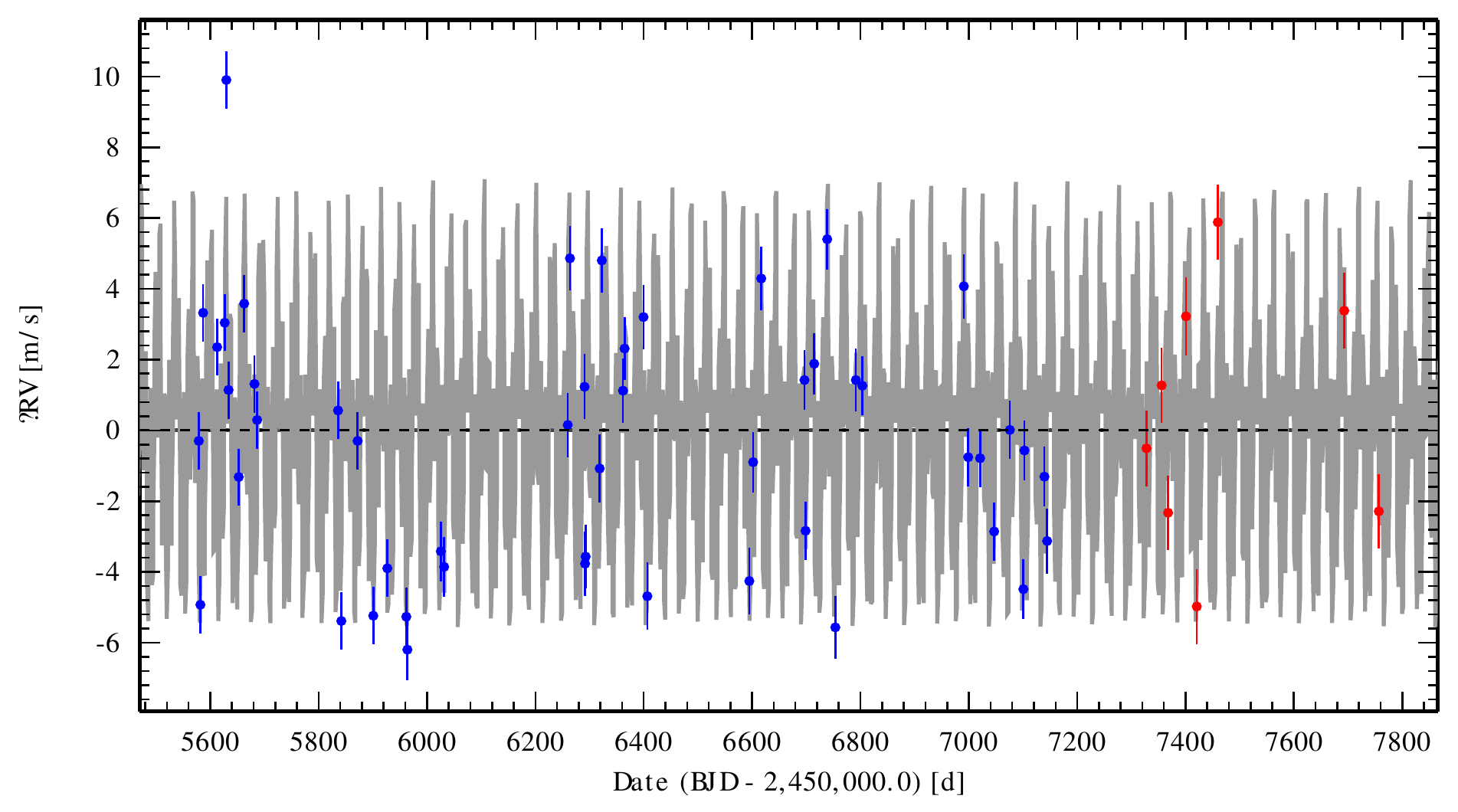} 
    \vspace{4ex}
  \end{minipage}
  \begin{minipage}[b]{0.5\linewidth}
    \centering
    \includegraphics[width=9cm]{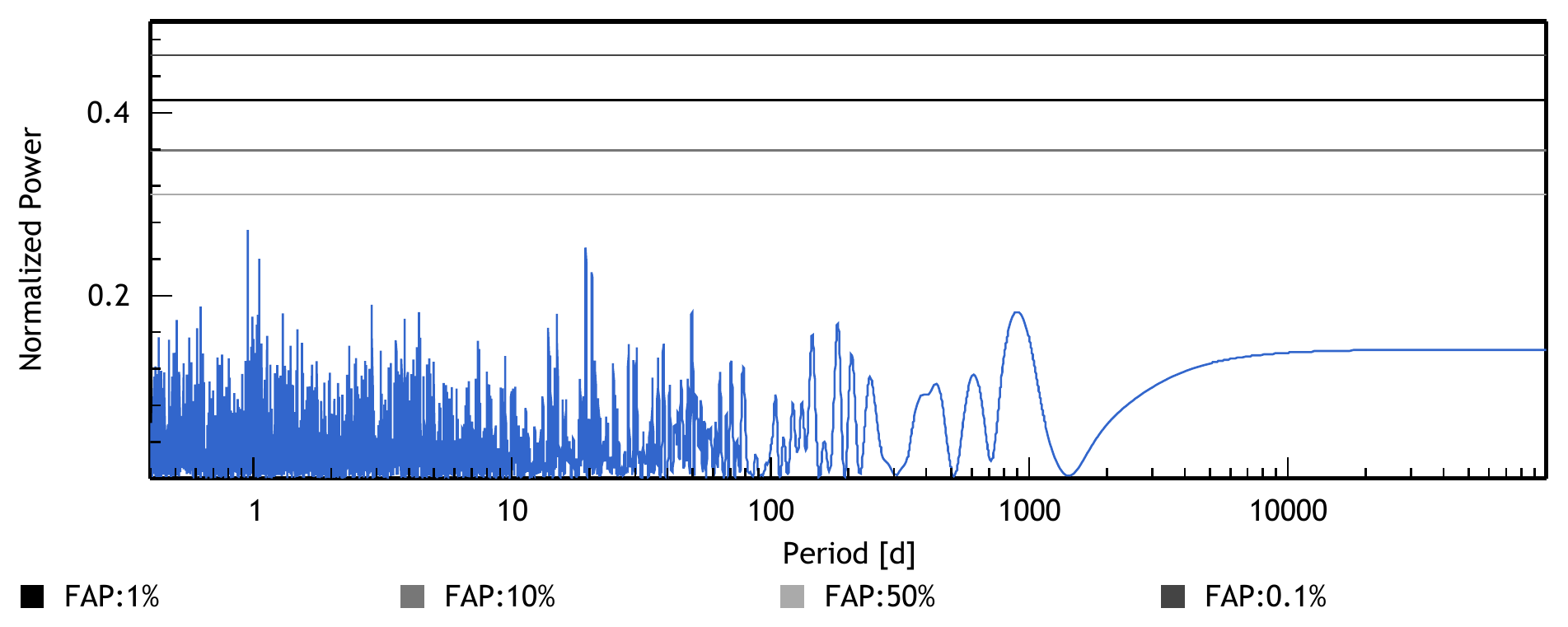} 
    \vspace{4ex}
  \end{minipage} 
  \begin{minipage}[b]{0.5\linewidth}
    \centering
    \includegraphics[width=6.8cm]{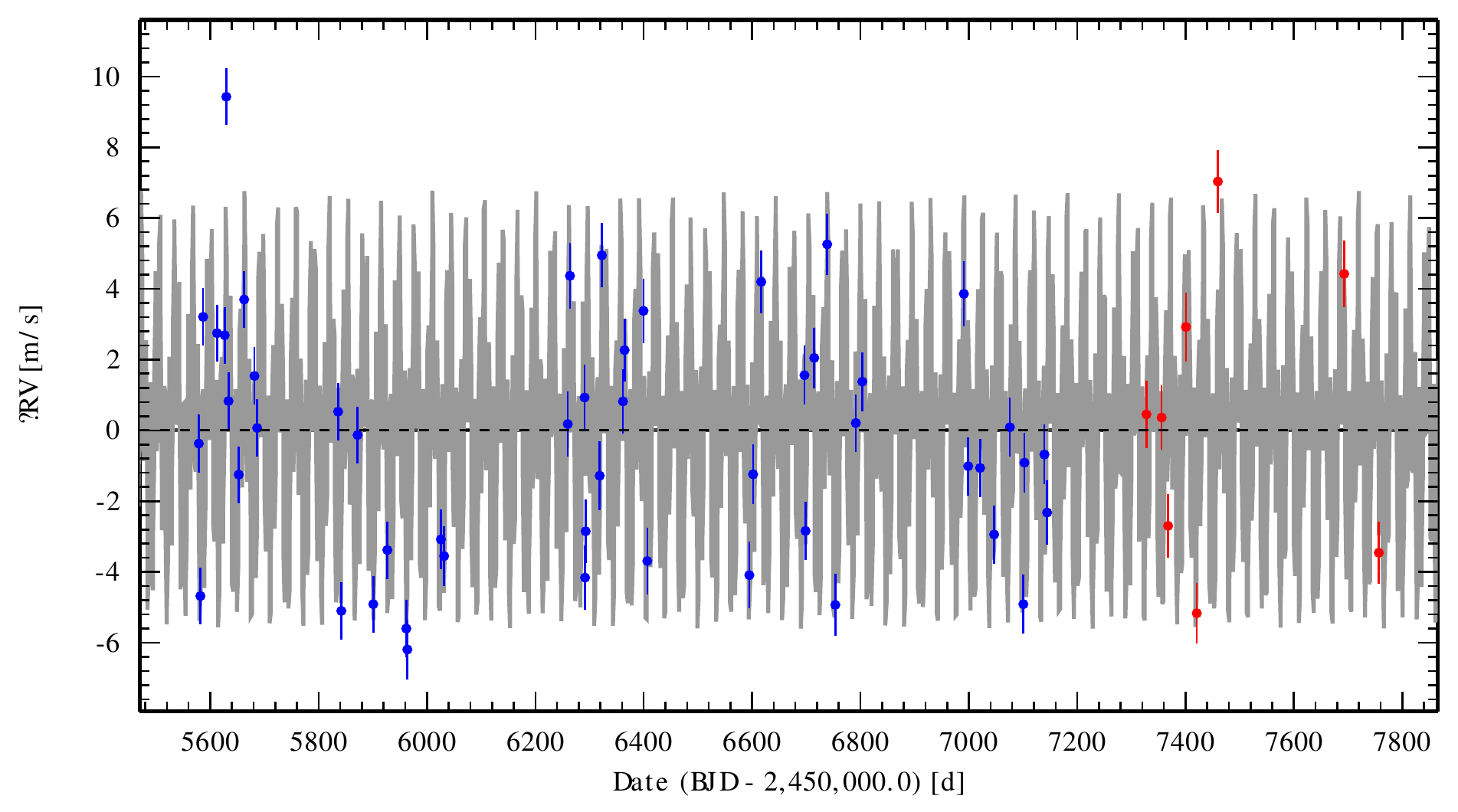} 
    \vspace{4ex}
  \end{minipage}
  \begin{minipage}[b]{0.5\linewidth}
    \centering
    \includegraphics[width=9cm]{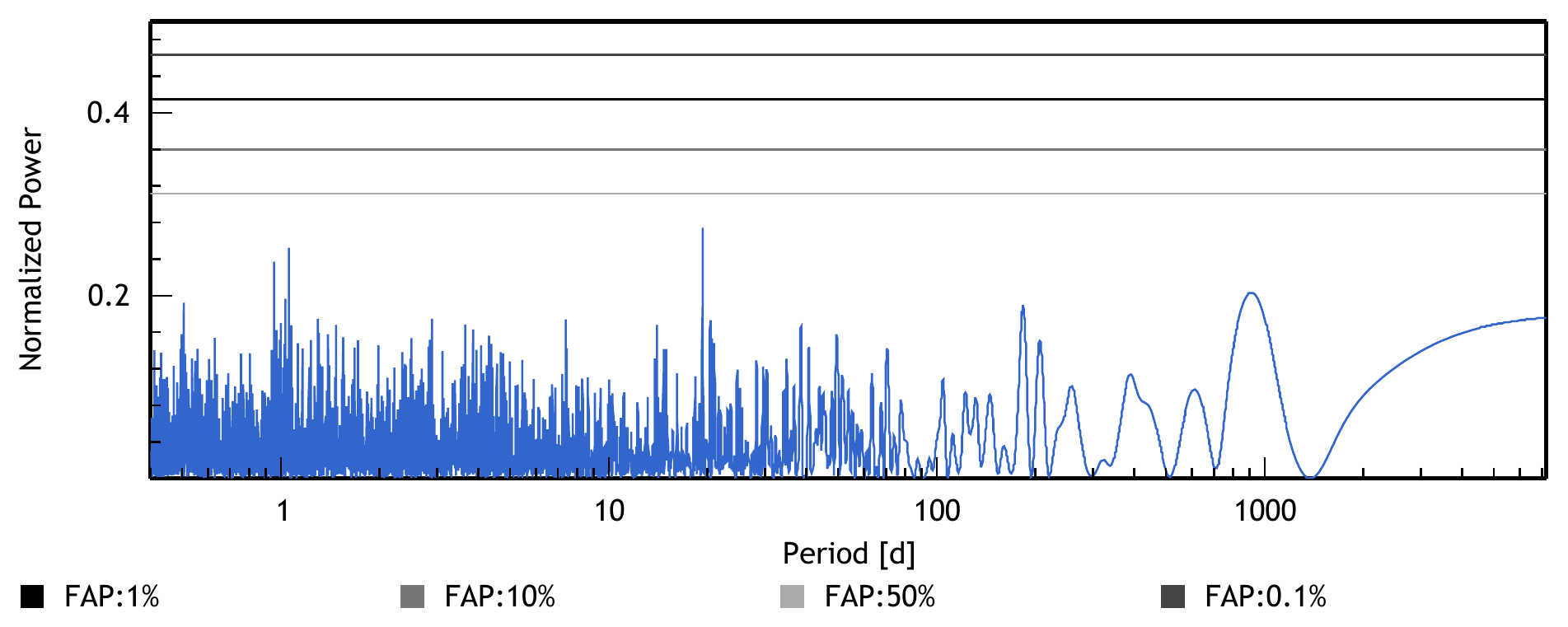} 
    \vspace{4ex}
  \end{minipage}
\begin{minipage}[b]{0.5\linewidth}
    \centering
    \includegraphics[width=6.8cm]{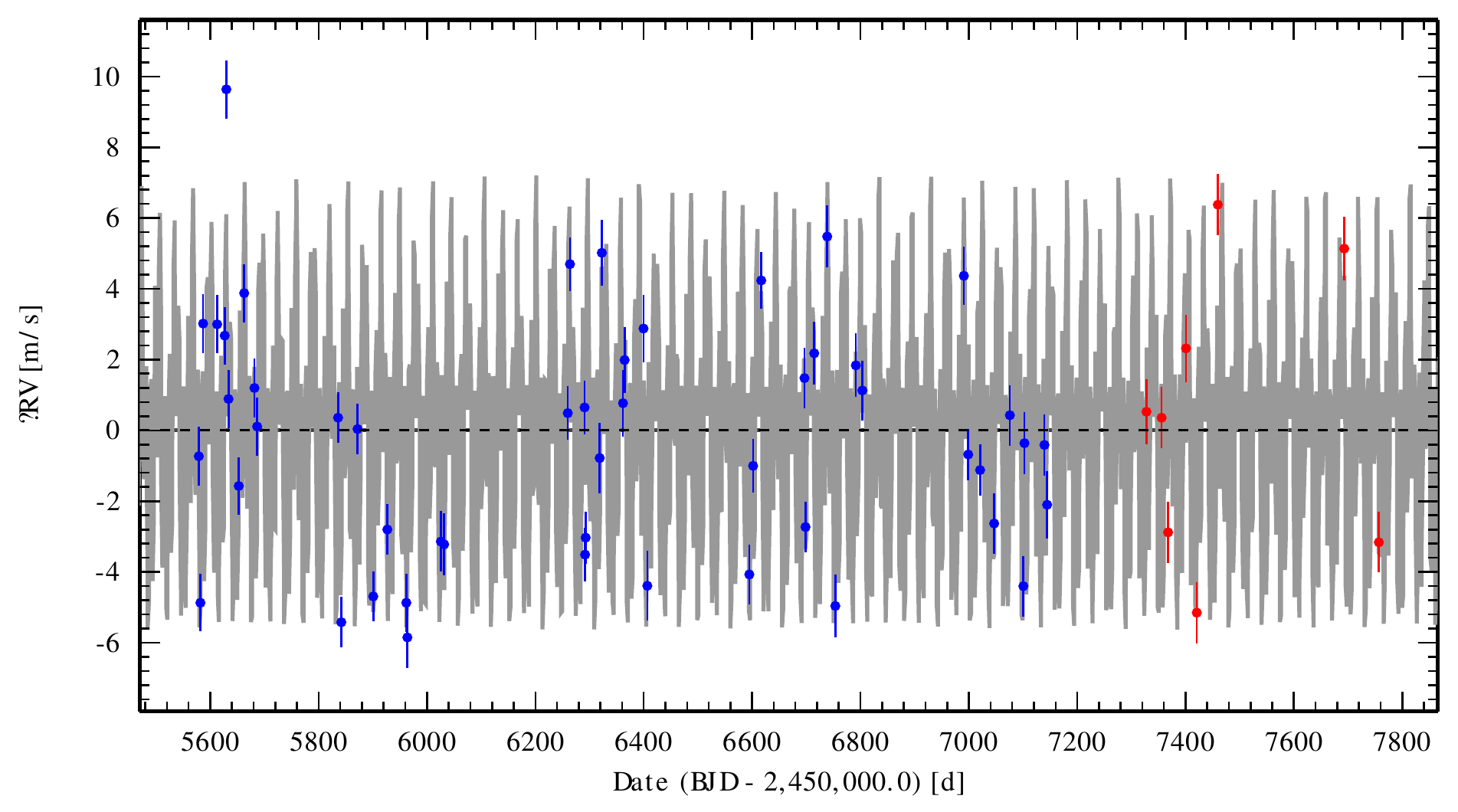} 
    \vspace{4ex}
  \end{minipage}
  \begin{minipage}[b]{0.5\linewidth}
    \centering
    \includegraphics[width=9cm]{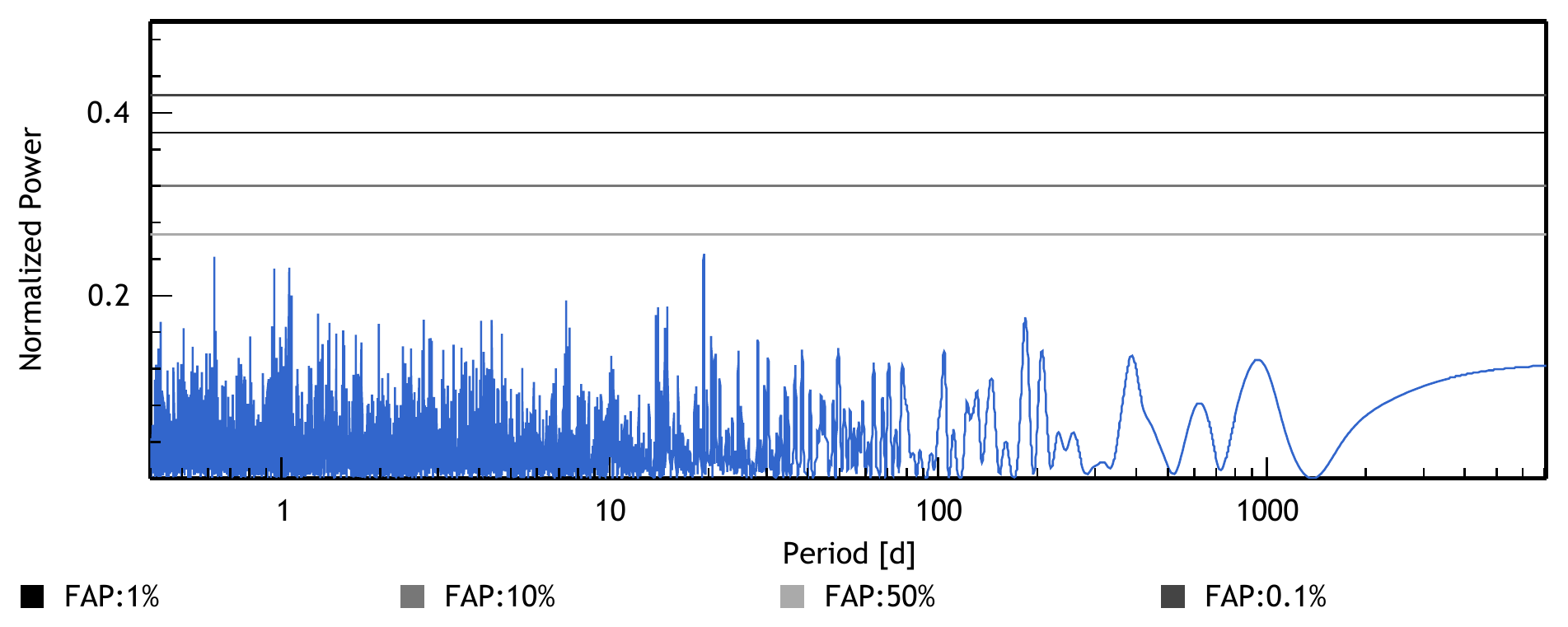} 
    \vspace{4ex}
  \end{minipage}
  \caption{On the left, the HARPS RV data of HD~69830. The plotted period, the order of the different DRS and the colour of the two different datasets (before and after the change of the fibres) is the same as in Fig.\ref{fig:7}. The right-hand plots show the respective LSP of the residuals to the fitted Keplerian model with the two known planets at 8.67 and 31.64 days period.}
    \label{fig:9}
\end{figure*}

The radial-velocity dispersion, after subtracting the 2-planet signal, is 1.56/1.32 \,m\,s$^{-1}$ (before/after change of fibres) for the standard DRS and 1.44/1.35  m\,s$^{-1}$ for the new DRS, respectively. The values obtained with C19 are slightly worse but similar to the new DRS.\\
The left side of Fig.\ref{fig:9} shows the radial velocities of HD~69830 and the fitted 2-planet model. The right side of the figure shows the LSP of the residuals to the fit for every version of the DRS, respectively, with the usual FAP levels. We have to note here that the improvement with the new DRS is still significant, but much less spectacular than for the other targets. The reason might be simply that the dispersion of this star is likely to be dominated by stellar jitter \citep{Tanner2014} rather than by ``calibration noise''.\\

\begin{table*}[h]
\begin{center}
\begin{tabular}{ccccc}
\hline\hline
\multicolumn{4}{c}{HD69830}\\
\hline
\hline
Keplerian&&HD69830\,b&HD69830\,c\\
\hline
P&[day]&8.672$\pm$0.001&31.64$\pm$0.01\\
T$_{\rm p}$&[day]&55492.40&55476.02\\
e&&0.15$\pm$0.04&0.31$\pm$0.05\\
$\omega$&[deg]&83.26$\pm$17.75&36.90$\pm$10.35\\
K&[m/s]&3.42$\pm$0.138&3.02$\pm$0.19\\
\hline
m\,$\sin{(i)}$&[M$_{\rm Earth}]$&10.87&14.20\\
a&[AU]&0.08&0.20\\
\hline
\end{tabular}
\end{center}
\caption{Model parameters of the Keplerian fit to the HD69830 radial-velocity data obtained with the new DRS}
\label{tab:4}
\end{table*}

The analysis of the three targets gives a clear indication for the improvements obtained with the new calibration strategy, although the results differ slightly depending on the star. We find however the following common aspects for all the targets:
\begin{enumerate}
    \item The standard DRS shows much higher residuals on data obtained after the fiber change. This is simply due to the fact that the standard DRS had not been optimized for but rather only adapted to the new spectral format. By consequence, the initial guess for the reference wavelengths of the thorium lines had not been updated, leading by a matter of fact to poorer calibration quality.
    \item The most important improvement of the C19 and new DRS with respect to the standard DRS is obtained on post-6/2015 data. This is a direct consequence of the previous point.
    \item The global dispersion is systematically lower for the C19 DRS and lowest for the new DRS.
    \item Even when considering only pre-6/2015 data, the dispersion is reduced by the C19 DRS and is lowest on data obtained with the new DRS.
    \item It is important to note that for the new DRS the dispersion is not only reduced but also similar in both pre- and post-6/2018 data.
\end{enumerate}

\section{Conclusion}
Hollow-cathode thorium lamps (HCL) have
successfully served for wavelength calibration and drift measurements in astronomical spectrographs despite their non-ideal characteristics. Fabry-P\'erot etalons combined with HCL are good alternatives that produce regularly spaced
calibration lines covering the entire
spectral range of the spectrograph. Since the cavity width of the etalon is not known accurately and considering possibly drifting over time, the peak wavelengths are uncertain. Therefore, Fabry-P\'erot interferometers cannot be used as an absolute calibrator. The combination with HCL provides the possibility of ``anchoring'' of the wavelength solution in absolute terms. We have developed and presented in our paper a method that allows to combining the precision provided by the etalon with the accuracy provided by the HCL.\\


We demonstrated that the developed calibration method works very well if we include also the CCD block-stitching correction developed by  \citet{Coffinet+2018}. We have first tested the new wavelength solution on the HARPS LFC, demonstrating that the dispersion of difference between the LFC-lines wavelengths after calibration and their theoretical wavelengths is reduced by more than a factor of 2 and that the residual to the wavelength solution is flatter along the spectral orders. The measured systematic offset of 38 m\,s$^{-1}$ of the LFC lines with respect to their nominal wavelength is not understood but at least it is identical for all versions of the DRS. We conclude that this offset arises either from the reference wavelengths of the thorium or of those of the LFC.\\
Finally, we applied the new wavelength calibration to three stable stars of the HARPS Program. Also this test demonstrated that the radial-velocity dispersion is reduced in all cases, although the improvement is obviously more visible on very quiet stars.\\
 
We conclude that the new DRS significantly improved the quality of the radial velocities and should be considered in future to be used as standard pipeline for HARPS after confirming the functionalities and the performances also in statistical terms over the full HARPS data set. The use of a ``calibrated'' FP can be considered less expensive alternative to broad-band LFCs for precise and accurate wavelength calibration and may be considered for existing instruments, such as e.g HARPS-N, CORALIE, ESPRESSO, SPIRou, as well as for future instruments such as NIRPS or HIRES@ELT.



\begin{acknowledgements}
    This publication makes use of the Data \& Analysis Center for Exoplanets (DACE), which is a facility based at the University of Geneva (CH) dedicated to extrasolar planets data visualisation, exchange and analysis. DACE is a platform of the Swiss National Centre of Competence in Research (NCCR) PlanetS, federating the Swiss expertise in Exoplanet research. The DACE platform is available at \url{https://dace.unige.ch}. The NCCR are a research instrument of the Swiss National Science Foundation. We acknowledge the Swiss National Science Foundation for their continuous support through project funding.
\end{acknowledgements}



\bibliographystyle{aa} 
\bibliography{bibliography.bib} 

\begin{thebibliography}{31}
\expandafter\ifx\csname natexlab\endcsname\relax\def\natexlab#1{#1}\fi

\bibitem[{{Baranne} {et~al.}(1996){Baranne}, {Queloz}, {Mayor}, {Adrianzyk},
  {Knispel}, {Kohler}, {Lacroix}, {Meunier}, {Rimbaud}, \& {Vin}}]{baranne96}
{Baranne}, A., {Queloz}, D., {Mayor}, M., {et~al.} 1996, \aaps, 119, 373

\bibitem[{{Bauer} {et~al.}(2015){Bauer}, {Zechmeister}, \& {Reiners}}]{Bauer}
{Bauer}, F.~F., {Zechmeister}, M., \& {Reiners}, A. 2015, \aap, 581, A117

\bibitem[{{Caballero} {et~al.}(2016){Caballero}, {Gu{\`a}rdia}, {L{\'o}pez del
  Fresno}, {Zechmeister}, {de Juan}, {Alonso-Floriano}, {Amado}, {Colom{\'e}},
  {Cort{\'e}s-Contreras}, {Garc{\'{\i}}a-Piquer}, {Gesa}, {de Guindos},
  {Hagen}, {Helmling}, {Hern{\'a}ndez Casta{\~n}o}, {K{\"u}rster},
  {L{\'o}pez-Santiago}, {Montes}, {Morales Mu{\~n}oz}, {Pavlov}, {Quirrenbach},
  {Reiners}, {Ribas}, {Seifert}, \& {Solano}}]{Caballero}
{Caballero}, J.~A., {Gu{\`a}rdia}, J., {L{\'o}pez del Fresno}, M., {et~al.}
  2016, in \procspie, Vol. 9910, Observatory Operations: Strategies, Processes,
  and Systems VI, 99100E

\bibitem[{{Cersullo} {et~al.}(2017){Cersullo}, {Wildi}, {Chazelas}, \&
  {Pepe}}]{cersullo2017}
{Cersullo}, F., {Wildi}, F., {Chazelas}, B., \& {Pepe}, F. 2017, \aap, 601,
  A102

\bibitem[{{Chazelas} {et~al.}(2012){Chazelas}, {Pepe}, \&
  {Wildi}}]{Chazelas2012}
{Chazelas}, B., {Pepe}, F., \& {Wildi}, F. 2012, in \procspie, Vol. 8450,
  Modern Technologies in Space- and Ground-based Telescopes and Instrumentation
  II, 845013

\bibitem[{{Coffinet} {et~al.}(2019){Coffinet}, {Lovis}, {Dumusque}, \&
  {Pepe}}]{Coffinet+2018}
{Coffinet}, A., {Lovis}, C., {Dumusque}, X., \& {Pepe}, F. 2019, {arXiv
  e-prints} (accepted by \aap), arXiv:1901.03294

\bibitem[{{Feng} {et~al.}(2017{\natexlab{a}}){Feng}, {Tuomi}, \&
  {Jones}}]{Feng2017pepe}
{Feng}, F., {Tuomi}, M., \& {Jones}, H.~R.~A. 2017{\natexlab{a}}, \aap, 605,
  A103

\bibitem[{{Feng} {et~al.}(2017{\natexlab{b}}){Feng}, {Tuomi}, {Jones},
  {Barnes}, {Anglada-Escud{\'e}}, {Vogt}, \& {Butler}}]{feng2017}
{Feng}, F., {Tuomi}, M., {Jones}, H.~R.~A., {et~al.} 2017{\natexlab{b}}, \aj,
  154, 135

\bibitem[{{Fischer} {et~al.}(2017){Fischer}, {Jurgenson}, {McCracken},
  {Sawyer}, {Blackman}, \& {Szymkowiak}}]{EXPRES}
{Fischer}, D., {Jurgenson}, C., {McCracken}, T., {et~al.} 2017, in American
  Astronomical Society Meeting Abstracts, Vol. 229, American Astronomical
  Society Meeting Abstracts \#229, 126.04

\bibitem[{{Ireland} {et~al.}(2016){Ireland}, {Artigau}, {Burley}, {Edgar},
  {Margheim}, {Robertson}, {Pazder}, {McDermid}, \& {Zhelem}}]{ghost}
{Ireland}, M.~J., {Artigau}, {\'E}., {Burley}, G., {et~al.} 2016, in \procspie,
  Vol. 9908, Ground-based and Airborne Instrumentation for Astronomy VI, 99087A

\bibitem[{{Kerber} {et~al.}(2007){Kerber}, {Nave}, {Sansonetti}, {Bristow}, \&
  {Rosa}}]{kerber2007}
{Kerber}, F., {Nave}, G., {Sansonetti}, C.~J., {Bristow}, P., \& {Rosa}, M.~R.
  2007, in Astronomical Society of the Pacific Conference Series, Vol. 364, The
  Future of Photometric, Spectrophotometric and Polarimetric Standardization,
  ed. C.~{Sterken}, 461

\bibitem[{{Lo Curto} {et~al.}(2012){Lo Curto}, {Manescau}, {Avila}, {Pasquini},
  {Wilken}, {Steinmetz}, {Holzwarth}, {Probst}, {Udem}, {H{\"a}nsch},
  {Gonz{\'a}lez Hern{\'a}ndez}, {Esposito}, {Rebolo}, {Canto Martins}, \& {de
  Medeiros}}]{Locurto2012}
{Lo Curto}, G., {Manescau}, A., {Avila}, G., {et~al.} 2012, in \procspie, Vol.
  8446, Ground-based and Airborne Instrumentation for Astronomy IV, 84461W

\bibitem[{{Lo Curto} {et~al.}(2015){Lo Curto}, {Pepe}, {Avila}, {Boffin},
  {Bovay}, {Chazelas}, {Coffinet}, {Fleury}, {Hughes}, {Lovis}, {Maire},
  {Manescau}, {Pasquini}, {Rihs}, {Sinclaire}, \& {Udry}}]{Locurto2015}
{Lo Curto}, G., {Pepe}, F., {Avila}, G., {et~al.} 2015, The Messenger, 162, 9

\bibitem[{{Lovis} {et~al.}(2006){Lovis}, {Mayor}, {Pepe}, {Alibert}, {Benz},
  {Bouchy}, {Correia}, {Laskar}, {Mordasini}, {Queloz}, {Santos}, {Udry},
  {Bertaux}, \& {Sivan}}]{Lovis2006}
{Lovis}, C., {Mayor}, M., {Pepe}, F., {et~al.} 2006, \nat, 441, 305

\bibitem[{{Marconi} {et~al.}(2016){Marconi}, {Di Marcantonio}, {D'Odorico},
  {Cristiani}, {Maiolino}, {Oliva}, {Origlia}, {Riva}, {Valenziano}, {Zerbi},
  {Abreu}, {Adibekyan}, {Allende Prieto}, {Amado}, {Benz}, {Boisse}, {Bonfils},
  {Bouchy}, {Buchhave}, {Buscher}, {Cabral}, {Canto Martins}, {Chiavassa},
  {Coelho}, {Christensen}, {Delgado-Mena}, {de Medeiros}, {Di Varano},
  {Figueira}, {Fisher}, {Fynbo}, {Glasse}, {Haehnelt}, {Haniff}, {Hansen},
  {Hatzes}, {Huke}, {Korn}, {Le{\~a}o}, {Liske}, {Lovis}, {Maslowski},
  {Matute}, {McCracken}, {Martins}, {Monteiro}, {Morris}, {Morris}, {Nicklas},
  {Niedzielski}, {Nunes}, {Palle}, {Parr-Burman}, {Parro}, {Parry}, {Pepe},
  {Piskunov}, {Queloz}, {Quirrenbach}, {Rebolo Lopez}, {Reiners}, {Reid},
  {Santos}, {Seifert}, {Sousa}, {Stempels}, {Strassmeier}, {Sun}, {Udry},
  {Vanzi}, {Vestergaard}, {Weber}, \& {Zackrisson}}]{Marconi2016}
{Marconi}, A., {Di Marcantonio}, P., {D'Odorico}, V., {et~al.} 2016, in
  \procspie, Vol. 9908, Ground-based and Airborne Instrumentation for Astronomy
  VI, 990823

\bibitem[{{Mayor} {et~al.}(2003){Mayor}, {Pepe}, {Queloz}, {Bouchy},
  {Rupprecht}, {Lo Curto}, {Avila}, {Benz}, {Bertaux}, {Bonfils}, {Dall},
  {Dekker}, {Delabre}, {Eckert}, {Fleury}, {Gilliotte}, {Gojak}, {Guzman},
  {Kohler}, {Lizon}, {Longinotti}, {Lovis}, {Megevand}, {Pasquini}, {Reyes},
  {Sivan}, {Sosnowska}, {Soto}, {Udry}, {van Kesteren}, {Weber}, \&
  {Weilenmann}}]{Mayor2003}
{Mayor}, M., {Pepe}, F., {Queloz}, D., {et~al.} 2003, The Messenger, 114, 20

\bibitem[{{Palmer} \& {Engleman}(1983)}]{PE83}
{Palmer}, B.~A. \& {Engleman}, R. 1983, {Atlas of the Thorium spectrum}

\bibitem[{{Pepe} {et~al.}(2011){Pepe}, {Lovis}, {S{\'e}gransan}, {Benz},
  {Bouchy}, {Dumusque}, {Mayor}, {Queloz}, {Santos}, \& {Udry}}]{Pepe2011}
{Pepe}, F., {Lovis}, C., {S{\'e}gransan}, D., {et~al.} 2011, \aap, 534, A58

\bibitem[{{Pepe} {et~al.}(2000){Pepe}, {Mayor}, {Delabre}, {Kohler}, {Lacroix},
  {Queloz}, {Udry}, {Benz}, {Bertaux}, \& {Sivan}}]{Pepe+2000}
{Pepe}, F., {Mayor}, M., {Delabre}, B., {et~al.} 2000, in Society of
  Photo-Optical Instrumentation Engineers (SPIE) Conference Series, Vol. 4008,
  Optical and IR Telescope Instrumentation and Detectors, ed. M.~{Iye} \& A.~F.
  {Moorwood}, 582--592

\bibitem[{{Pepe} {et~al.}(2014){Pepe}, {Molaro}, {Cristiani}, {Rebolo},
  {Santos}, {Dekker}, {M{\'e}gevand}, {Zerbi}, {Cabral}, {Di Marcantonio},
  {Abreu}, {Affolter}, {Aliverti}, {Allende Prieto}, {Amate}, {Avila},
  {Baldini}, {Bristow}, {Broeg}, {Cirami}, {Coelho}, {Conconi}, {Coretti},
  {Cupani}, {D'Odorico}, {De Caprio}, {Delabre}, {Dorn}, {Figueira}, {Fragoso},
  {Galeotta}, {Genolet}, {Gomes}, {Gonz{\'a}lez Hern{\'a}ndez}, {Hughes},
  {Iwert}, {Kerber}, {Landoni}, {Lizon}, {Lovis}, {Maire}, {Mannetta},
  {Martins}, {Monteiro}, {Oliveira}, {Poretti}, {Rasilla}, {Riva}, {Santana
  Tschudi}, {Santos}, {Sosnowska}, {Sousa}, {Span{\'o}}, {Tenegi}, {Toso},
  {Vanzella}, {Viel}, \& {Zapatero Osorio}}]{Pepe2014}
{Pepe}, F., {Molaro}, P., {Cristiani}, S., {et~al.} 2014, Astronomische
  Nachrichten, 335, 8

\bibitem[{{Queloz} {et~al.}(2001){Queloz}, {Mayor}, {Udry}, {Burnet},
  {Carrier}, {Eggenberger}, {Naef}, {Santos}, {Pepe}, {Rupprecht}, {Avila},
  {Baeza}, {Benz}, {Bertaux}, {Bouchy}, {Cavadore}, {Delabre}, {Eckert},
  {Fischer}, {Fleury}, {Gilliotte}, {Goyak}, {Guzman}, {Kohler}, {Lacroix},
  {Lizon}, {Megevand}, {Sivan}, {Sosnowska}, \& {Weilenmann}}]{Queloz2001}
{Queloz}, D., {Mayor}, M., {Udry}, S., {et~al.} 2001, The Messenger, 105, 1

\bibitem[{{Ravi} {et~al.}(2017){Ravi}, {Phillips}, {Beck}, {Martin}, {Cecconi},
  {Ghedina}, {Molinari}, {Bartels}, {Sasselov}, {Szentgyorgyi}, \&
  {Walsworth}}]{ravi2017}
{Ravi}, A., {Phillips}, D.~F., {Beck}, M., {et~al.} 2017, Journal of
  Astronomical Telescopes, Instruments, and Systems, 3, 045003

\bibitem[{{Redman} {et~al.}(2014){Redman}, {Nave}, \& {Sansonetti}}]{RNS14}
{Redman}, S.~L., {Nave}, G., \& {Sansonetti}, C.~J. 2014, \apjs, 211, 4

\bibitem[{{Seemann} {et~al.}(2014){Seemann}, {Anglada-Escude}, {Baade},
  {Bristow}, {Dorn}, {Follert}, {Gojak}, {Grunhut}, {Hatzes}, {Heiter}, {Ives},
  {Jeep}, {Jung}, {K{\"a}ufl}, {Kerber}, {Klein}, {Lizon}, {Lockhart},
  {L{\"o}winger}, {Marquart}, {Oliva}, {Paufique}, {Piskunov}, {Pozna},
  {Reiners}, {Smette}, {Smoker}, {Stempels}, \& {Valenti}}]{Seemann2014}
{Seemann}, U., {Anglada-Escude}, G., {Baade}, D., {et~al.} 2014, in \procspie,
  Vol. 9147, Ground-based and Airborne Instrumentation for Astronomy V, 91475G

\bibitem[{{Szentgyorgyi} {et~al.}(2016){Szentgyorgyi}, {Baldwin}, {Barnes},
  {Bean}, {Ben-Ami}, {Brennan}, {Budynkiewicz}, {Chun}, {Conroy}, {Crane},
  {Epps}, {Evans}, {Evans}, {Foster}, {Frebel}, {Gauron}, {Guzm{\'a}n}, {Hare},
  {Jang}, {Jang}, {Jordan}, {Kim}, {Kim}, {Mendes de Oliveira},
  {Lopez-Morales}, {McCracken}, {McMuldroch}, {Miller}, {Mueller}, {Oh},
  {Onyuksel}, {Ordway}, {Park}, {Park}, {Park}, {Paxson}, {Phillips},
  {Plummer}, {Podgorski}, {Seifahrt}, {Stark}, {Steiner}, {Uomoto},
  {Walsworth}, \& {Yu}}]{gmt}
{Szentgyorgyi}, A., {Baldwin}, D., {Barnes}, S., {et~al.} 2016, in \procspie,
  Vol. 9908, Ground-based and Airborne Instrumentation for Astronomy VI, 990822

\bibitem[{{Tanner} {et~al.}(2014){Tanner}, {Boyajian}, {von Braun}, {van
  Belle}, {Beichman}, {Fischer}, {Brewer}, \& {GSU CHARA Team}}]{Tanner2014}
{Tanner}, A.~M., {Boyajian}, T.~S., {von Braun}, K., {et~al.} 2014, in American
  Astronomical Society Meeting Abstracts, Vol. 223, American Astronomical
  Society Meeting Abstracts \#223, 347.27

\bibitem[{{Tuomi} {et~al.}(2013){Tuomi}, {Jones}, {Jenkins}, {Tinney},
  {Butler}, {Vogt}, {Barnes}, {Wittenmyer}, {O'Toole}, {Horner}, {Bailey},
  {Carter}, {Wright}, {Salter}, \& {Pinfield}}]{Tuomi2013}
{Tuomi}, M., {Jones}, H.~R.~A., {Jenkins}, J.~S., {et~al.} 2013, \aap, 551, A79

\bibitem[{{van Leeuwen}(2007)}]{van2007}
{van Leeuwen}, F., ed. 2007, Astrophysics and Space Science Library, Vol. 350,
  {Hipparcos, the New Reduction of the Raw Data}

\bibitem[{{Wildi} {et~al.}(2010){Wildi}, {Pepe}, {Chazelas}, {Lo Curto}, \&
  {Lovis}}]{wildi2010}
{Wildi}, F., {Pepe}, F., {Chazelas}, B., {Lo Curto}, G., \& {Lovis}, C. 2010,
  in \procspie, Vol. 7735, Ground-based and Airborne Instrumentation for
  Astronomy III, 77354X

\bibitem[{{Wildi} {et~al.}(2011){Wildi}, {Pepe}, {Chazelas}, {Lo Curto}, \&
  {Lovis}}]{wildi2011}
{Wildi}, F., {Pepe}, F., {Chazelas}, B., {Lo Curto}, G., \& {Lovis}, C. 2011,
  in Techniques and Instrumentation for Detection of Exoplanets V, Vol. 8151,
  81511F

\bibitem[{{Wilken} {et~al.}(2010){Wilken}, {Lovis}, {Manescau}, {Steinmetz},
  {Pasquini}, {Lo Curto}, {H{\"a}nsch}, {Holzwarth}, \& {Udem}}]{wilken2010}
{Wilken}, T., {Lovis}, C., {Manescau}, A., {et~al.} 2010, \mnras, 405, L16

\end{thebibliography}

\appendix
\section{Additional Data}
\label{app:c} 
Fig. \ref{fig:3}  shows  the  difference  between  the  theoretical  wavelengths  and  the  ones  of  the  laser’s  lines  along  a subset of  spectral order of HARPS covered by the LFC. In this appendix, we report the comparison between the solution obtained after  calibration  of  the spectra using the new DRS (combination of both thorium and Fabry-P\'erot) and the standard DRS (thorium only) along each spectral order of HARPS covered by the LFC (orders 25$^{th}$ to 69$^{th}$).

\begin{figure*}
  \centering
 \subcaptionbox*{25th order index}[.3\linewidth][c]{%
    \includegraphics[width=.32\linewidth]{25.pdf}}\quad
  \subcaptionbox*{26th order index}[.3\linewidth][c]{%
    \includegraphics[width=.32\linewidth]{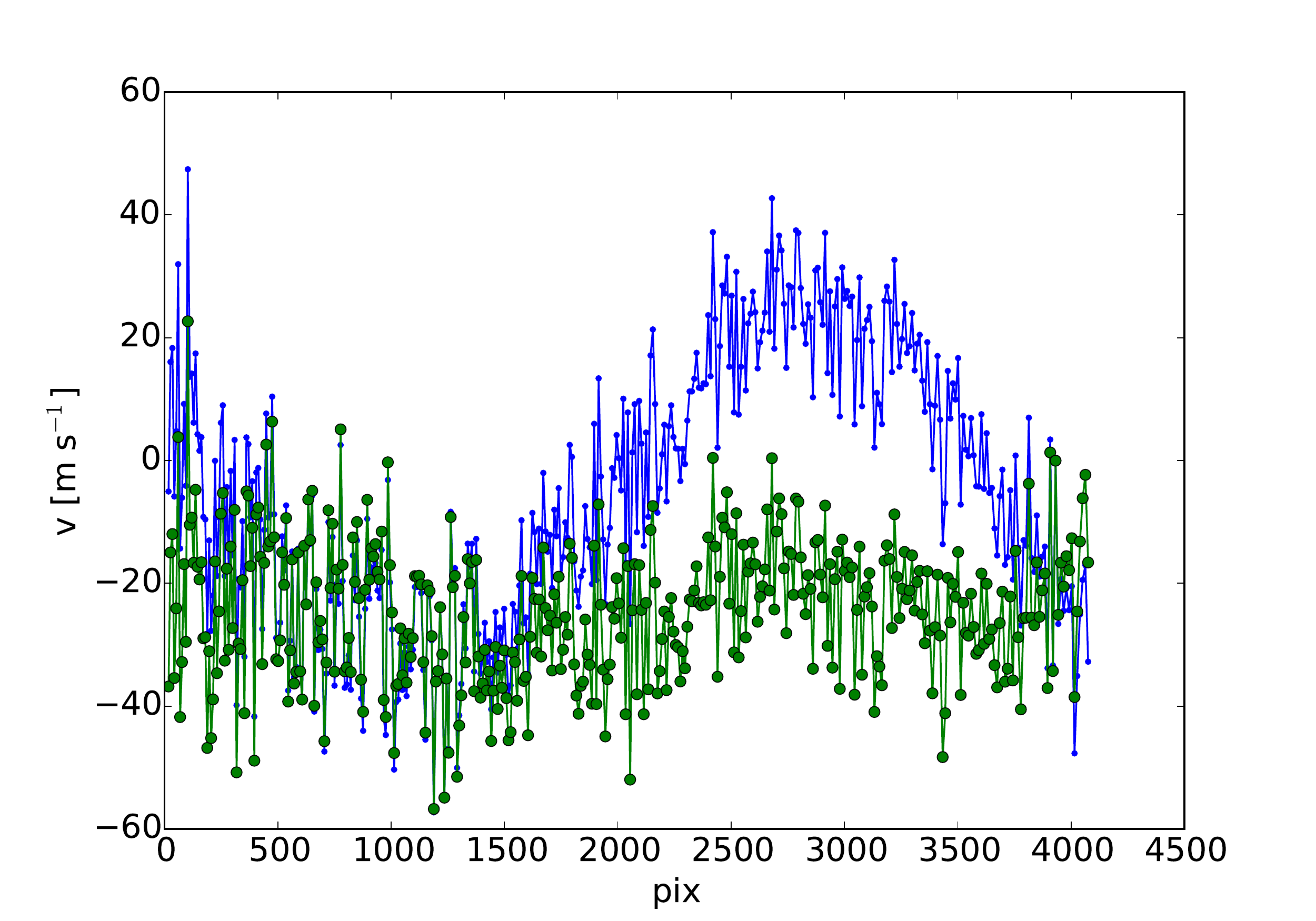}}\quad
  \subcaptionbox*{27th order index}[.3\linewidth][c]{%
    \includegraphics[width=.32\linewidth]{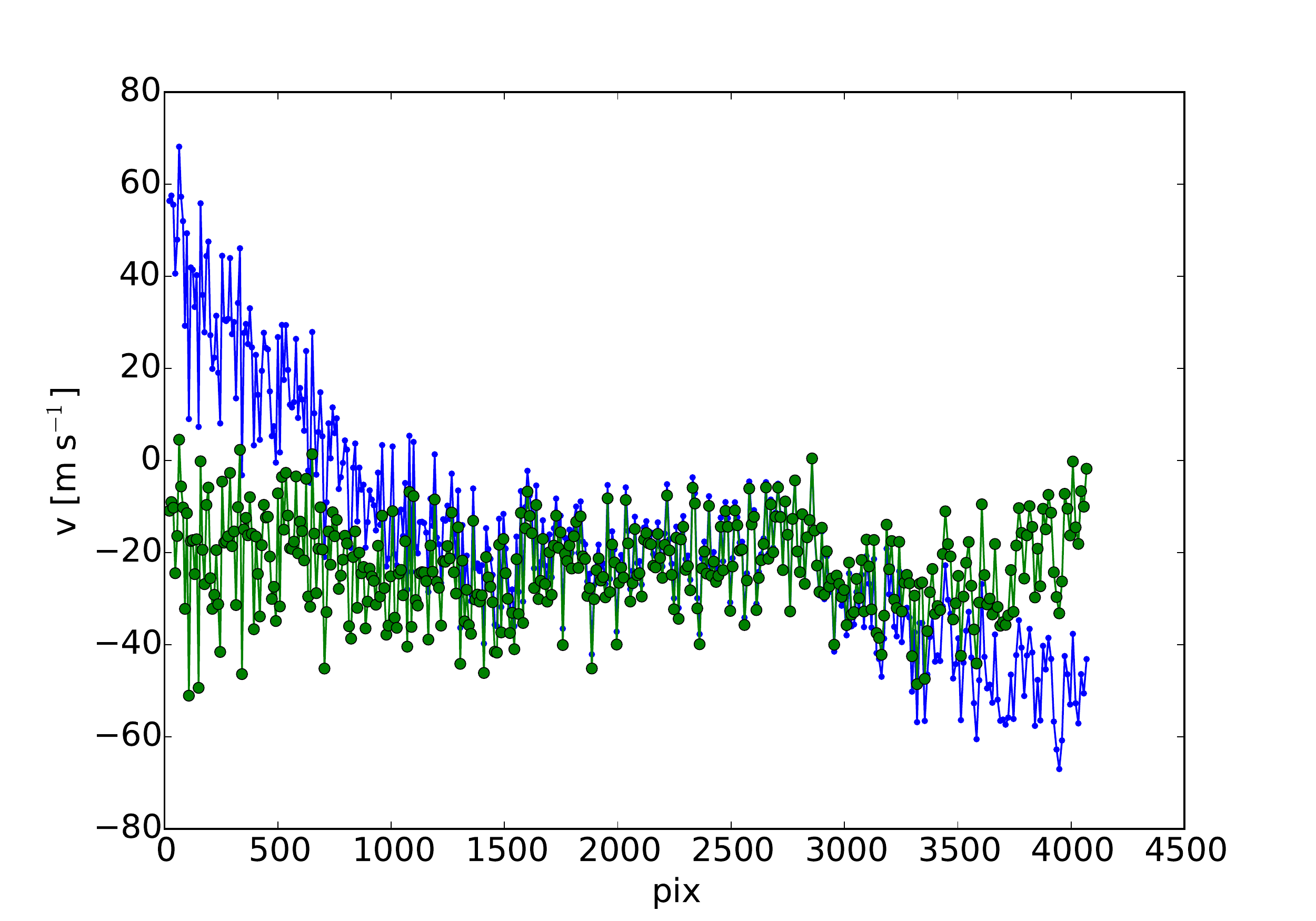}}

  \bigskip

   \subcaptionbox*{28th order index}[.3\linewidth][c]{%
    \includegraphics[width=.32\linewidth]{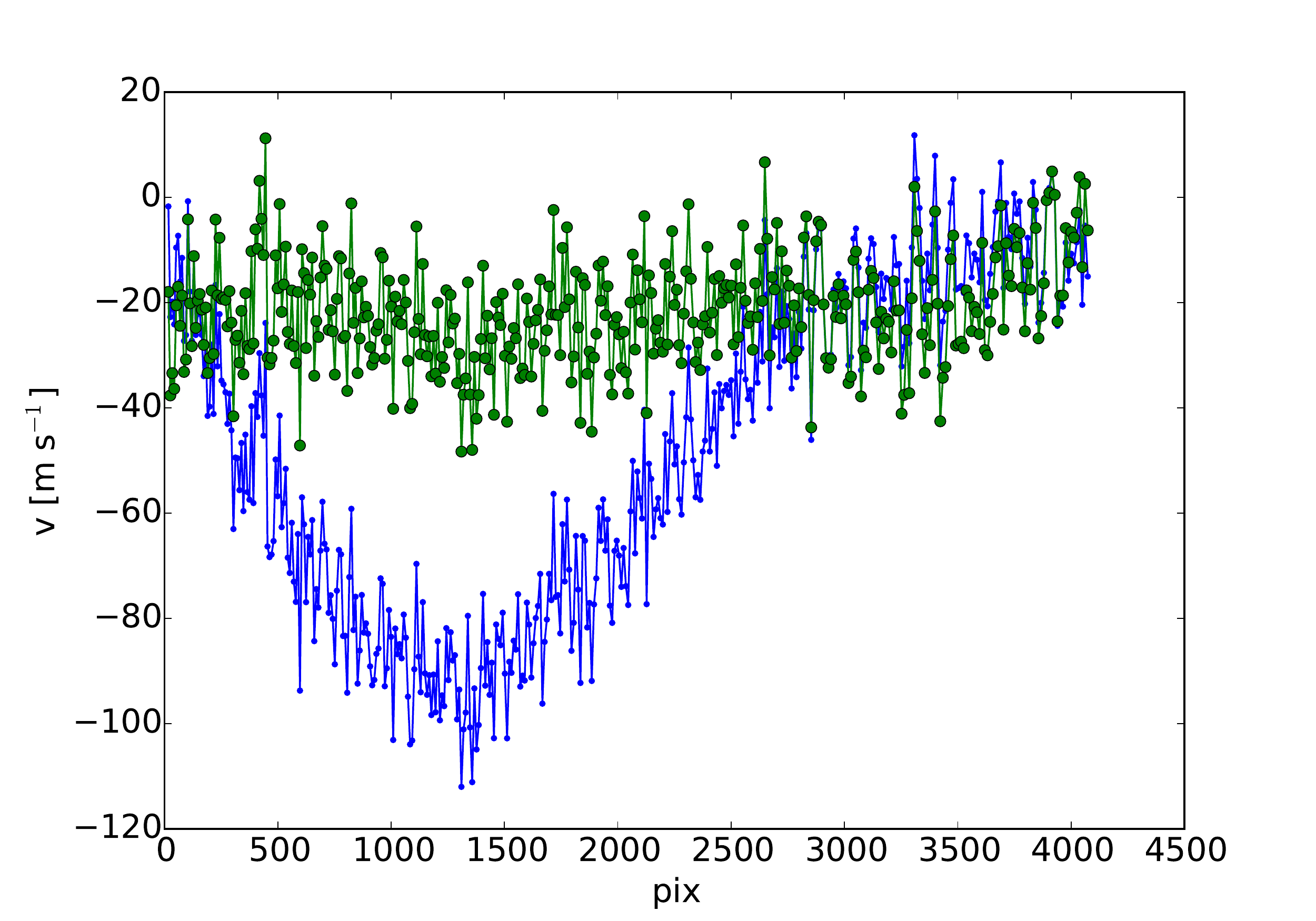}}\quad
  \subcaptionbox*{29th order index}[.3\linewidth][c]{%
    \includegraphics[width=.32\linewidth]{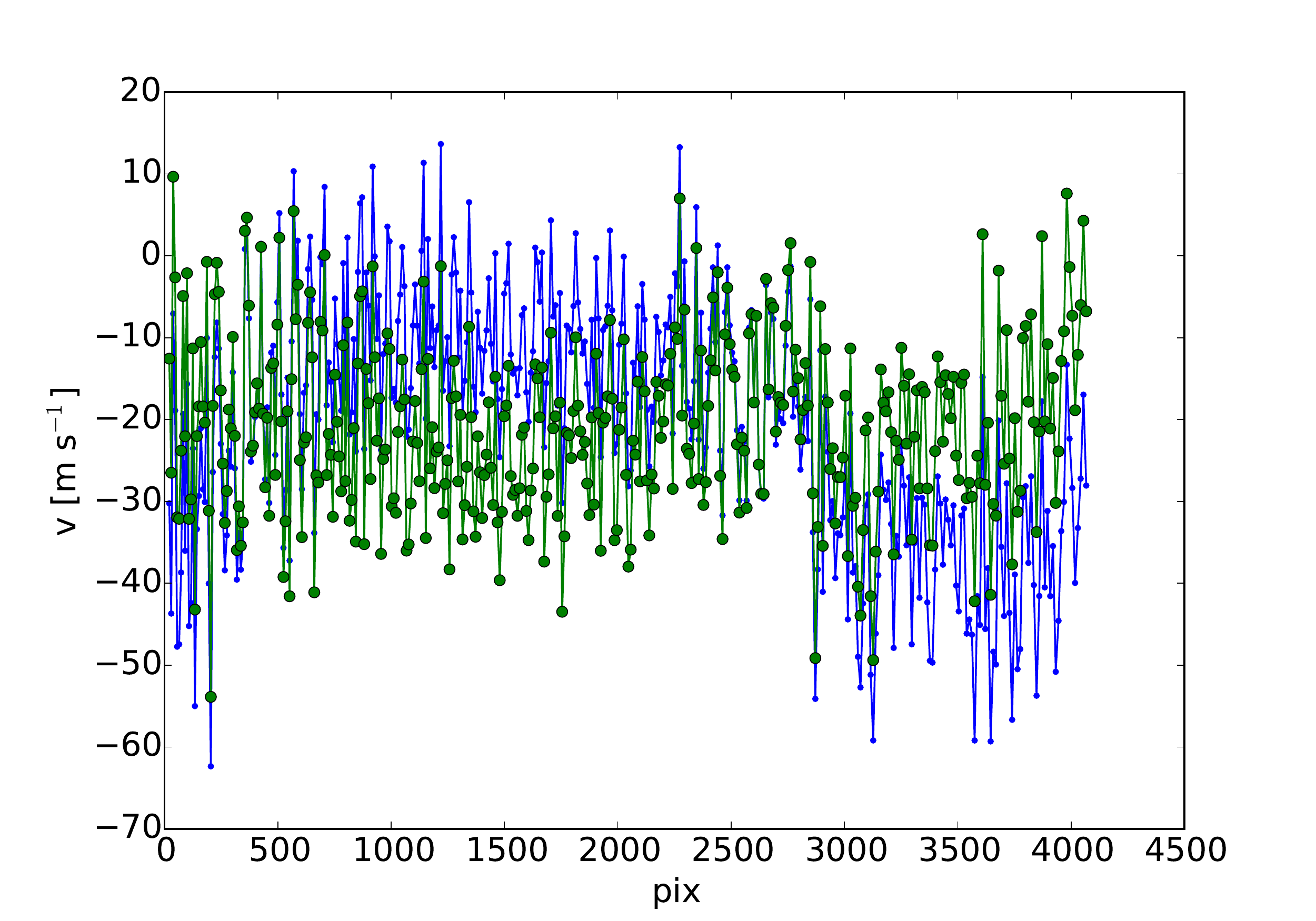}}\quad
  \subcaptionbox*{30th order index}[.3\linewidth][c]{%
    \includegraphics[width=.32\linewidth]{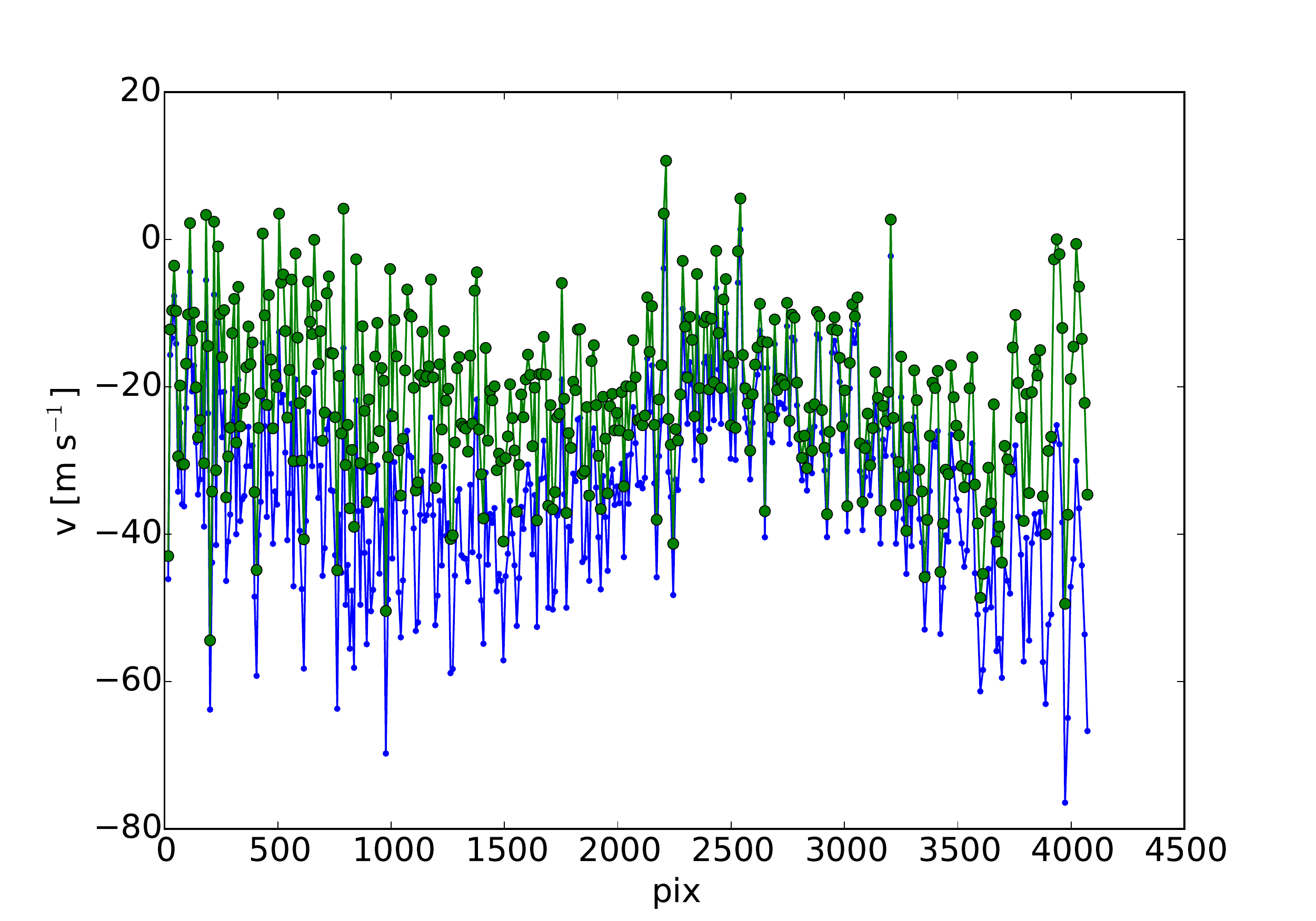}}
  \bigskip

   \subcaptionbox*{31st order index}[.3\linewidth][c]{%
    \includegraphics[width=.32\linewidth]{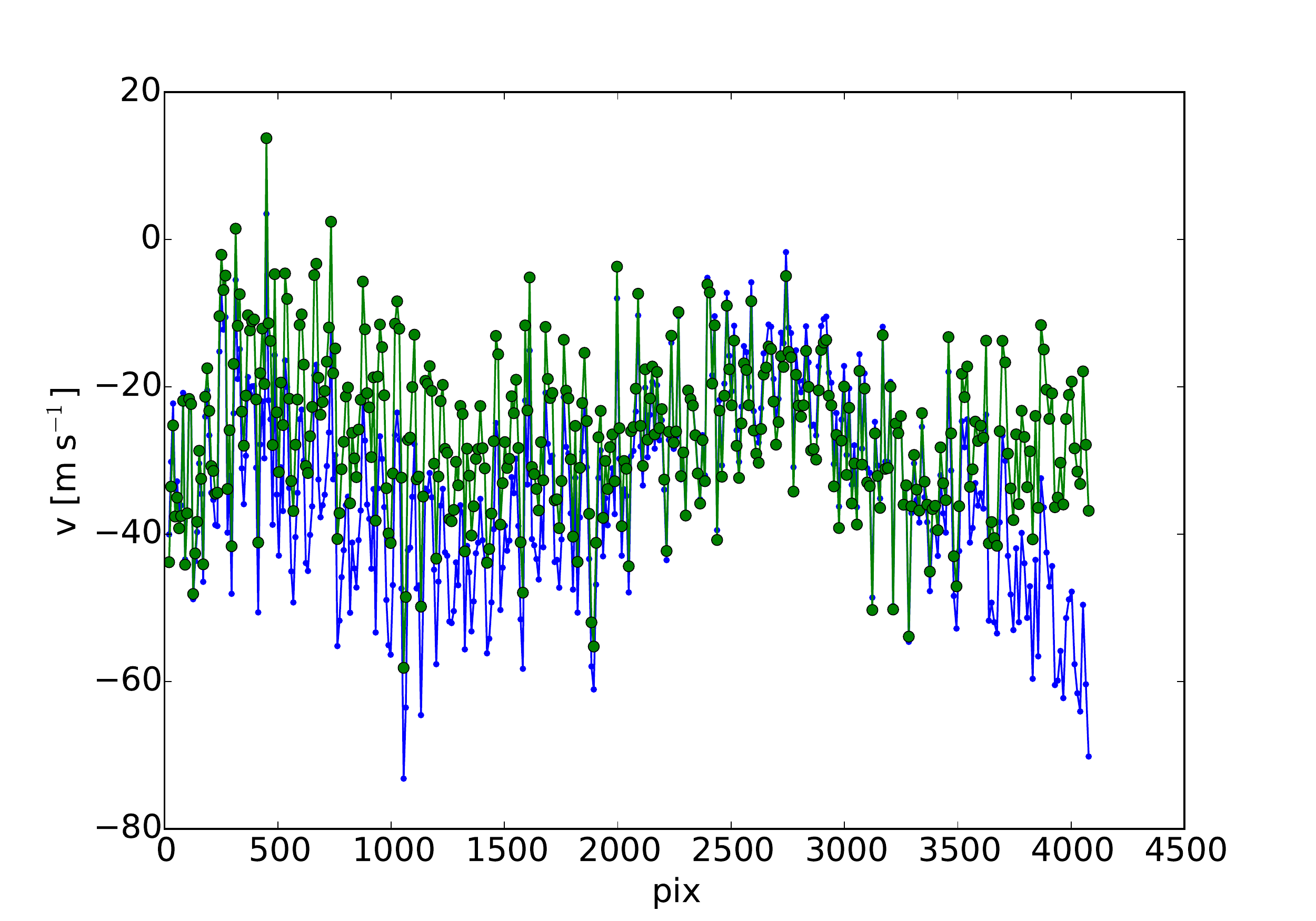}}\quad
  \subcaptionbox*{32nd order index}[.3\linewidth][c]{%
    \includegraphics[width=.32\linewidth]{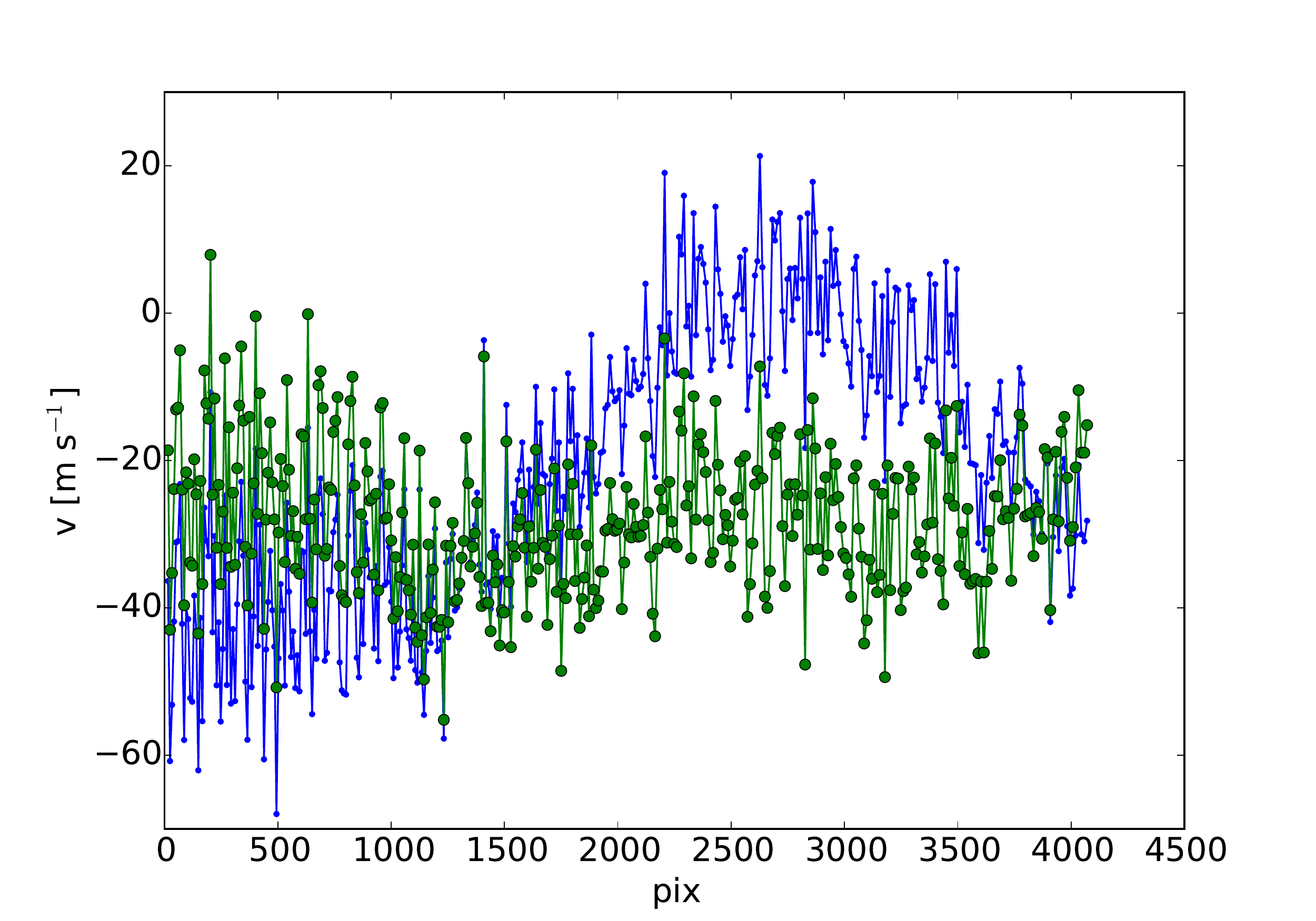}}\quad
  \subcaptionbox*{33rd order index}[.3\linewidth][c]{%
    \includegraphics[width=.32\linewidth]{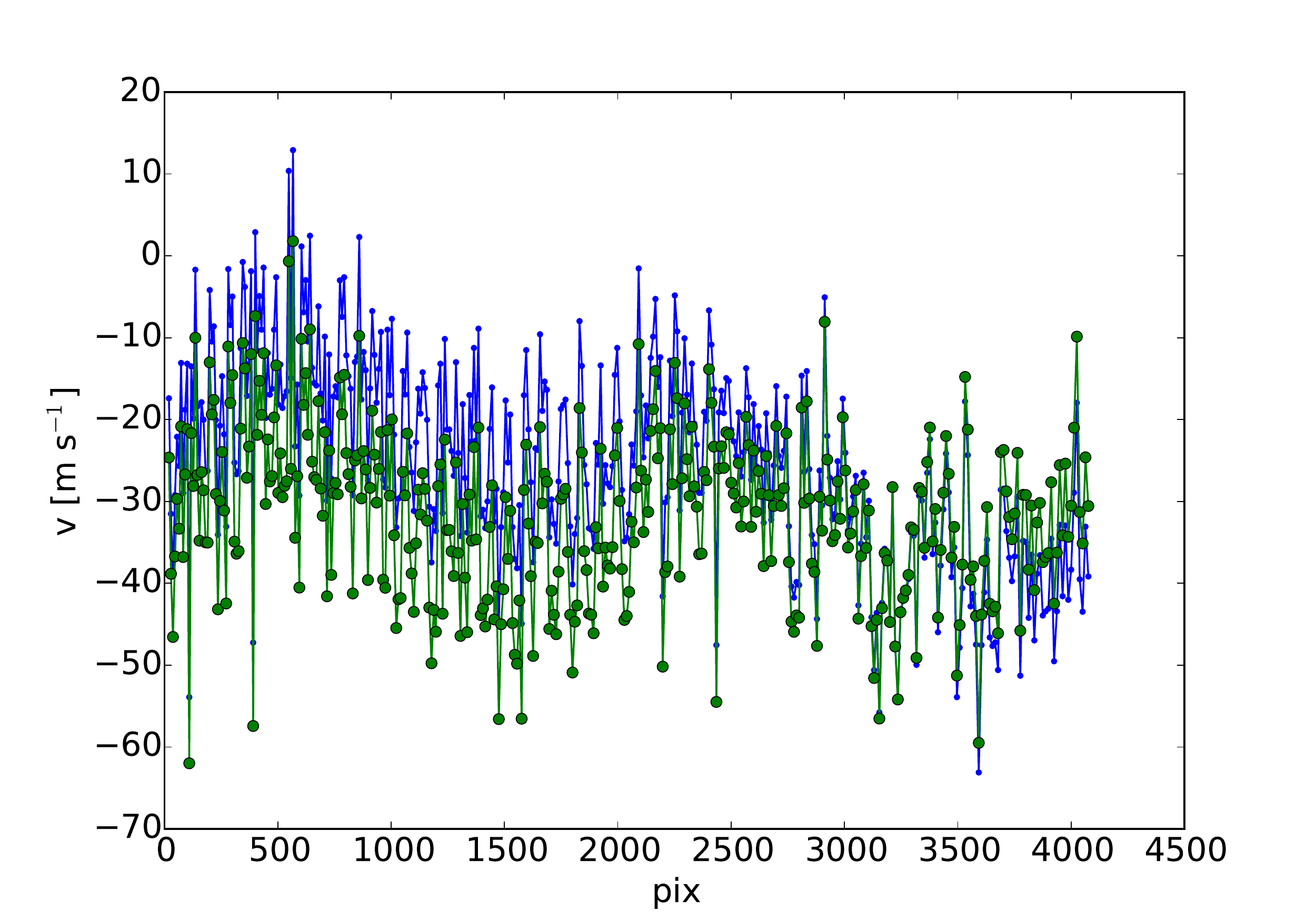}}
    \bigskip
   \subcaptionbox*{34th order index}[.3\linewidth][c]{%
    \includegraphics[width=.32\linewidth]{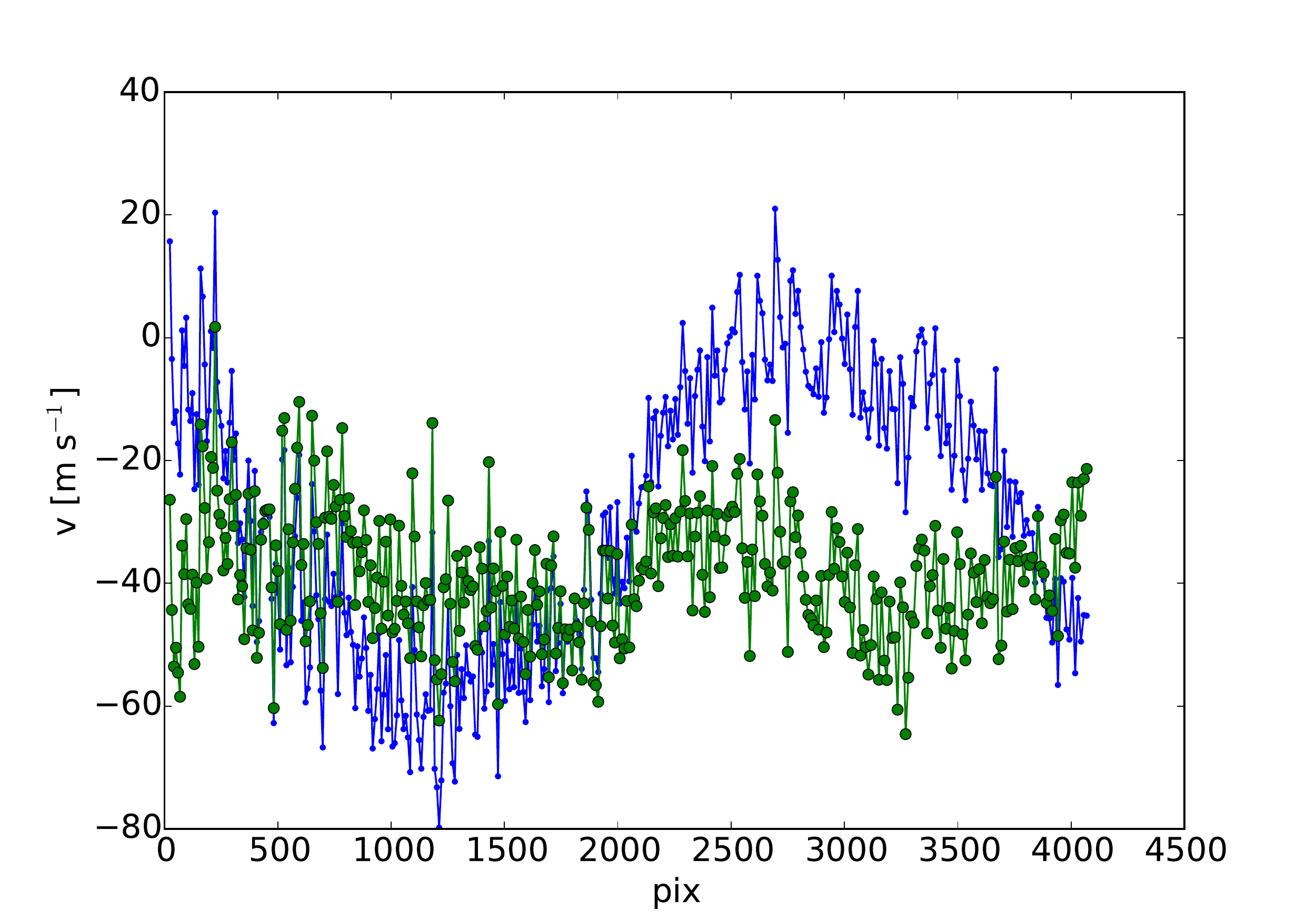}}\quad
  \subcaptionbox*{35th order index}[.3\linewidth][c]{%
    \includegraphics[width=.32\linewidth]{35.pdf}}\quad
  \subcaptionbox*{36th order index}[.3\linewidth][c]{%
    \includegraphics[width=.32\linewidth]{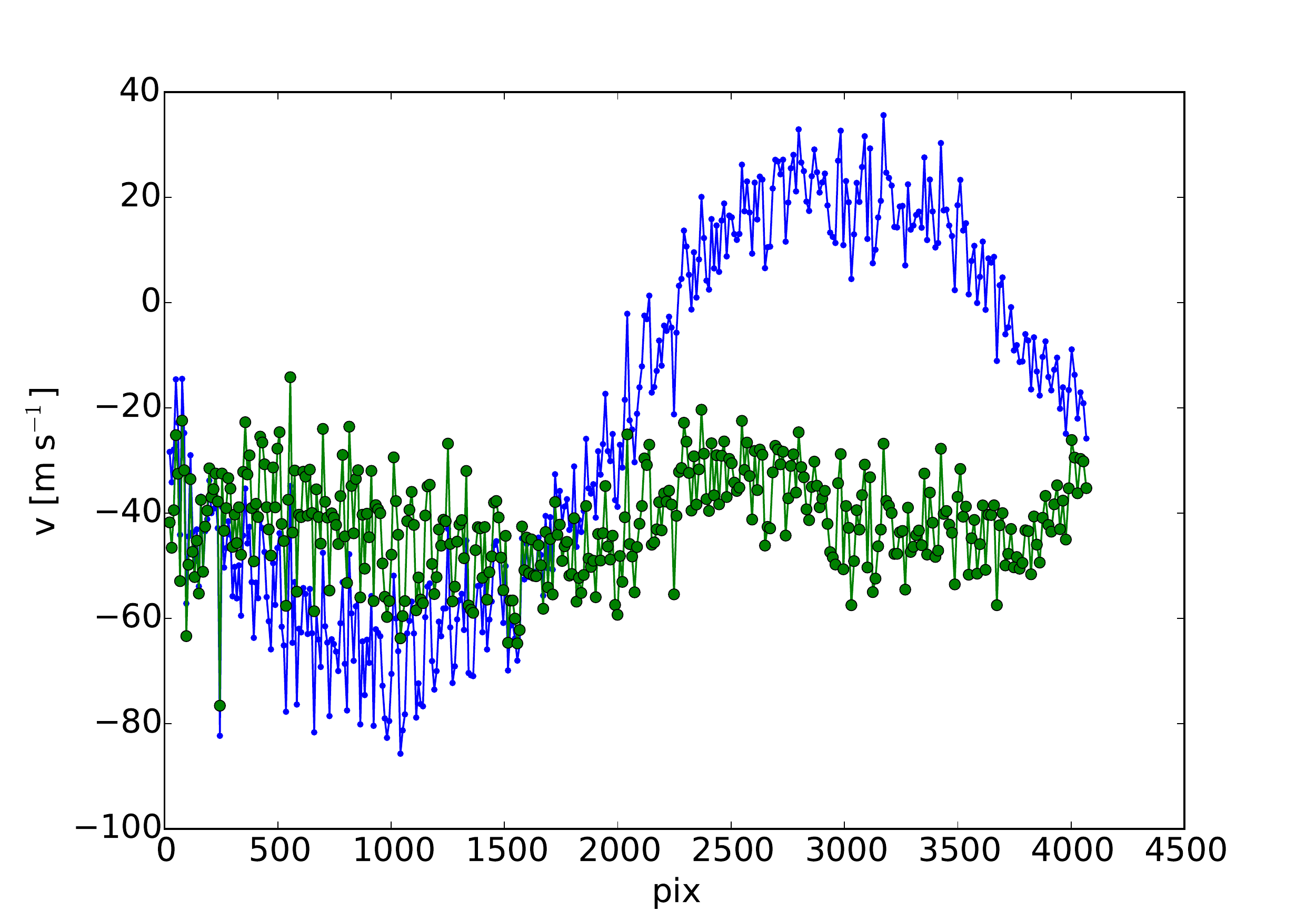}}
    \bigskip 
     \subcaptionbox*{37th order index}[.3\linewidth][c]{%
    \includegraphics[width=.32\linewidth]{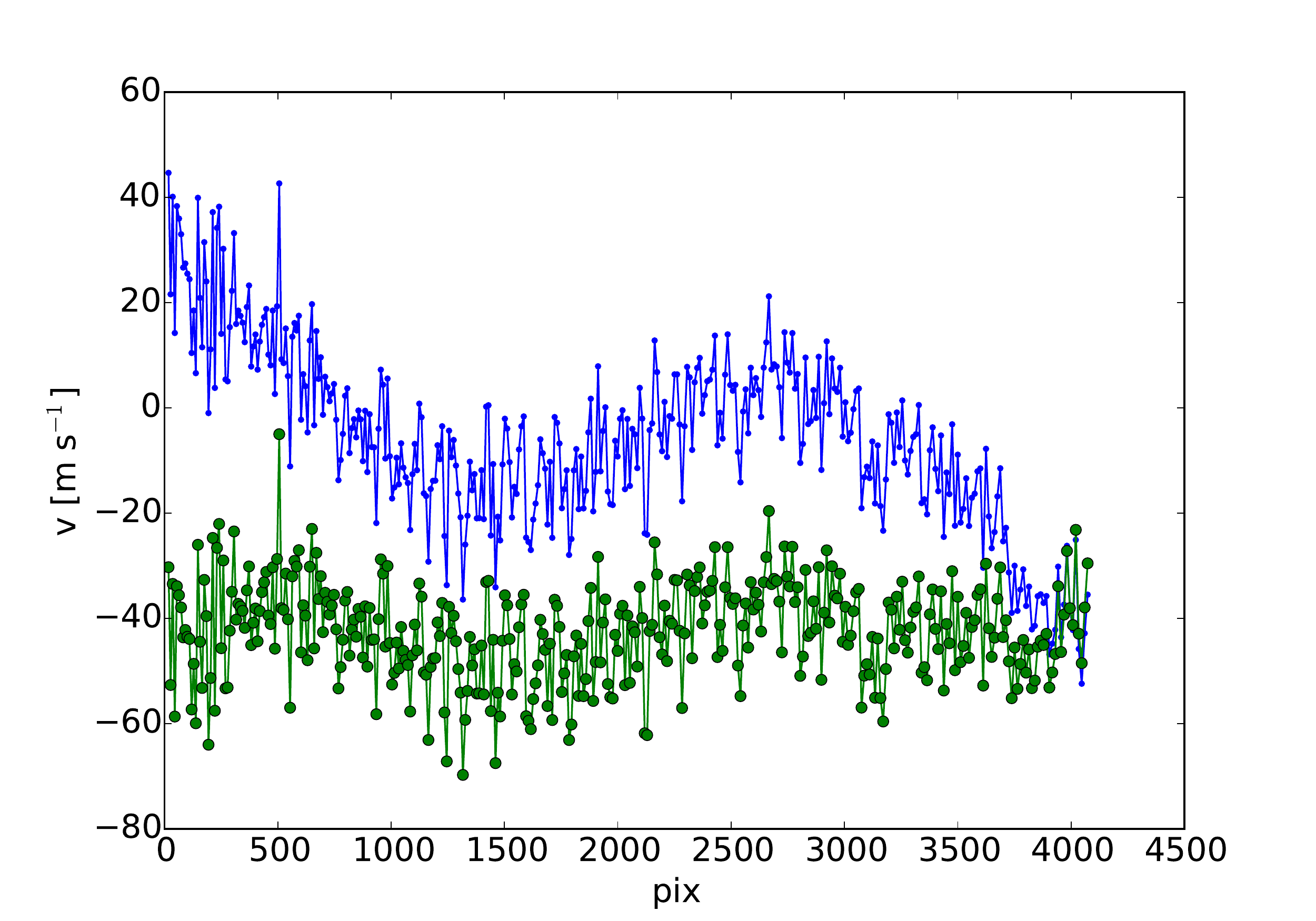}}\quad
  \subcaptionbox*{38th order index}[.3\linewidth][c]{%
    \includegraphics[width=.32\linewidth]{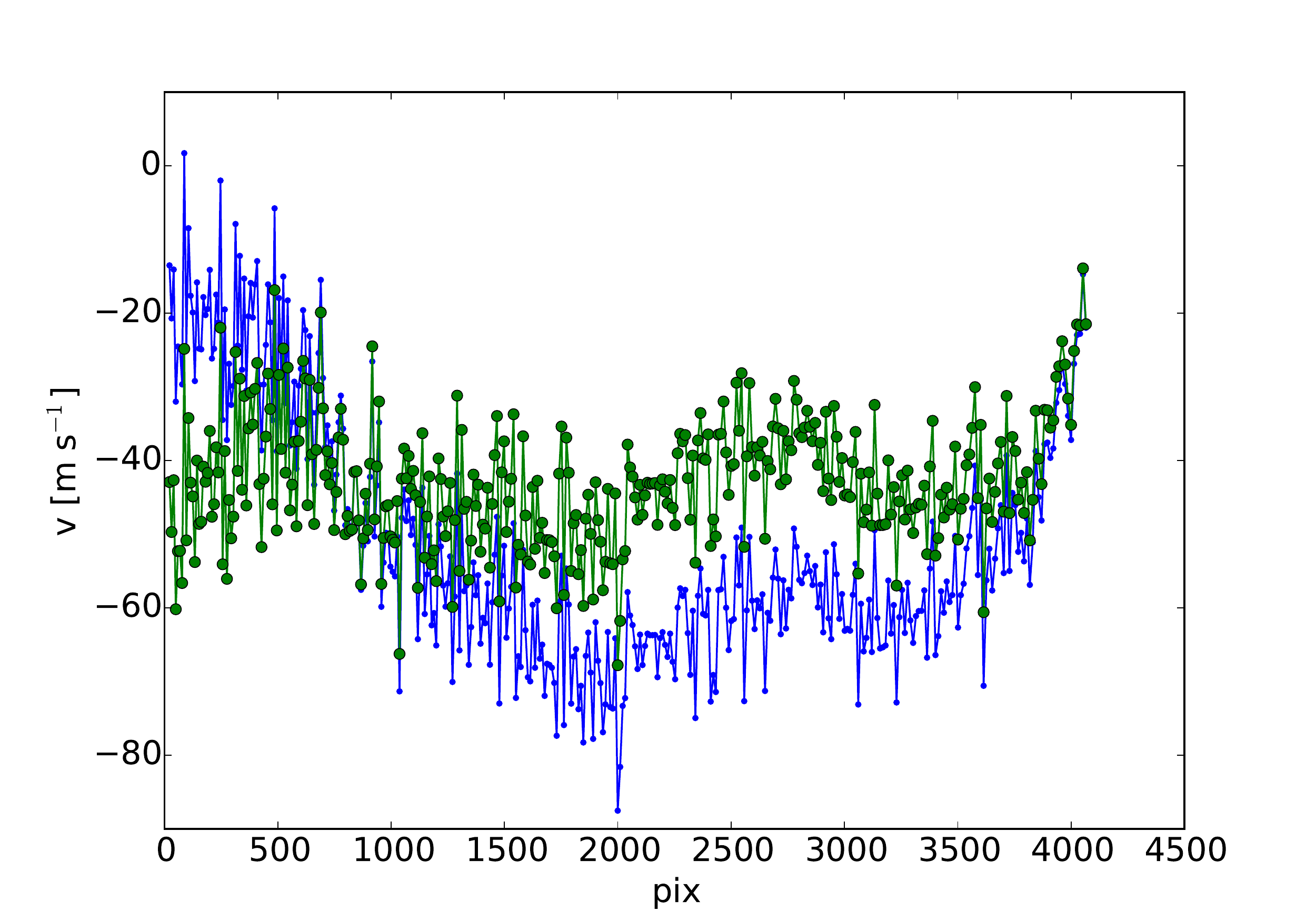}}\quad
  \subcaptionbox*{39th order index}[.3\linewidth][c]{%
    \includegraphics[width=.32\linewidth]{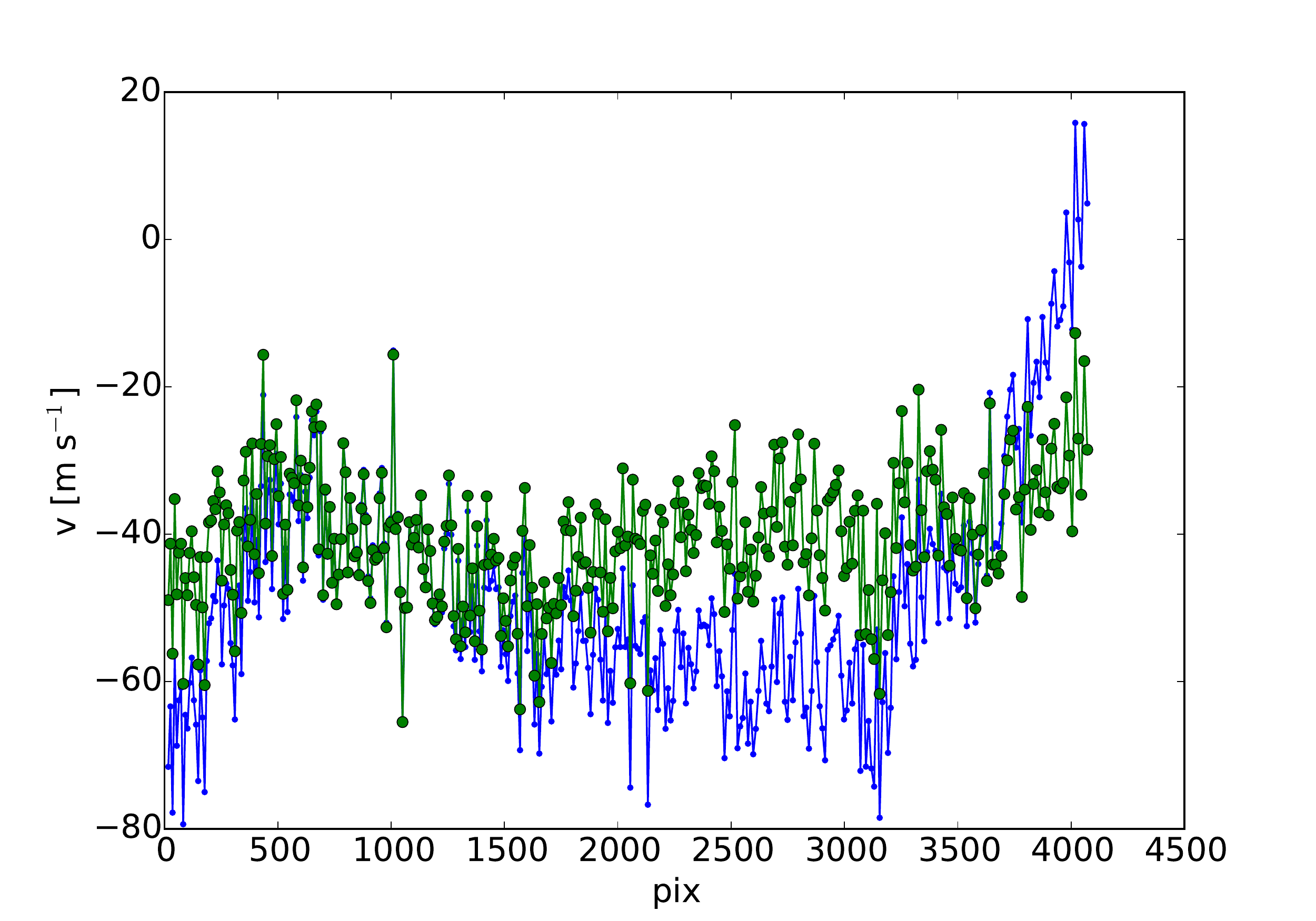}}
    \bigskip
      \caption{Difference between the theoretical wavelengths of the LFC lines and the wavelengths along the spectral orders for HARPS obtained after calibration with jumps gaps corrected \citep{Coffinet+2018} using the "pure" thorium-wavelength solution (blue curve) and the TH+FP wavelength solution of the new DRS (green curve).}
      \label{fig:10}
\end{figure*}

\begin{figure*}
\renewcommand{\figurename}{Supplementary Figure}  \centering
  \subcaptionbox*{40th order index}[.3\linewidth][c]{%
    \includegraphics[width=.32\linewidth]{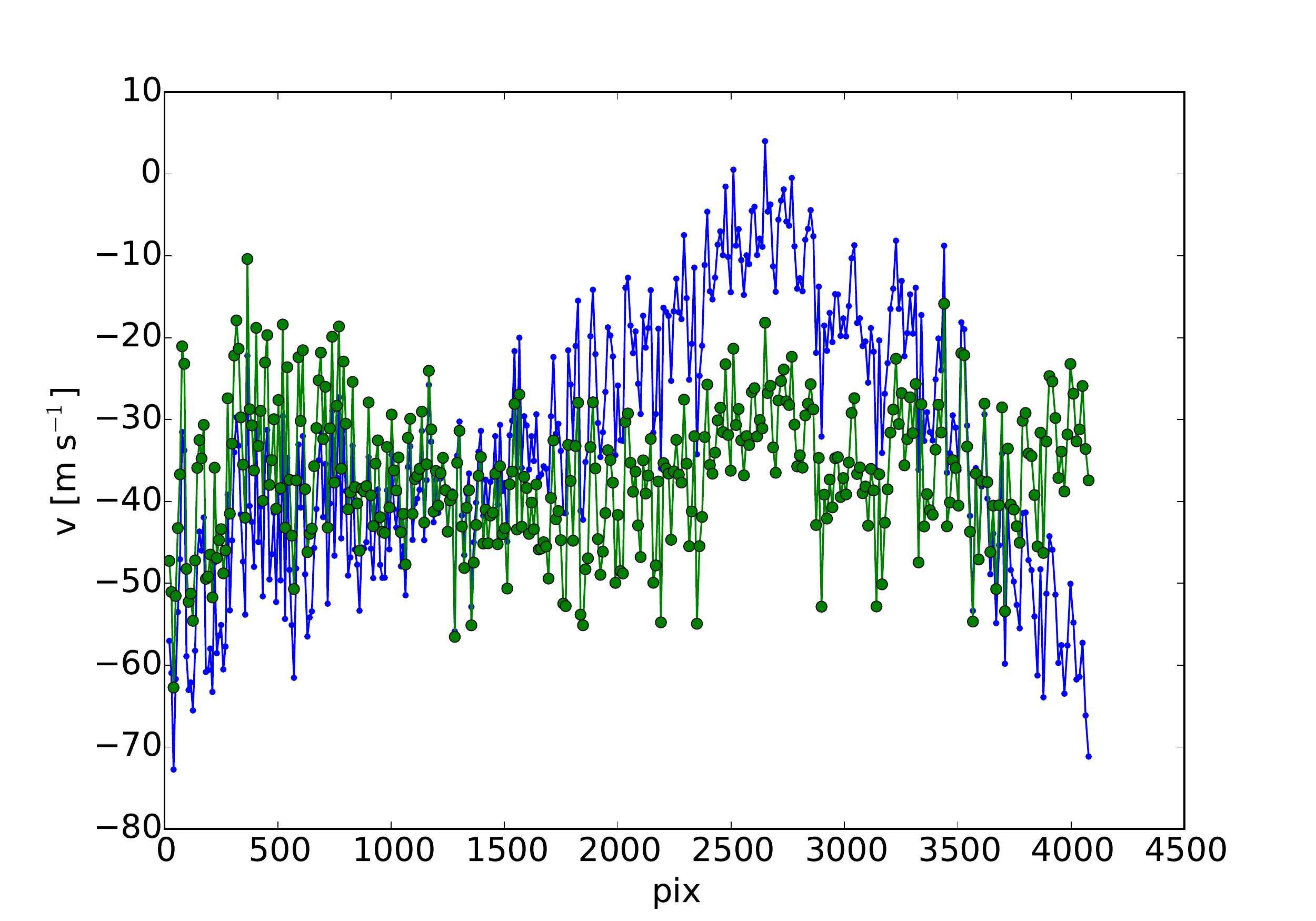}}\quad
  \subcaptionbox*{41st order index}[.3\linewidth][c]{%
    \includegraphics[width=.32\linewidth]{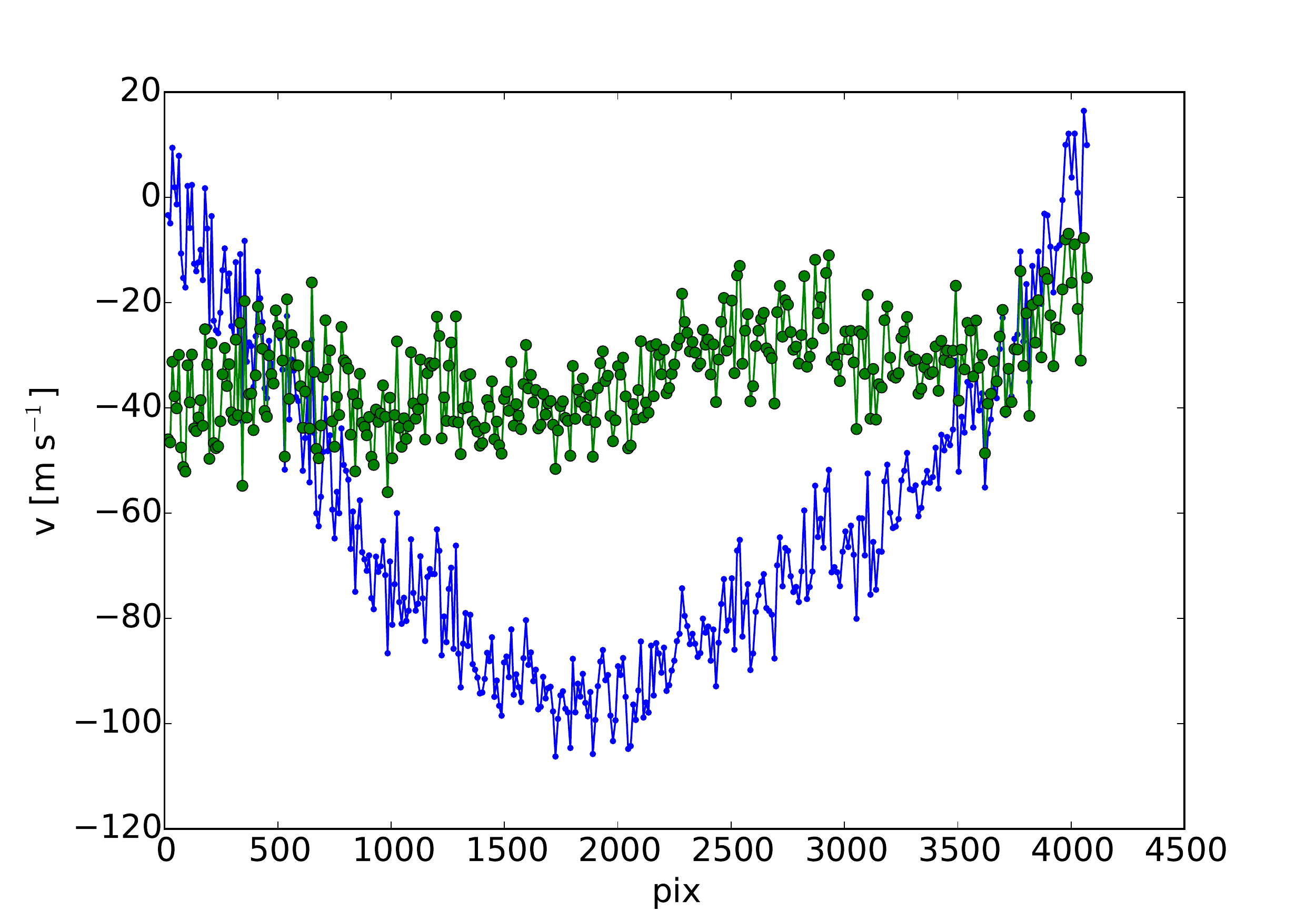}}\quad
  \subcaptionbox*{42nd order index}[.3\linewidth][c]{%
    \includegraphics[width=.32\linewidth]{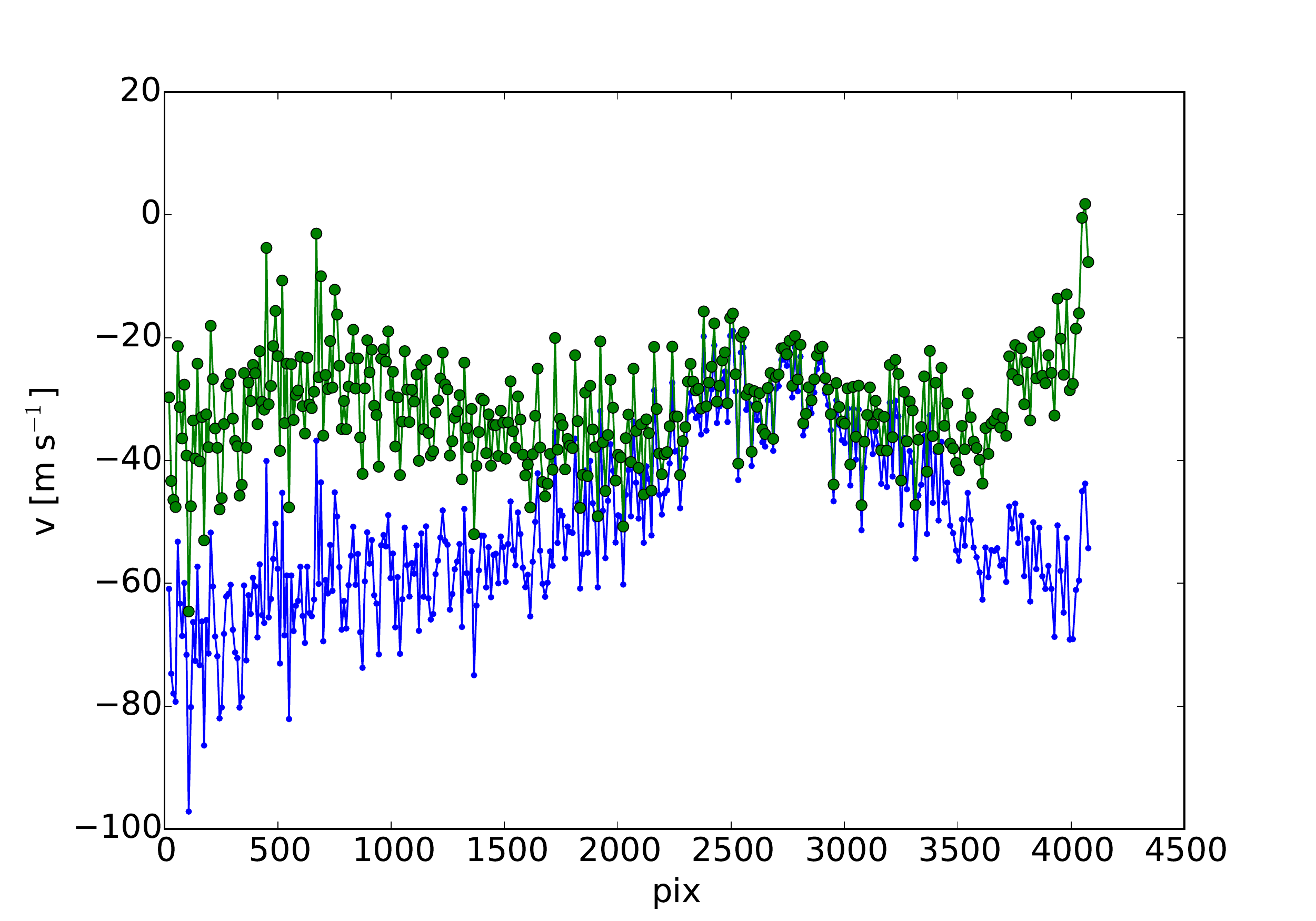}}

  \bigskip

   \subcaptionbox*{43rd order index}[.3\linewidth][c]{%
    \includegraphics[width=.32\linewidth]{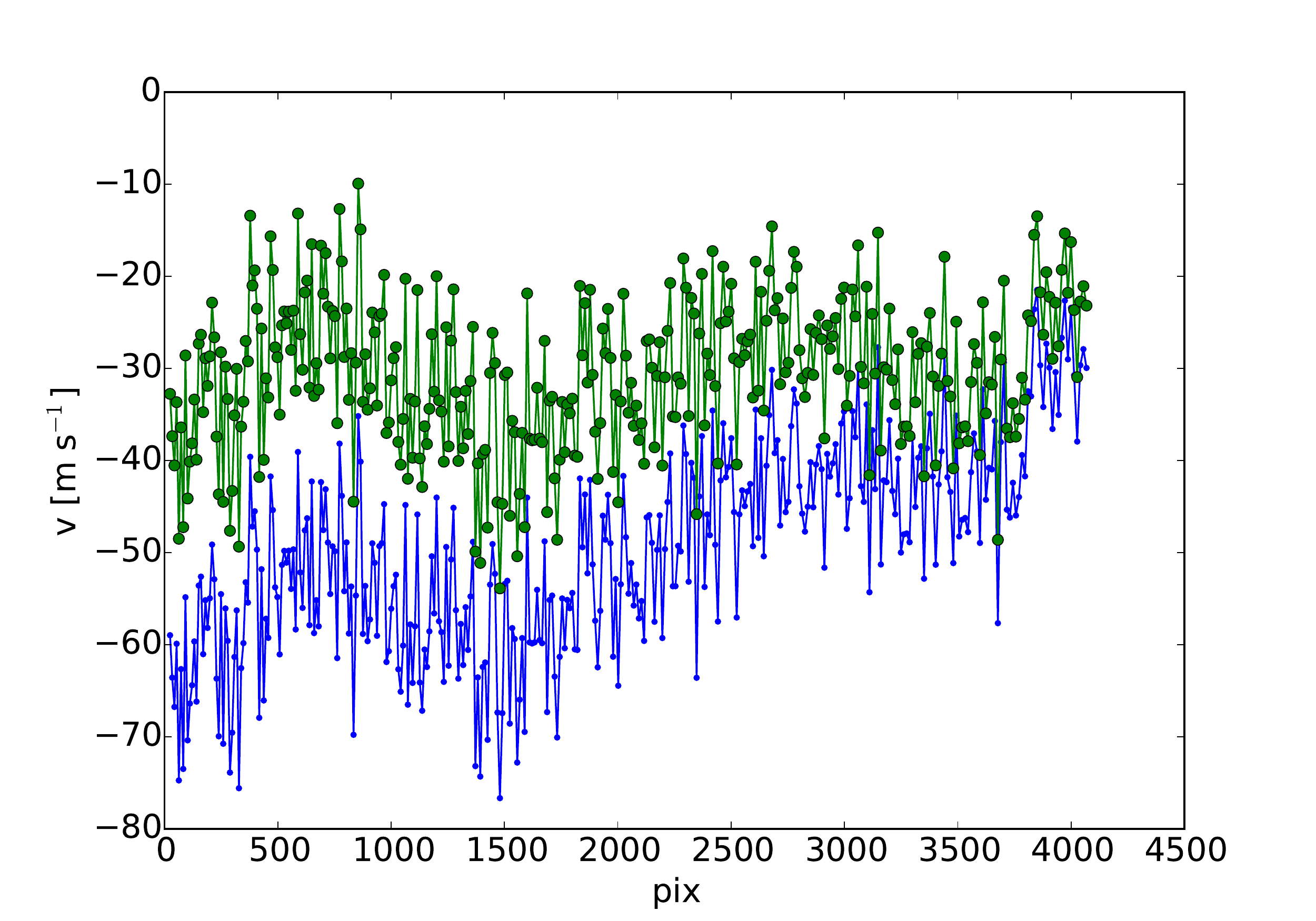}}\quad
  \subcaptionbox*{44th order index}[.3\linewidth][c]{%
    \includegraphics[width=.32\linewidth]{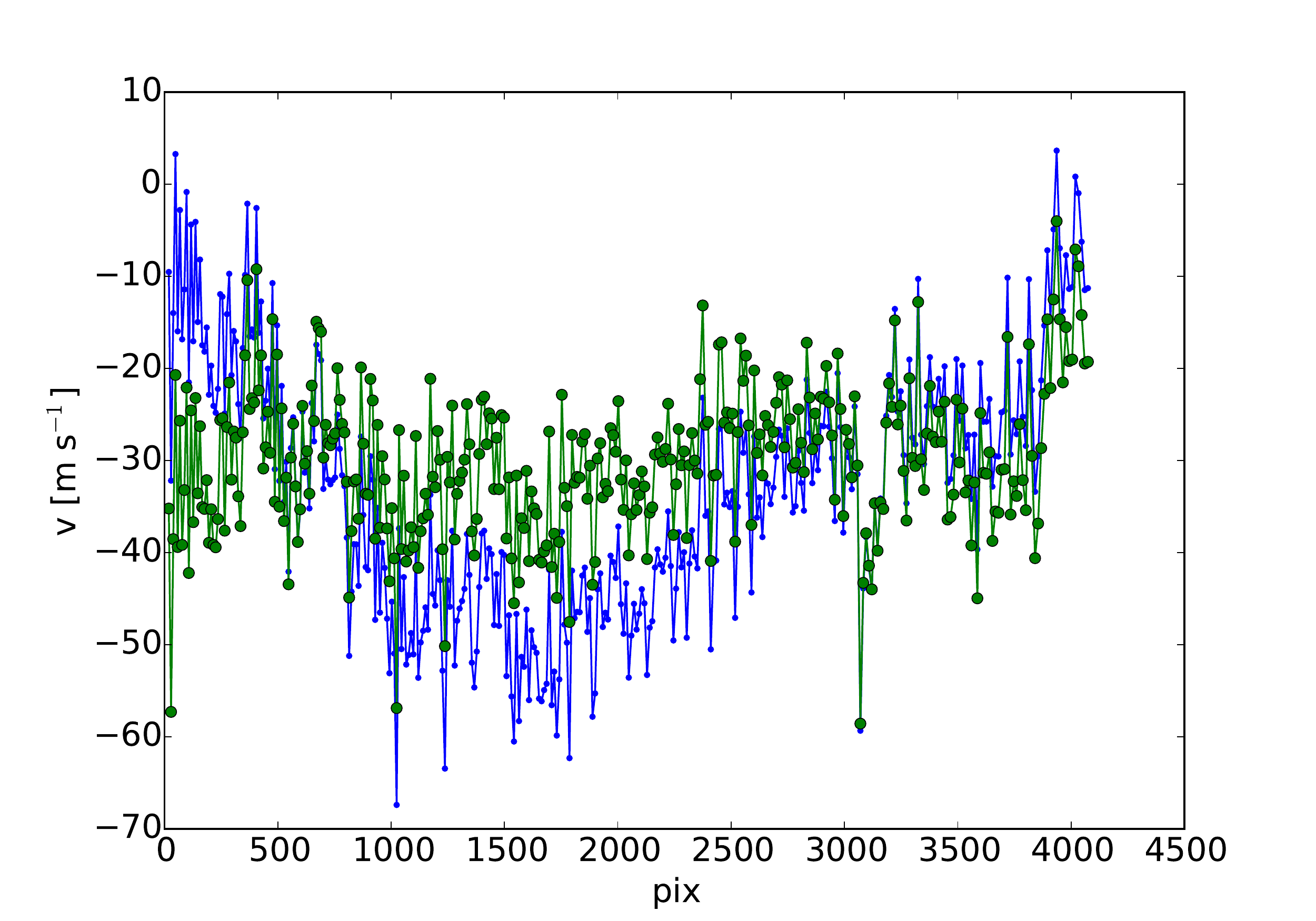}}\quad
  \subcaptionbox*{45th order index}[.3\linewidth][c]{%
    \includegraphics[width=.32\linewidth]{45.pdf}}
  \bigskip

   \subcaptionbox*{46th order index}[.3\linewidth][c]{%
    \includegraphics[width=.32\linewidth]{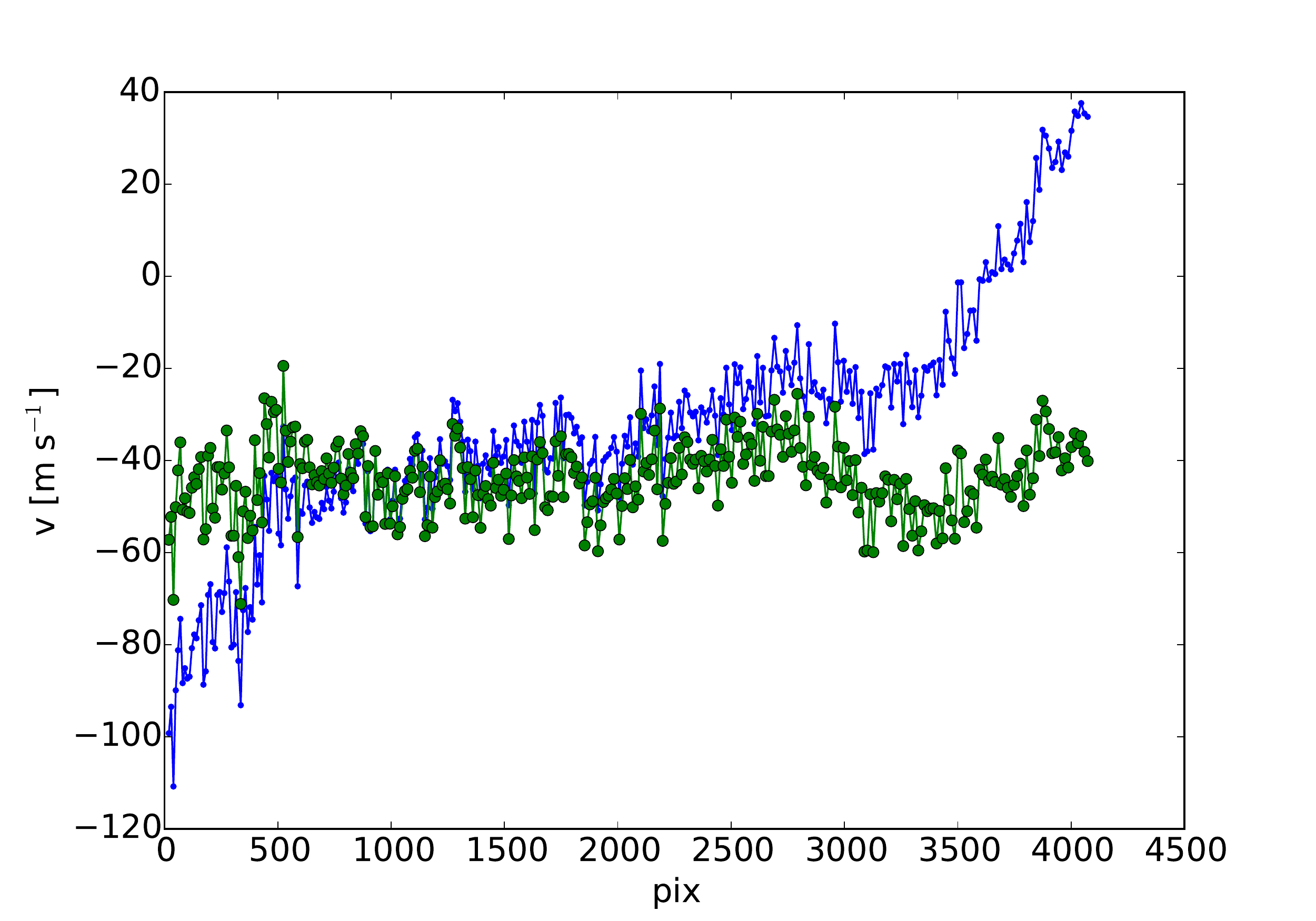}}\quad
  \subcaptionbox*{47th order index}[.3\linewidth][c]{%
    \includegraphics[width=.32\linewidth]{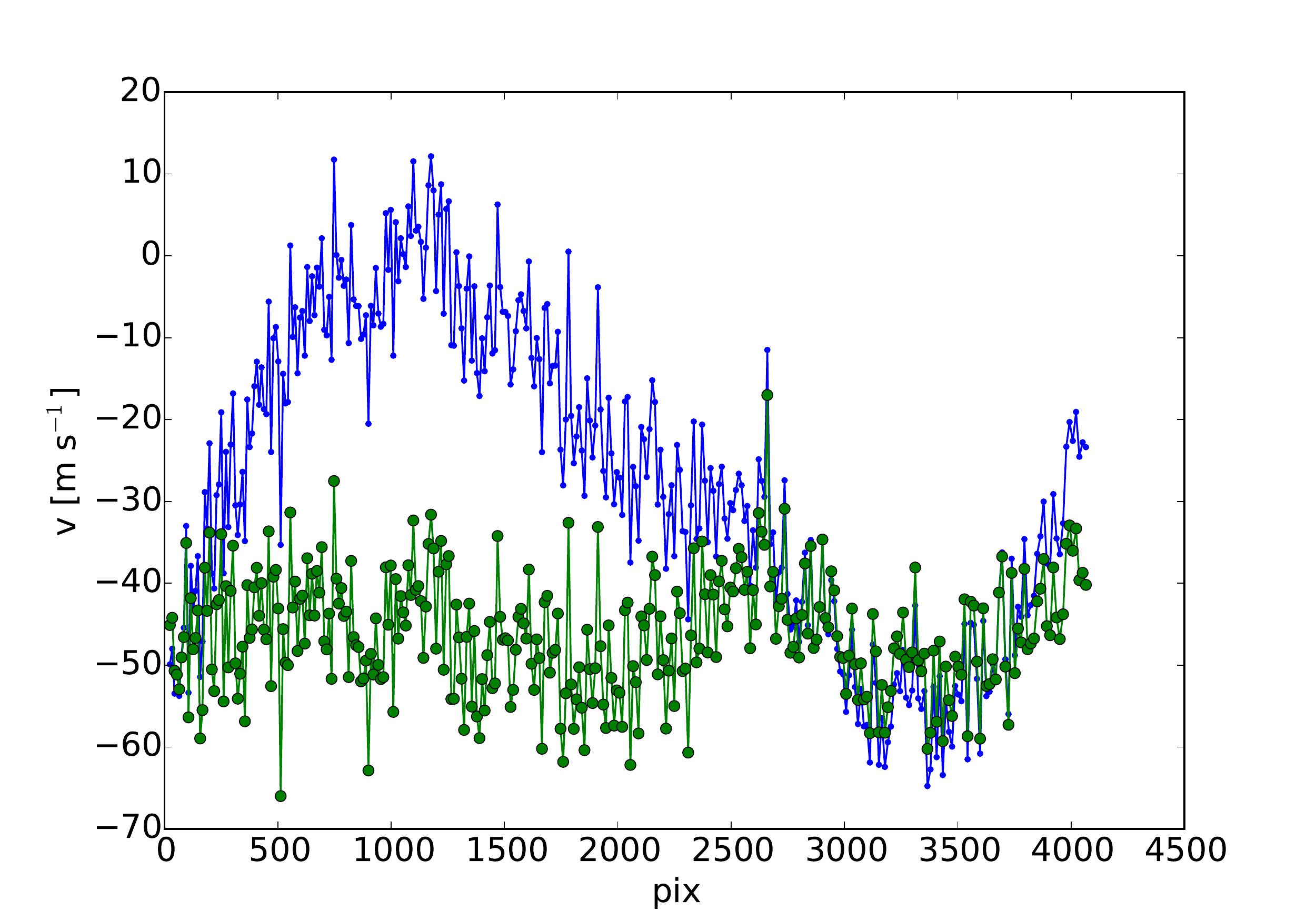}}\quad
  \subcaptionbox*{48th order index}[.3\linewidth][c]{%
    \includegraphics[width=.32\linewidth]{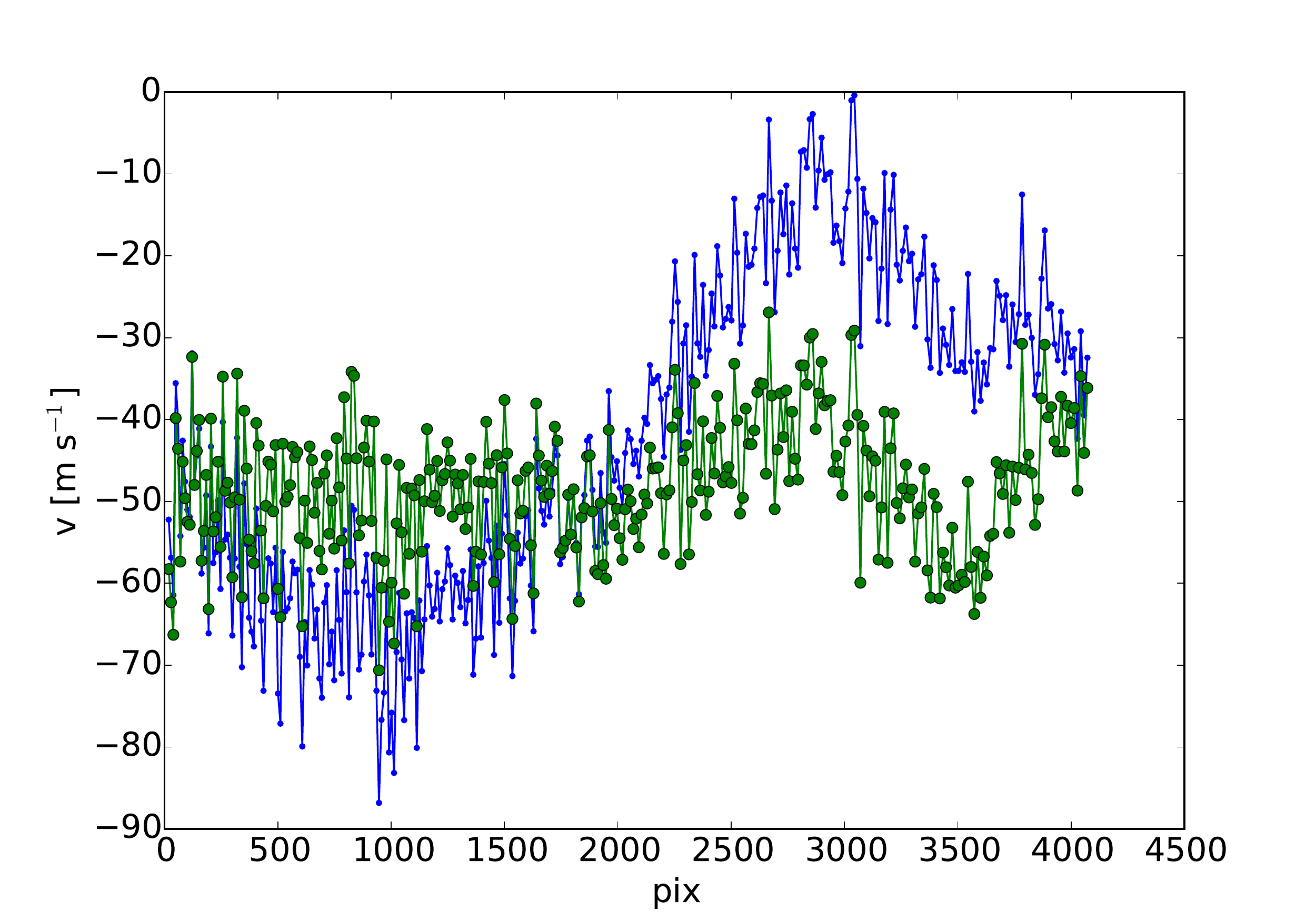}}
    \bigskip
   \subcaptionbox*{49th order index}[.3\linewidth][c]{%
    \includegraphics[width=.32\linewidth]{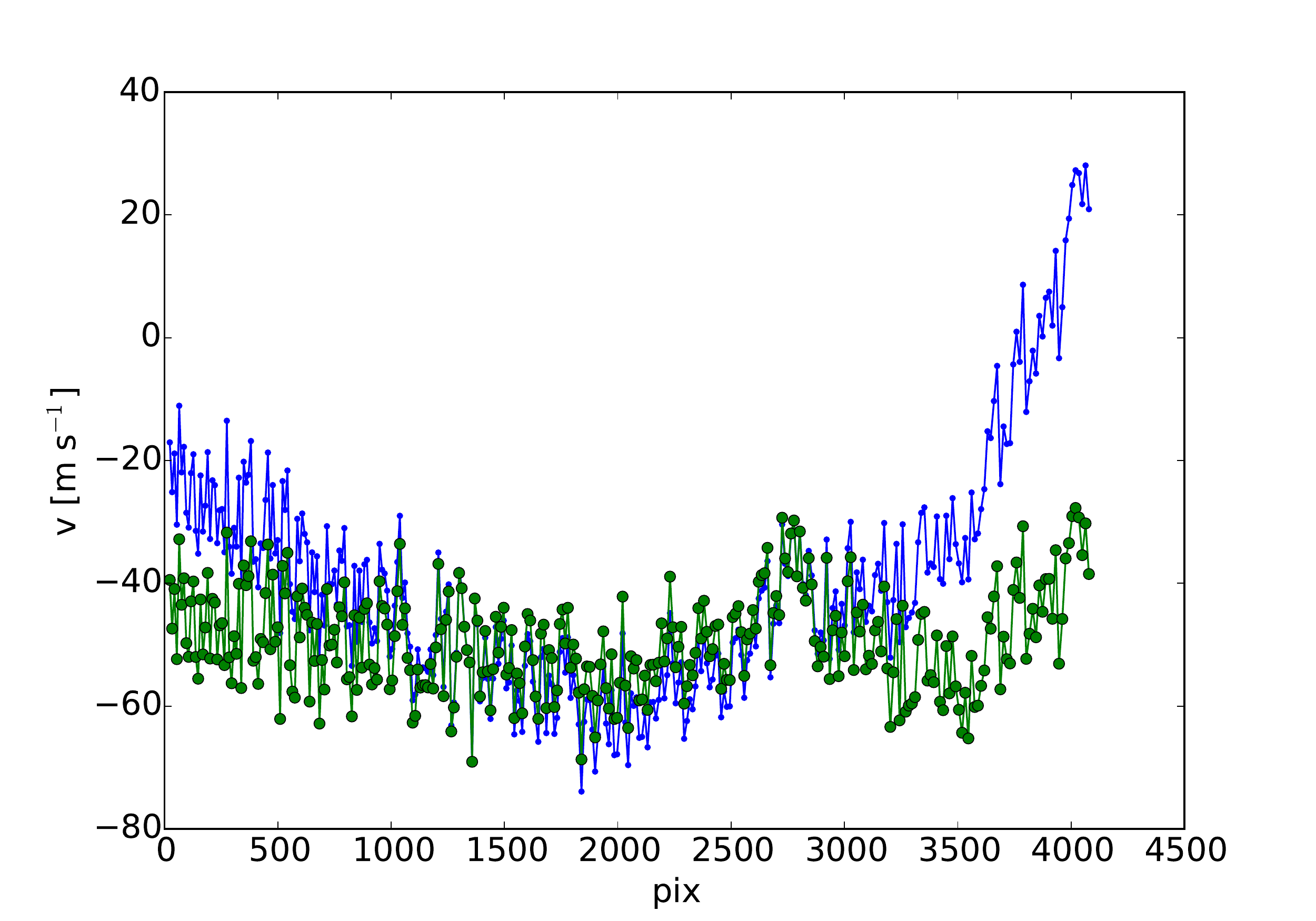}}\quad
  \subcaptionbox*{50th order index}[.3\linewidth][c]{%
    \includegraphics[width=.32\linewidth]{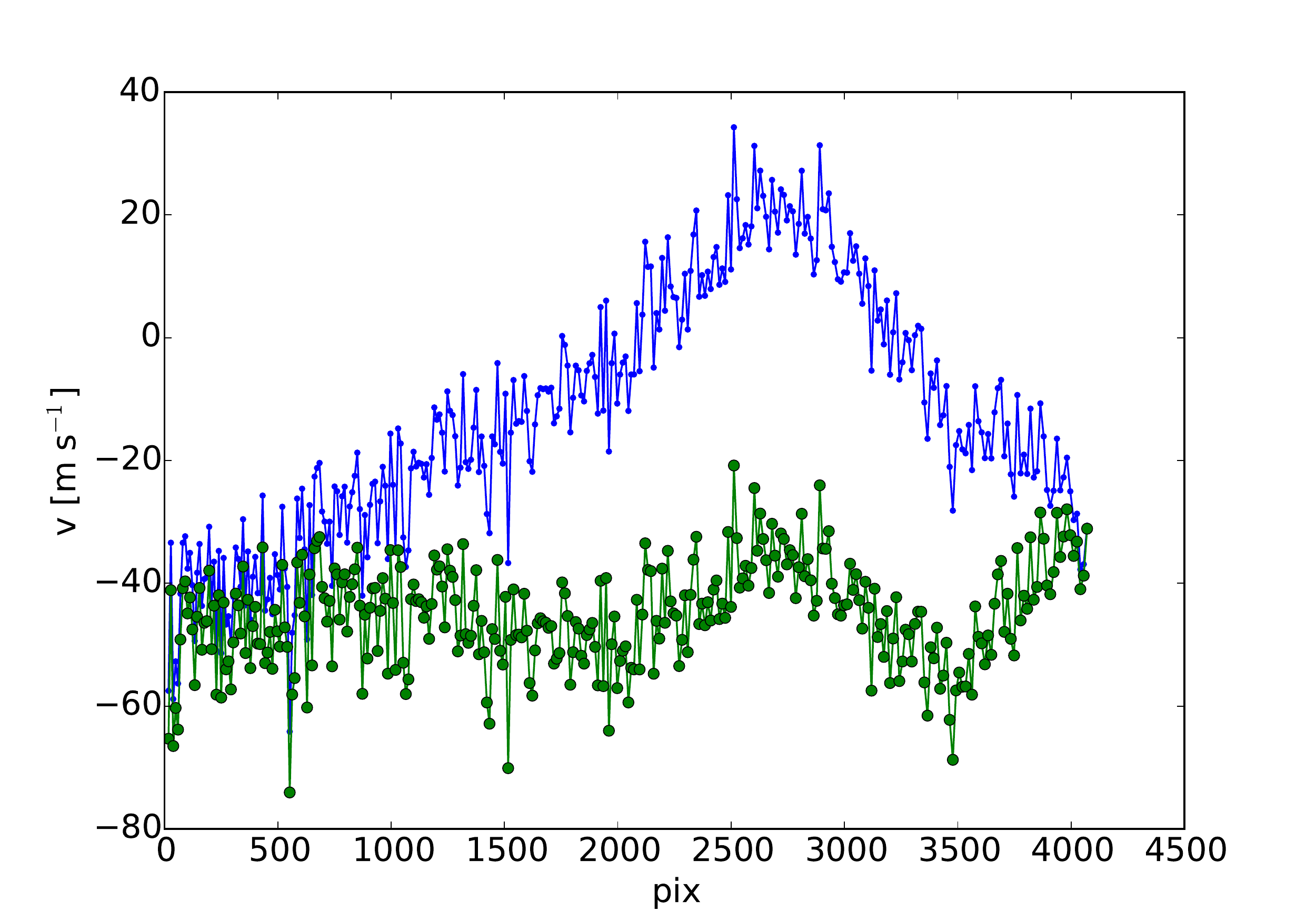}}\quad
  \subcaptionbox*{51st order index}[.3\linewidth][c]{%
    \includegraphics[width=.32\linewidth]{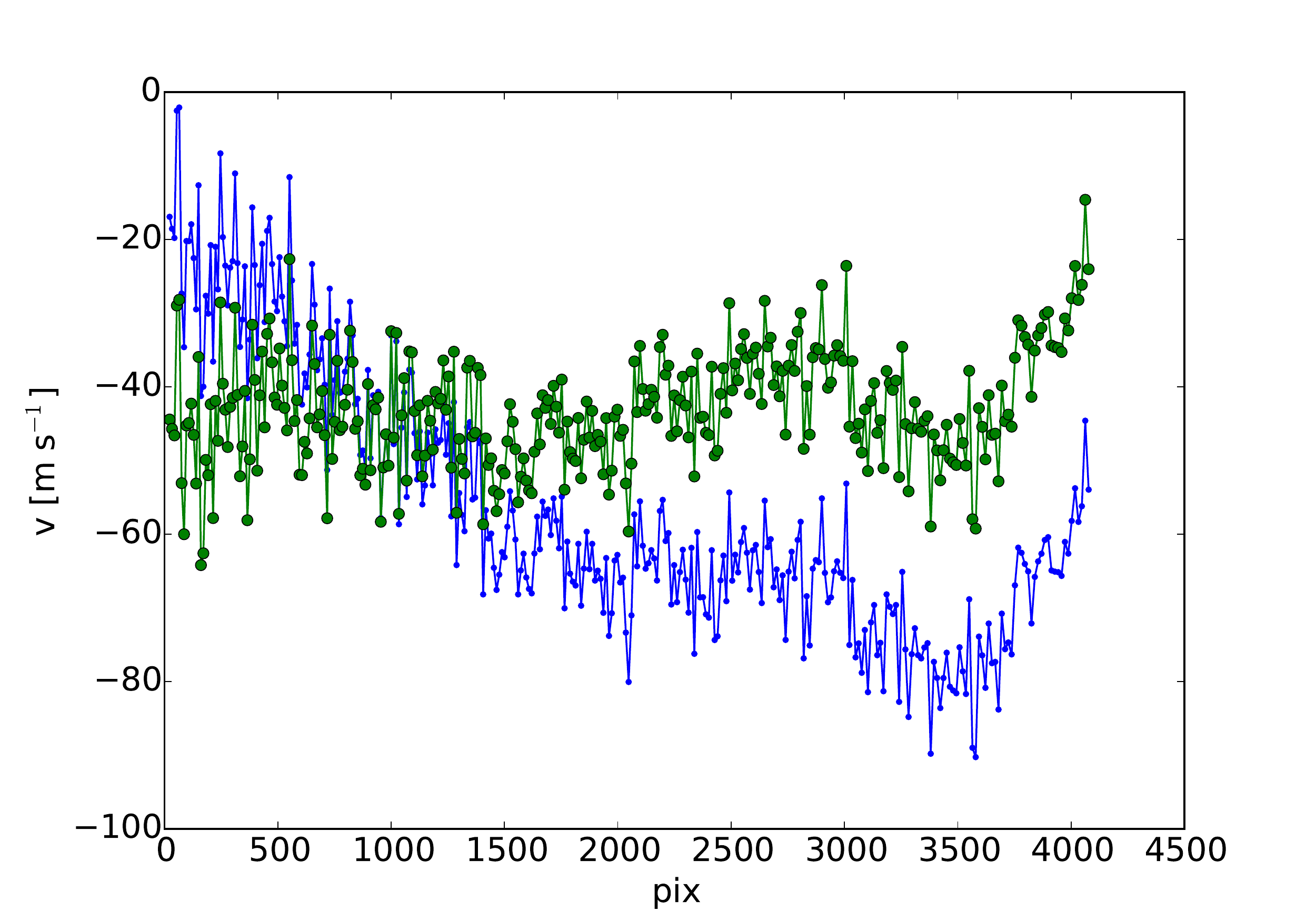}}
    \bigskip 
     \subcaptionbox*{52nd order index}[.3\linewidth][c]{%
    \includegraphics[width=.32\linewidth]{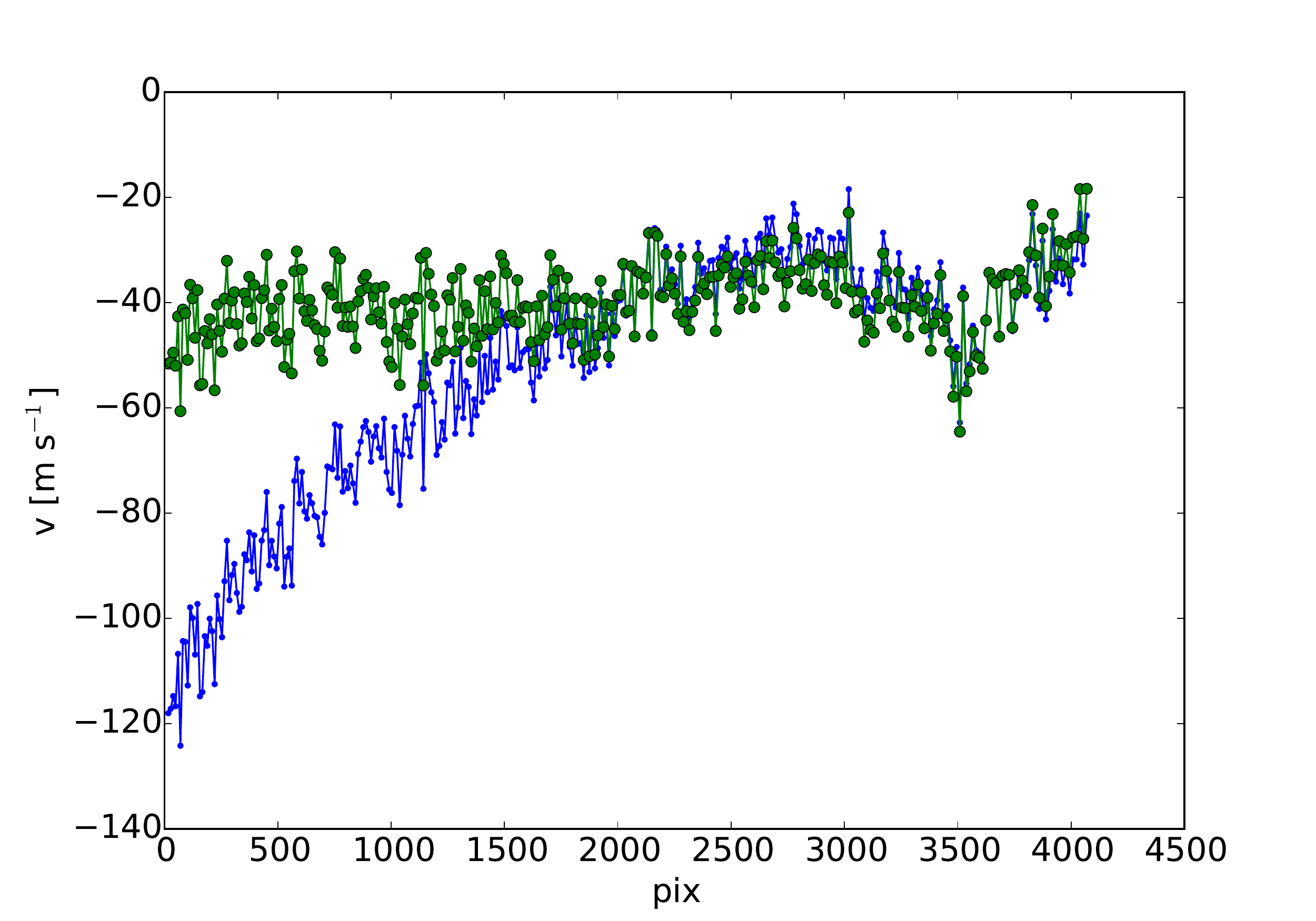}}\quad
  \subcaptionbox*{53rd order index}[.3\linewidth][c]{%
    \includegraphics[width=.32\linewidth]{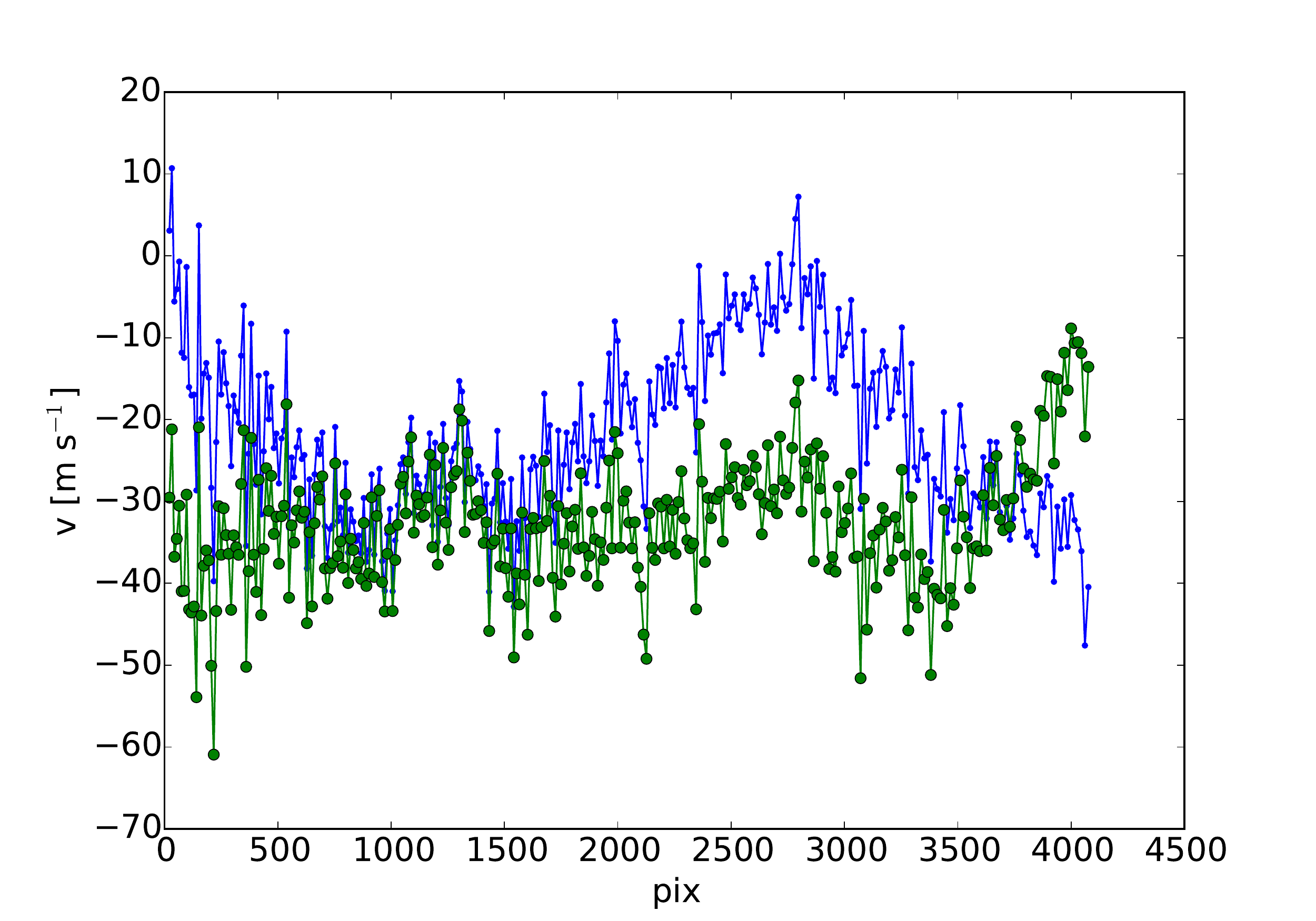}}\quad
  \subcaptionbox*{54th order index}[.3\linewidth][c]{%
    \includegraphics[width=.32\linewidth]{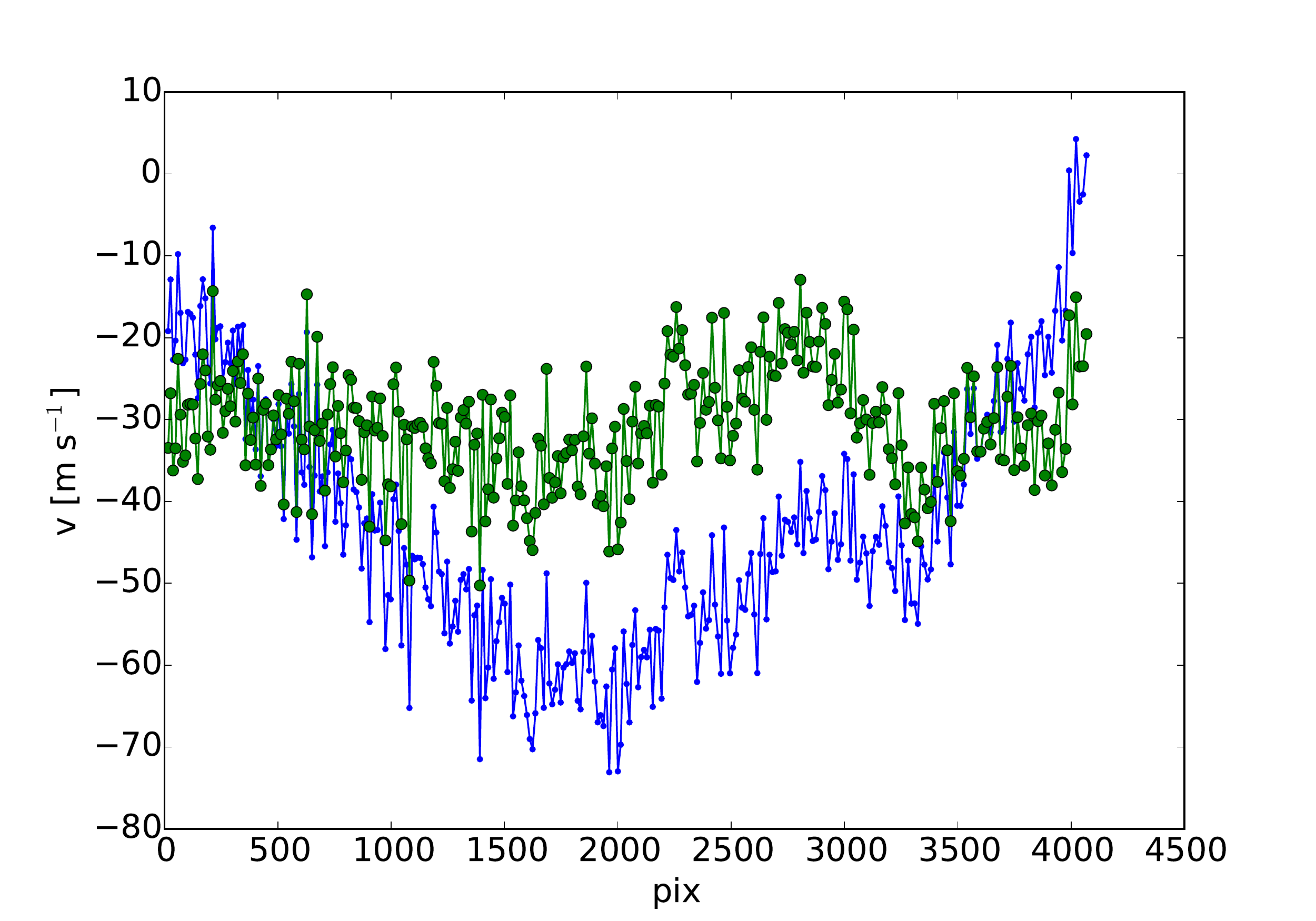}}
    \bigskip
      \label{fig:11}
\end{figure*}

\begin{figure*}
\renewcommand{\figurename}{Supplementary Figure}
  \centering
  \subcaptionbox*{55th order index}[.3\linewidth][c]{%
    \includegraphics[width=.33\linewidth]{55.pdf}}\quad
  \subcaptionbox*{56th order index}[.3\linewidth][c]{%
    \includegraphics[width=.33\linewidth]{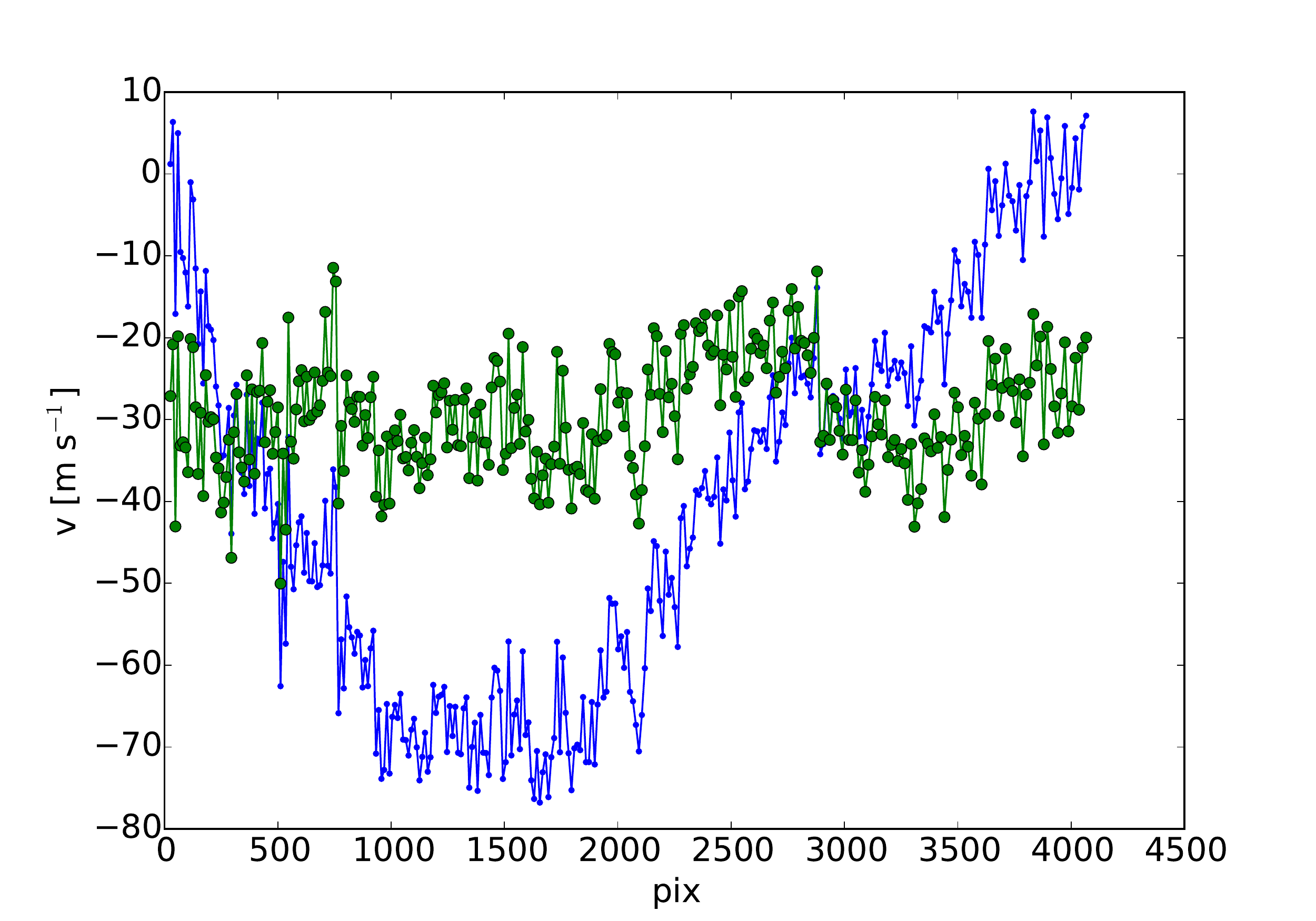}}\quad
  \subcaptionbox*{57th order index}[.3\linewidth][c]{%
    \includegraphics[width=.33\linewidth]{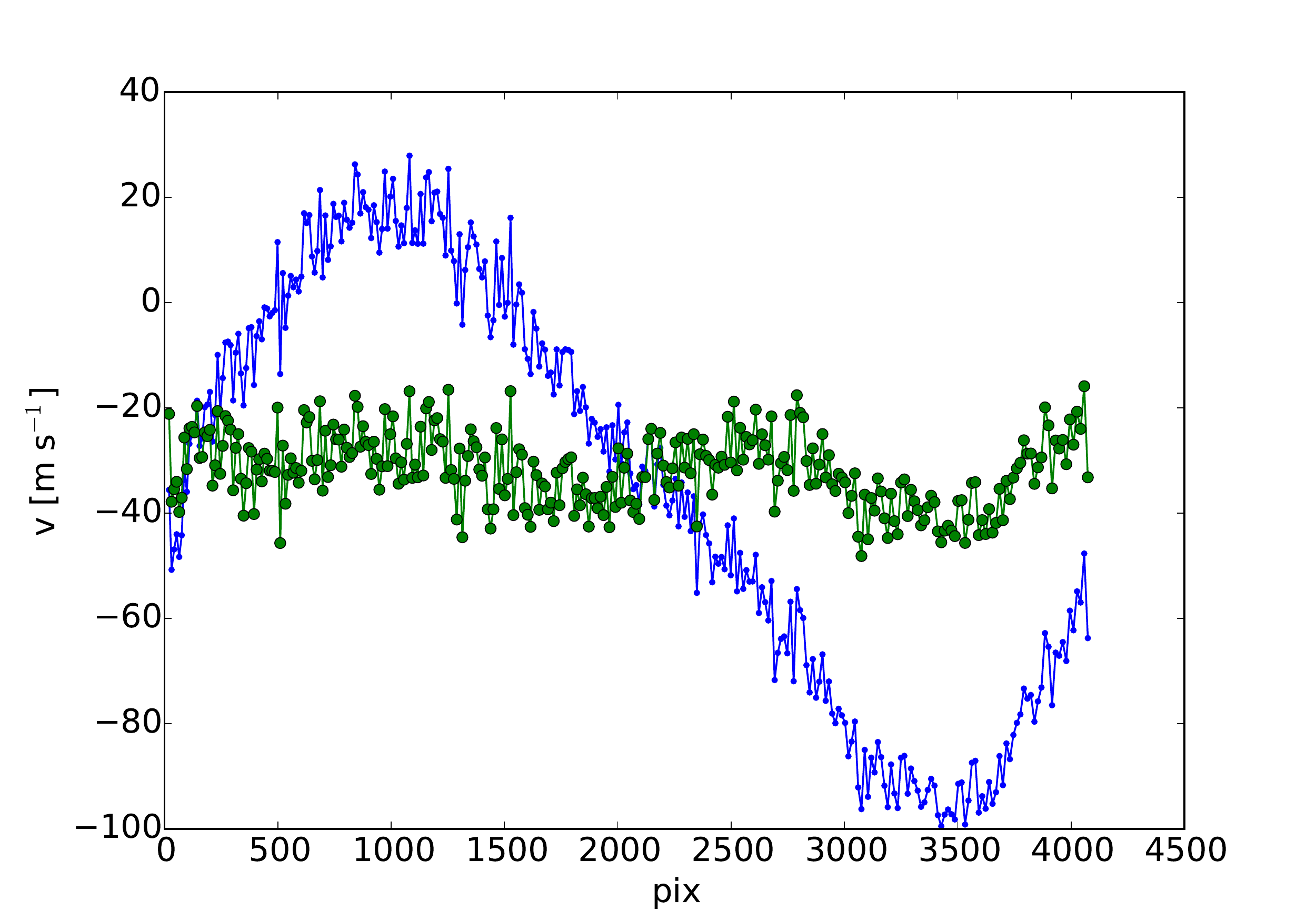}}

  \bigskip

   \subcaptionbox*{58th order index}[.3\linewidth][c]{%
    \includegraphics[width=.33\linewidth]{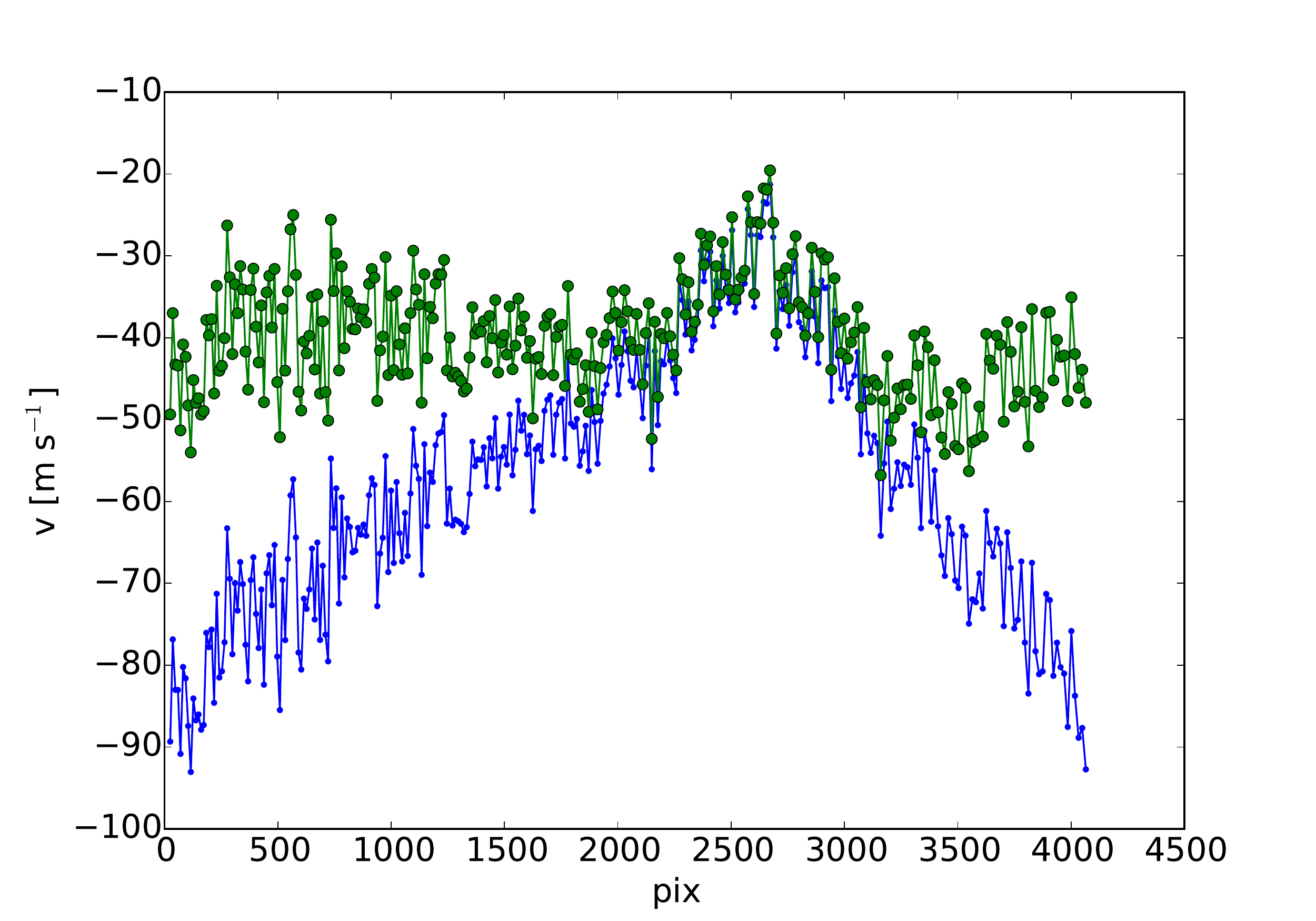}}\quad
  \subcaptionbox*{59th order index}[.3\linewidth][c]{%
    \includegraphics[width=.33\linewidth]{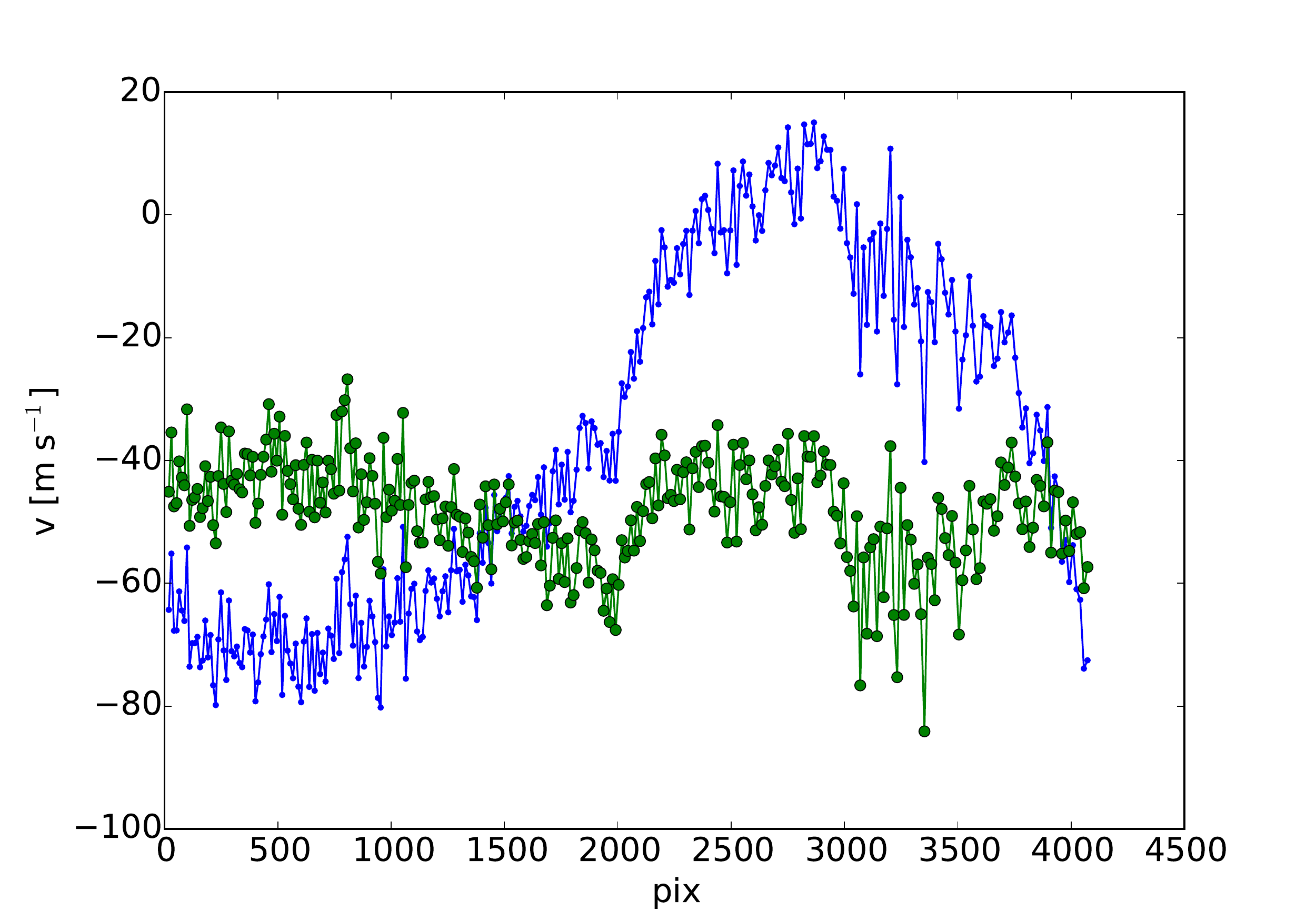}}\quad
  \subcaptionbox*{60th order index}[.3\linewidth][c]{%
    \includegraphics[width=.33\linewidth]{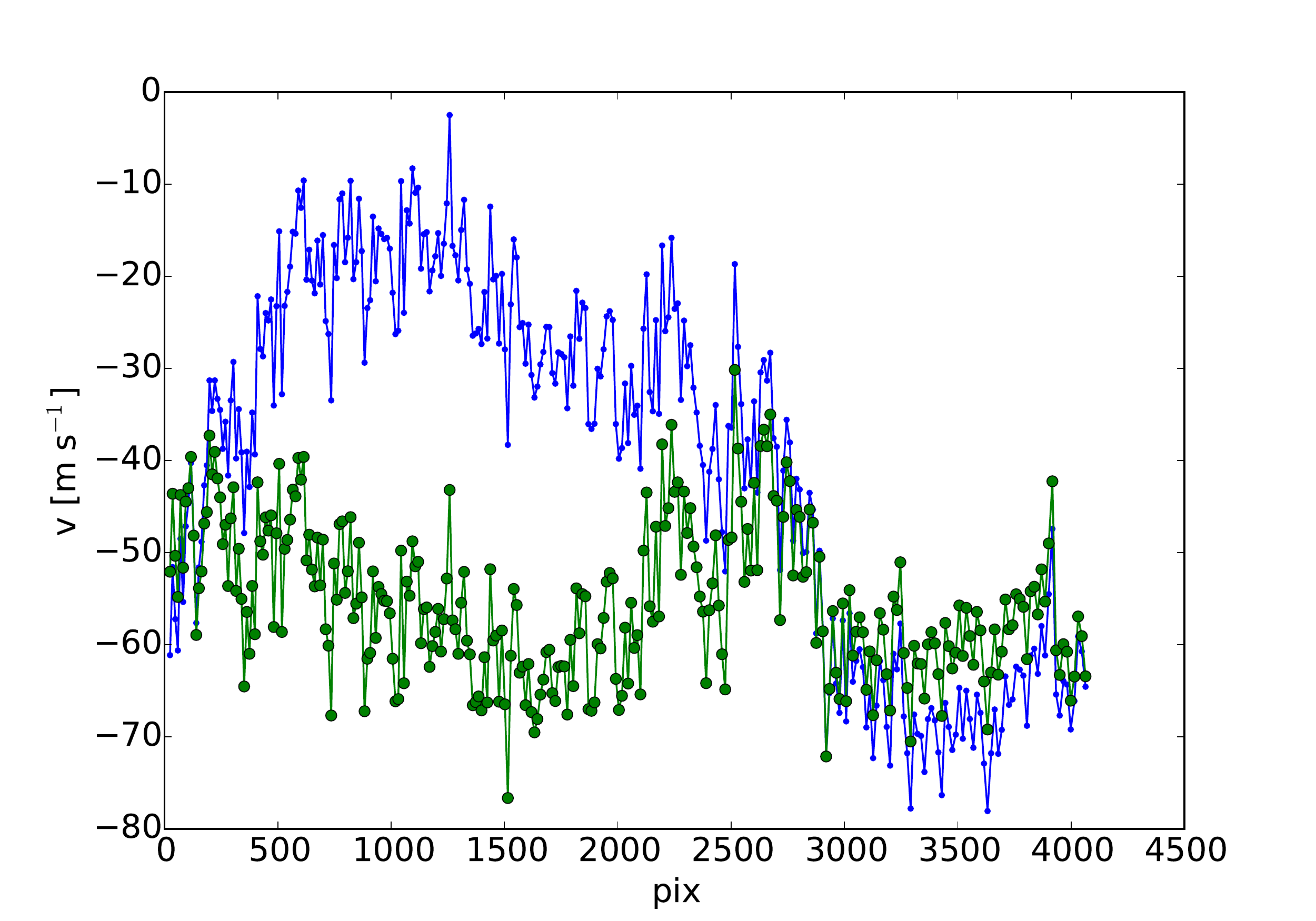}}
  \bigskip

   \subcaptionbox*{61st order index}[.3\linewidth][c]{%
    \includegraphics[width=.32\linewidth]{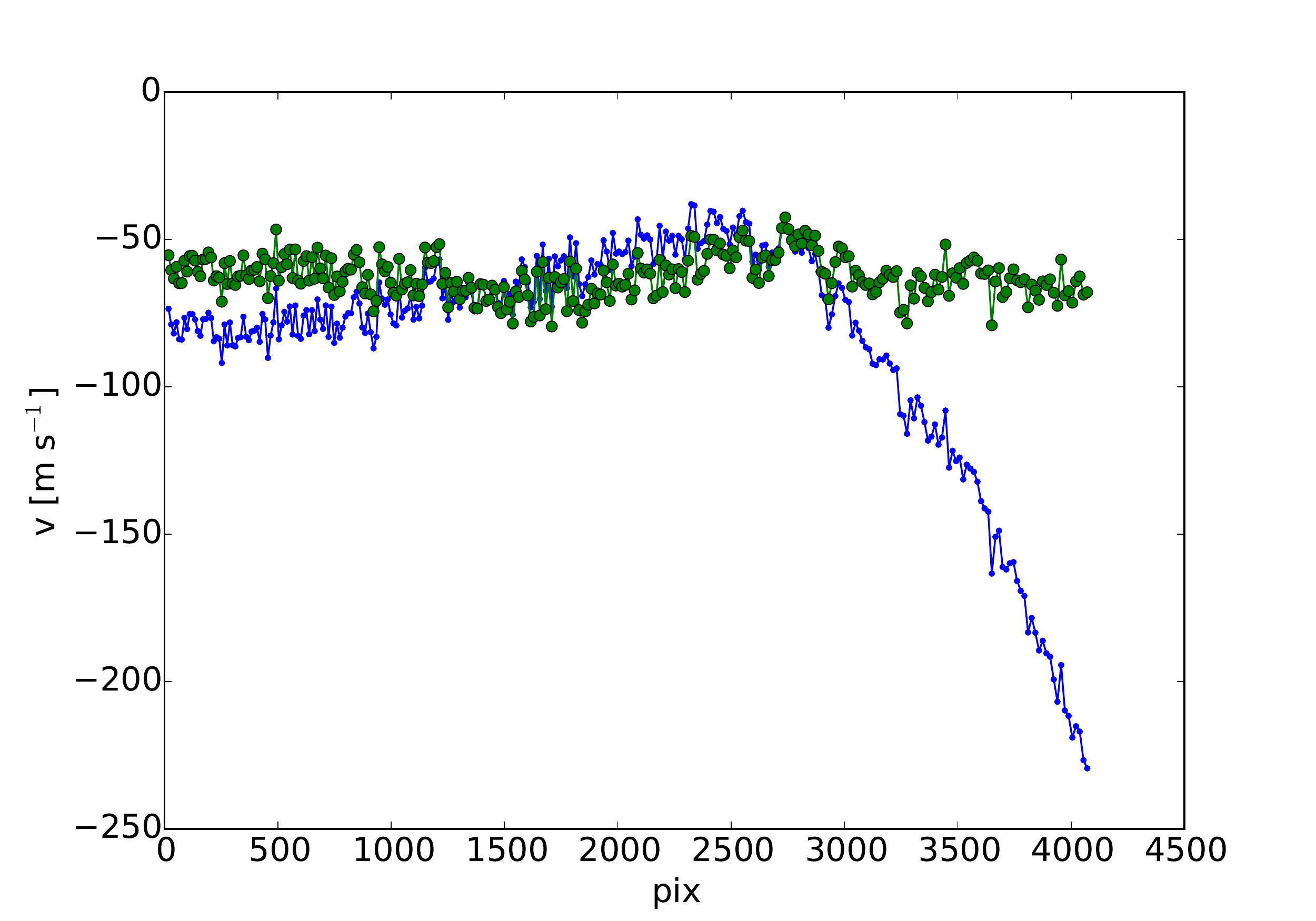}}\quad
  \subcaptionbox*{62nd order index}[.3\linewidth][c]{%
    \includegraphics[width=.32\linewidth]{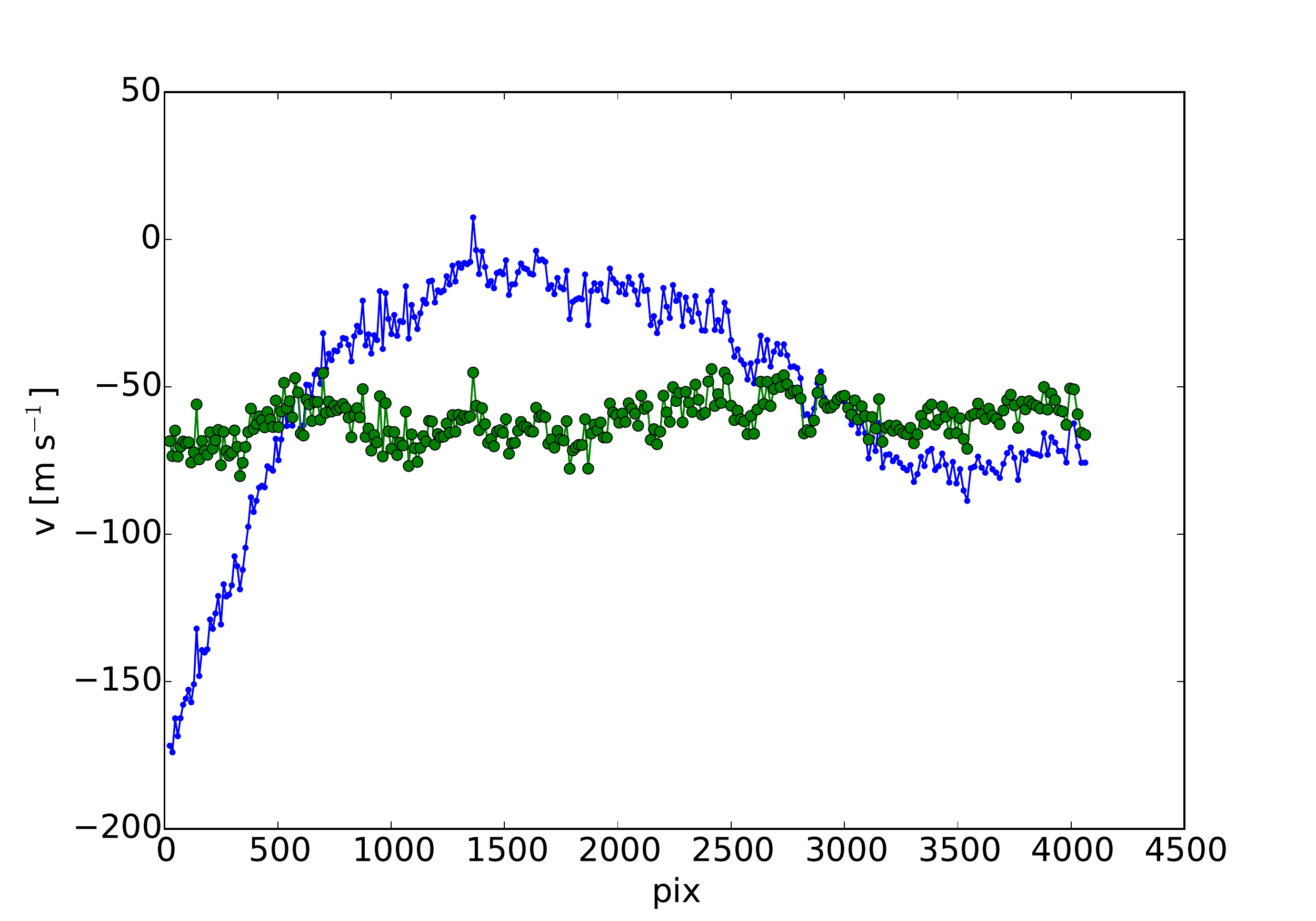}}\quad
  \subcaptionbox*{63rd order index}[.3\linewidth][c]{%
    \includegraphics[width=.32\linewidth]{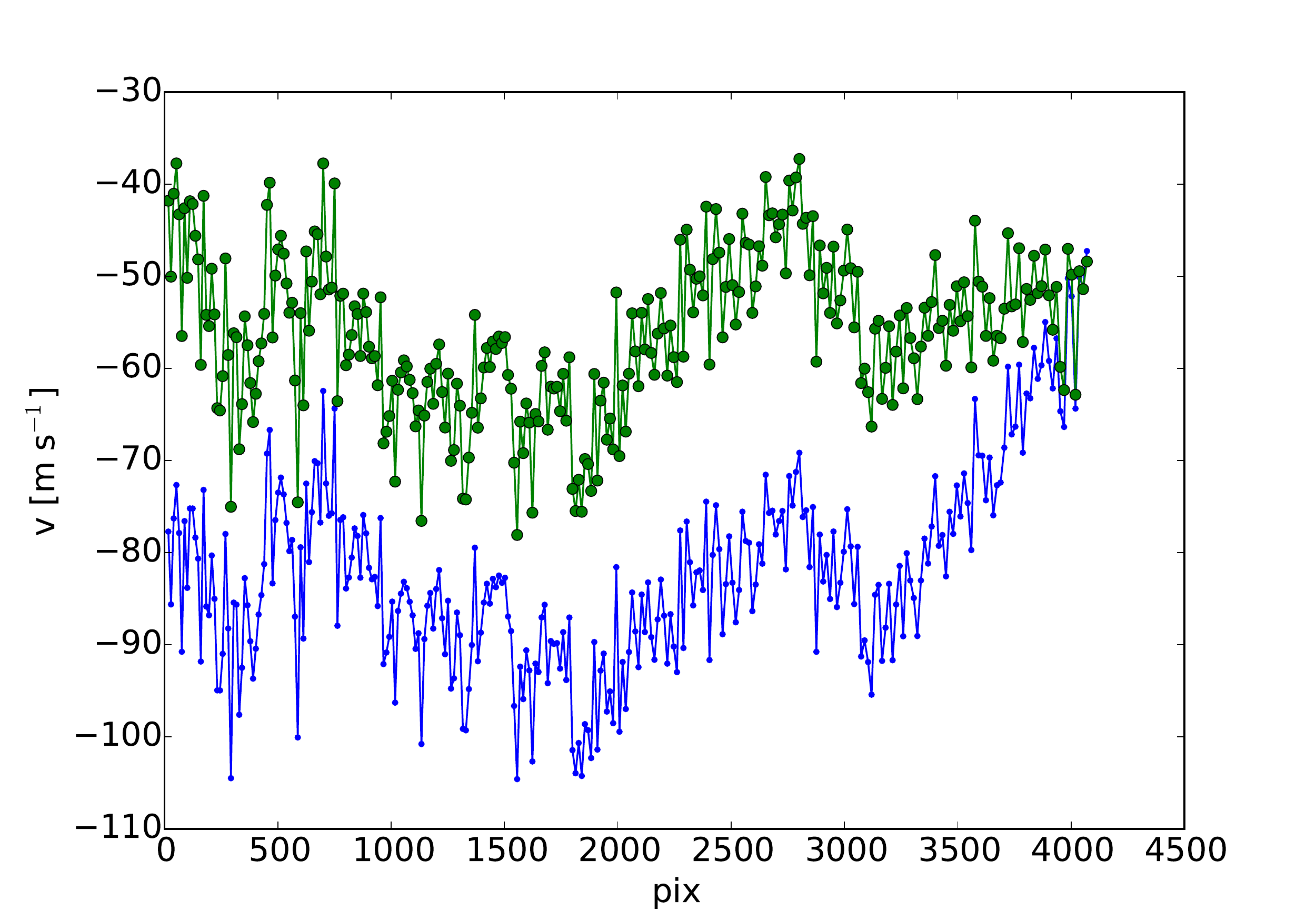}}
    \bigskip
   \subcaptionbox*{64th order index}[.3\linewidth][c]{%
    \includegraphics[width=.32\linewidth]{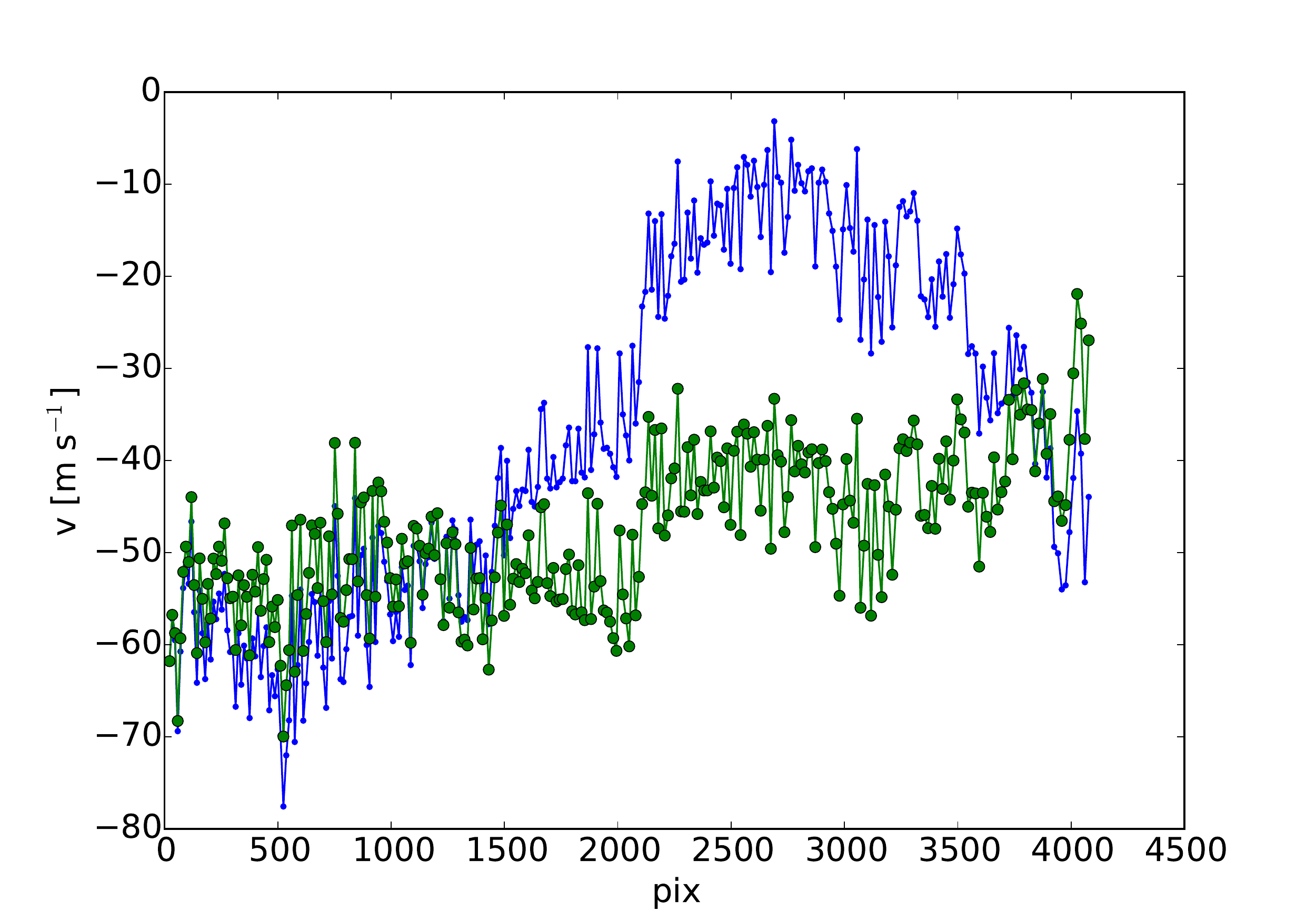}}\quad
  \subcaptionbox*{65th order index}[.3\linewidth][c]{%
    \includegraphics[width=.32\linewidth]{65.pdf}}\quad
  \subcaptionbox*{66th order index}[.3\linewidth][c]{%
    \includegraphics[width=.32\linewidth]{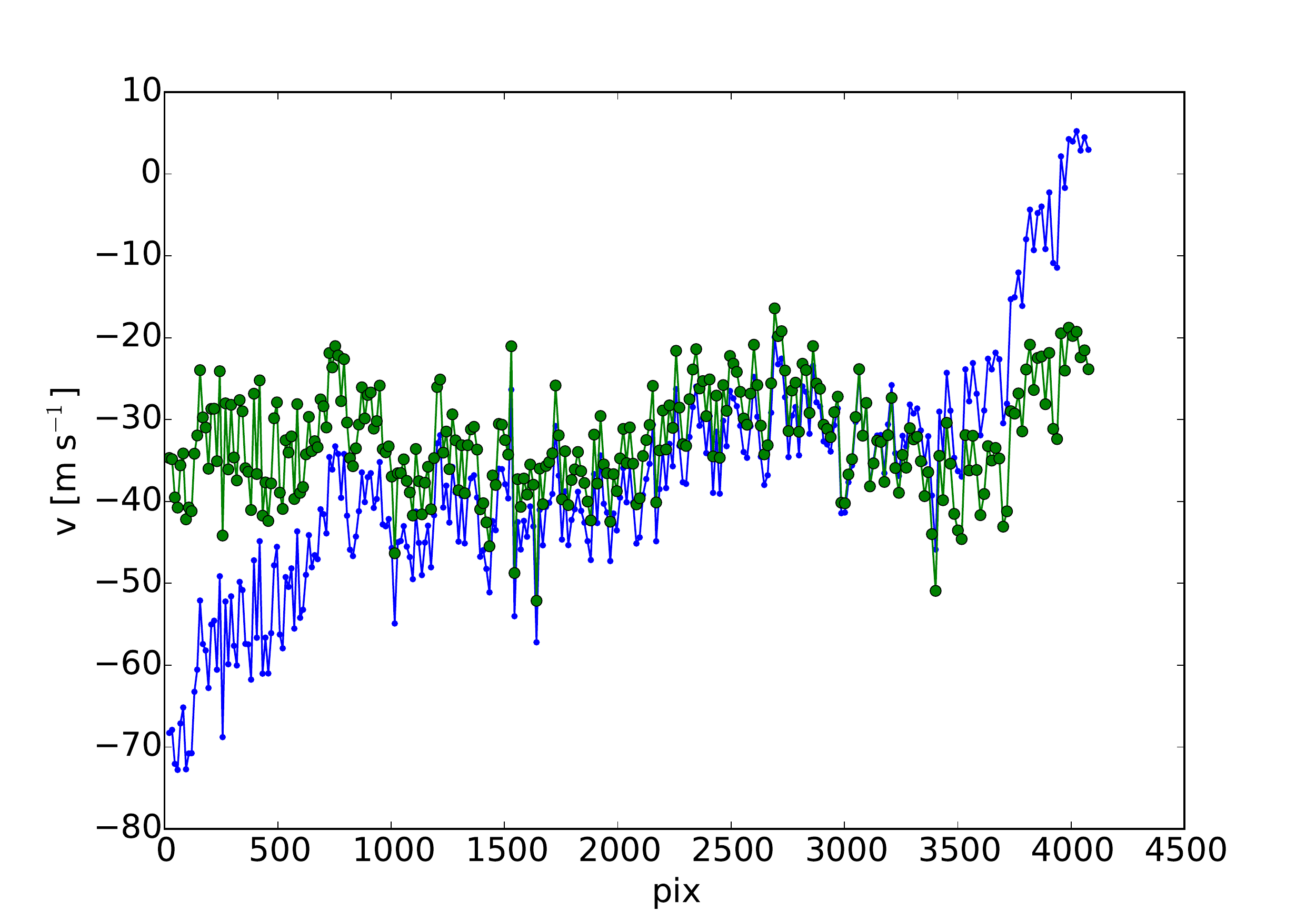}}
    \bigskip 
     \subcaptionbox*{67th order index}[.3\linewidth][c]{%
    \includegraphics[width=.32\linewidth]{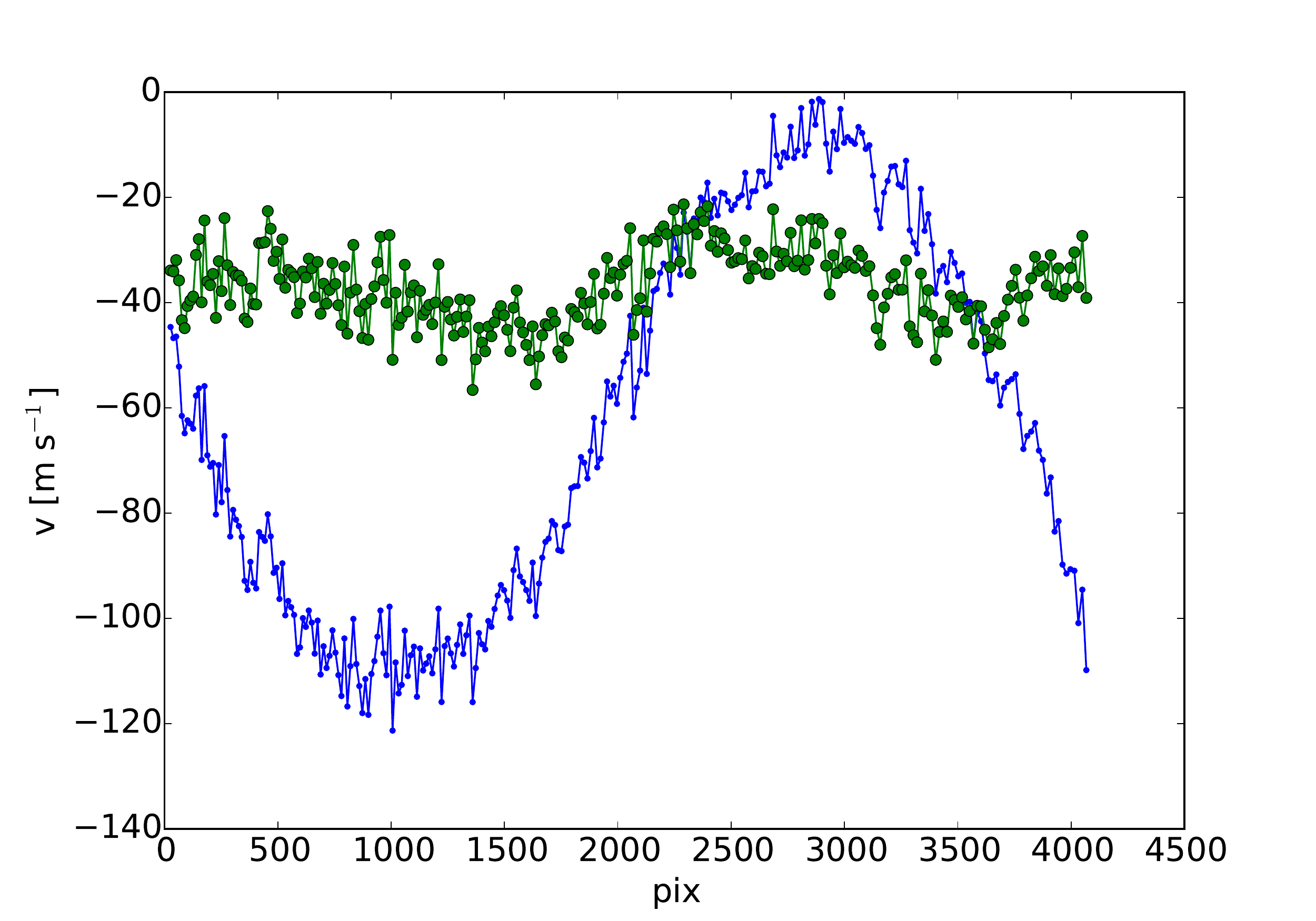}}\quad
  \subcaptionbox*{68th order index}[.3\linewidth][c]{%
    \includegraphics[width=.32\linewidth]{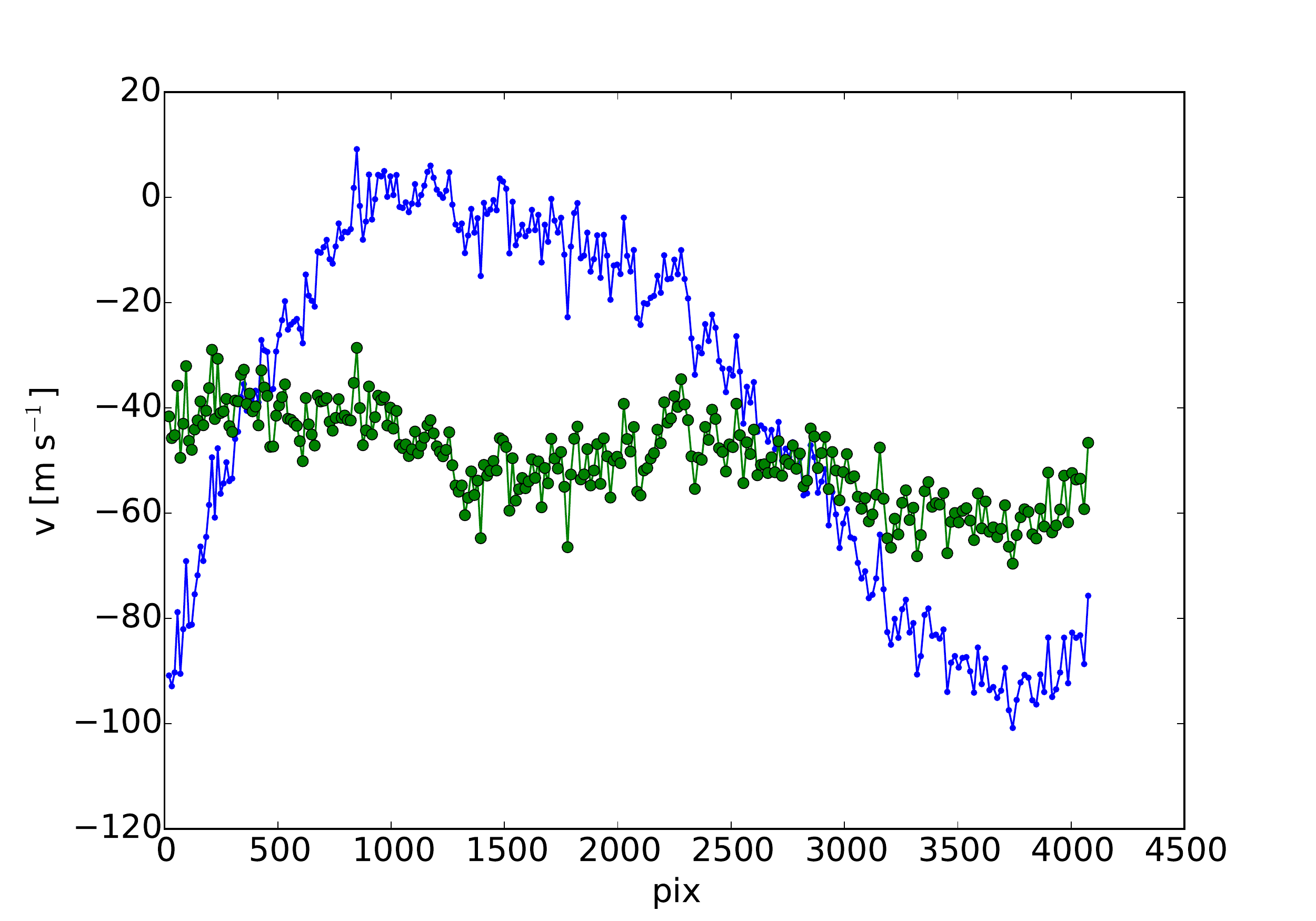}}\quad
  \subcaptionbox*{69th order index}[.3\linewidth][c]{%
    \includegraphics[width=.32\linewidth]{69.pdf}}
    \bigskip
      \label{fig:12}
\end{figure*}


\end{document}